\documentclass[12pt,amsmath,amssymb,aps]{revtex4}
\usepackage[colorlinks=true, citecolor=red, linkcolor=blue,urlcolor=blue ]{hyperref}
\usepackage{graphicx}
\usepackage{dcolumn}
\usepackage{bm}
\usepackage{epstopdf}
\usepackage{amssymb}
\usepackage{commath}
\usepackage{eqnarray,amsmath}
\usepackage{gensymb}
\usepackage{color}
\usepackage{physics}

\renewcommand{\vec}[1]{\mathbf{#1}}

\begin{document}
\title{Leading-order Stokes flows near a corner}
\author{Justas Dauparas}
\author{Eric Lauga}
\email{e.lauga@damtp.cam.ac.uk}

\affiliation{Department of Applied Mathematics and Theoretical Physics, University of Cambridge, CB3 0WA, United Kingdom}
\date{\today}
\begin{abstract}
Singular solutions of the Stokes equations play important roles in a variety of fluid dynamics problems. They allow the calculation of exact flows,  are the basis of  the boundary integral methods  used in numerical computations, and can be exploited  to derive  asymptotic flows in a wide range of physical problems. The most fundamental singular solution is the flow's Green function due to a point force, termed the Stokeslet. Its expression is classical both in free space and near a flat surface. Motivated by problems in biological physics occurring near corners, we derive in this paper the asymptotic behaviour for the Stokeslet both near and far from a corner geometry by using complex analysis on a known double integral solution for corner flows. We investigate all possible orientations of the point force relative to the corner and all corner geometries from acute to obtuse. The case of salient corners is also addressed for point forces aligned with both walls. We use experiments on  beads sedimenting in corn syrup to qualitatively test the applicability of our results. The final results and scaling laws will allow to address the role of hydrodynamic interactions in  problems  from colloidal science to microfluidics and biological physics. 
\end{abstract}

\maketitle

%\tableofcontents
%%%%%%%%%%%%%%%%%%%%%%%%%%%%%%%
%%%%%%%%%%%%%%%%%%%%%%%%%%%%%%%
%%%%%%%%%%%%%%%%%%%%%%%%%%%%%%%
\section{Introduction}

The Stokes equations of motion, which are linear and time-reversible, govern the dynamics of incompressible flow at low Reynolds number \cite{happelbook,kimbook}. The Green's function for the Stokes equations is interpreted physically as the flow due to a point force and called the Stokeslet \cite{pozrikidis_BI}.  Since most flows of practical interest happen near boundaries, a lot of work has focused on the modification of  the Stokeslet flow in simple geometries. Blake \cite{blake1971note} derived the flow near a flat no-slip boundary and interpreted the solution as due to a superposition of  image flow singularities; further work from the same group focused on higher-order singularities near no-slip walls \cite{blake1974fundamental}.  
Liron \& Mochon \cite{liron1976stokes} derived the flow due to a point force between two parallel walls. The solution is a complicated integral, but they  found that the leading-order flow decays exponentially if the force is perpendicular to the walls; if instead the force is parallel to the walls then the flow component along the  walls behaves as a source dipole while the perpendicular one decays exponentially. Mathijssen \cite{mathijssen2016hydrodynamics} et al. derived a numerically tractable approximation to the three-dimensional flow fields of a Stokeslet within a viscous film between a parallel no-slip surface and a no-shear interface and applied it to understand flow fields generated by swimmers.  Liron \& Shahar worked out the flow due to a Stokeslet placed at arbitrary location and orientation inside a pipe \cite{liron1978stokes}. 
 They obtained  that all velocity components decay exponentially to zero upstream or downstream away from the Stokeslet for all locations and orientations of the point force. The same results also hold for pressure fields of Stokeslets perpendicular to the axis of the pipe. In contrast, a Stokeslet which is parallel to the axis of the pipe raises the pressure difference between $-\infty$ to $+\infty$ by a finite  amount. The flow in infinite, semi-infinite and finite cylinders was further analysed by Blake \cite{blake1979generation} while the flow inside a cone was addressed by Ackerberg \cite{ackerberg1965viscous}.

Flows in corner geometries have attracted significant interest throughout the years. Dean \& Montagnon as well as Moffatt considered the general Stokes flow near a two-dimensional (2D) corner \cite{dean1949steady,moffatt1964viscous}. It was found that when either one,  or both,  of the corner boundaries is a  no-slip wall and when the corner angle is less than a  critical value, any flow sufficiently near the corner consists of a sequence of eddies of decreasing size and rapidly decreasing intensity. These eddies do not appear in  the corner flow  bounded by two plane free surfaces.
Recently, Crowdy \& Brzezicki \cite{crowdy2017analytical} used the complex variable formulation of Stokes flows in 2D to solve a biharmonic equation for the   streamfunction  generated by a   stresslet singularity   satisfying no-slip boundary conditions on a wedge. Unfortunately, such   techniques do not work in the case of   three-dimensional (3D) flows.

The flows near   3D corners has been analysed by Sano \& Hasimoto. They considered the slow motion of a spherical particle in a viscous fluid bounded by two walls \cite{sano1976slow,sano1977slow,sano1978effect} and used the method of reflections to obtain the leading-order effect of the wall. In their analysis, Sano \& Hasimoto used the general solution of the Stokes equations of motion \cite{imai1973ryutai} and reduced the whole problem to a mixed boundary-value problem determining at most four harmonic functions. Their 1980 review paper  summarised the work up to that point \cite{hasimoto1980stokeslets}. Later, these calculations were extended to the motion of a small particle in a viscous fluid in a salient wedge \cite{kim1983effect} and near a semi-infinite plane \cite{hasimoto1983effect}.  A special case when the point force is applied perpendicularly to the bisector of the wedge was addressed  by Sano \& Hasimoto \cite{sano1980three}.  If two plane walls intersecting at an angle $2\alpha$ which is less than $146.31^{\circ}$ then an infinite sequence of eddies appear. The intensities and sizes of adjacent eddies decrease in geometric progression at the same common ratios as those in the two-dimensional case analysed by Moffatt \cite{moffatt1964viscous}, but the streamlines are different from the two-dimensional ones. 
Further analysis on a class of three-dimensional corner eddies was done by Shankar who analysed flows which are symmetric and antisymmetric about the plane bisecting the corner \cite{shankar1998class}. It was found that the eigenvalues, $\lambda$, which determine the structure of the corner eddy fields satisfy the same equation, $\sin{\lambda\alpha}=\pm\lambda\sin{2\alpha}$, that arises in the corresponding two-dimensional case, but the three-dimensional velocity fields are quite different. A weakly three-dimensional stirring mechanism at some distance from the corner was considered by Moffatt \& Mak \cite{moffatt1999corner}. They showed that, when the stirring is antisymmetric about the bisecting plane $\theta=0$, the flow near a corner has the same eddy structure as in the two-dimensional case, but when the stirring is symmetric, a non-oscillatory component is in general present which is driven by conditions far from the corner. Further analysis was done by Shankar on Stokes flow in a semi-infinite wedge \cite{shankar2000stokes} and on Moffatt eddies in the cone \cite{shankar2005moffatt}. Viscous eddies in a circular cone \cite{malyuga2005viscous} and Moffatt-type flows in a trihedral cone \cite{scott2013moffatt} were analysed too.

The present work was originally instigated by questions arising from  the world of biological physics.  Recent experiments showed how swimming bacteria stuck in corners can lead to net flows and can perform net mechanical work  on microscopic gears \cite{sokolov2010swimming}. More generally, swimming bacteria are seen to swim along corners \cite{diluzio05} and might also get stuck there. If a bacterium is near a corner but does not swim while its flagella continue to rotate and to impose net forces on the surrounding fluid \cite{berg2008coli,lauga2016bacterial}, what is the net flow produced?
At  leading order, if the bacteria are stuck, the problem would be equivalent to considering the flow due to a point force and a point torque applied to the fluid in a corner geometry. 
In a different setting, the Stokeslet approximation is also important in studying hydrodynamic coupling of Brownian particles. It has been recently demonstrated that a confining surface can influence colloidal dynamics even at large separations, and that this coupling is accurately described by a leading-order Stokeslet approximation \cite{dufresne2000hydrodynamic}.

In the classical flow solution by Blake  near a no-slip wall, the total flow is given by superposing an original Stokeslet with hydrodynamic images consisting of a Stokeslet, a source doublet, and a Stokeslet doublet  \cite{blake1971note}. When the point force is parallel to the surface,   the net flow rate, $Q$, through any plane perpendicular to the direction of the Stokeslet is equal to the magnitude of the  point force, $F$, times the height above the wall, $h$, and divided by fluid viscosity, $\mu$, i.e.~$Q=Fh/(\pi\mu)$ \cite{blake1971note}.  In this paper, we pose the question: can we calculate the dominant flows due to a point force near any corner? Using cylindrical coordinates with $z$ along the edge direction, what is the flux $Q=\int_S u_z rdr d\theta$ generated by a point force, $(F_r,F_{\theta},F_z)$, located in a corner with angle $2\alpha$? By symmetry, only the axial force component, $F_z$, can generate a flux and by  conservation of  mass, it is independent of $z$, i.e.~$Q$ is either zero, a finite constant, or infinite. Therefore, it is sufficient to determine  the value of $u_z$ for large values of $z$ in order to calculate the flux. 
%Once we know the leading order flow, can we build an image system in terms of fundamental singularities? 

Exact solutions derived \cite{sano1978effect} for a point force of arbitrary orientation and at arbitrary location near a corner  are in a double integral form, involving the Fourier transform and the Kontorovich-Lebedev transform \cite{jones1980kontorovich, jones1986acoustic, sneddon1972use}.  In this paper we combine asymptotic expansions with complex analysis in order to describe the leading-order flows close and far away from the corner, for  all possible orientations of the point force relative to the orientation of the corner and all corner geometries from acute to obtuse (we also include the case of salient corners when the point force is parallel to both walls). Furthermore, we conduct  experiments by dropping small spheres in corn syrup next to a corner along both the axial and radial directions and obtain good qualitative agreement with our asymptotic results. The final results and scaling laws, summarised in Fig.~\ref{fig:12}, will allow to investigate the role of hydrodynamic interactions in a variety of physical and biological  problems.

%%%%%%%%%%%%%%%%%%%%%%%%%%%%%%%
%%%%%%%%%%%%%%%%%%%%%%%%%%%%%%%
%%%%%%%%%%%%%%%%%%%%%%%%%%%%%%%
\section{Integral solution for a point force in a corner}
\begin{figure}[t] 
 \centering      
 \includegraphics[width=0.65\textwidth]{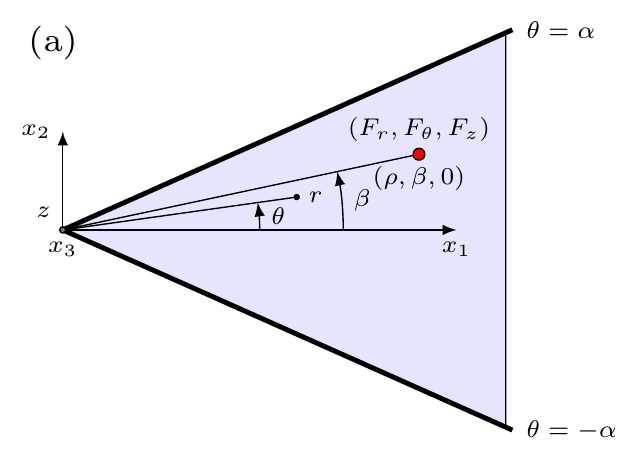}
 \includegraphics[width=0.343\textwidth]{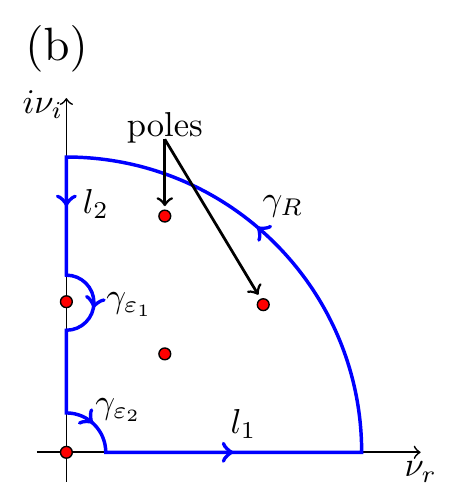}
       \caption{(a) Definition sketch for a point force $\vec{F}=(F_r,F_{\theta},F_z)$ located at $(r=\rho, \theta=\beta, z=0)$ and two plane no-slip walls at $\theta=\pm \alpha$. Cartesian co-ordinates $(x_1,x_2,x_3)$ and cylindrical co-ordinates $(r,\theta,z)$ are defined. (b) Contour integration in the complex plane which is used to evaluate the harmonic functions $\phi_n~ (n=1,...,4)$. The full Stokes flow is decomposed into flows due to individual poles with increasing imaginary part, each of them satisfying the no-slip and incompressibility conditions.}
   \label{fig:1}
\end{figure}

Throughout our analysis, the Reynolds number stays small as shown in Appendix~\ref{reynoldsn}, and thus it is appropriate to focus solely on the Stokes equations. We  consider a Stokeslet located in an incompressible viscous fluid bounded by two plane walls.   The exact solution is given in a double integral form by  Sano \&   Hasimoto \cite{sano1978effect}. We use Cartesian co-ordinates $(x_1,x_2,x_3)$ so that the $x_3$ axis coincides with the edge, and  the $x_1$ axis lies on the bisector of the wedge, as illustrated in Fig.~\ref{fig:1}a. We also introduce cylindrical co-ordinates $(r,\theta,z)$, where the $z$ axis correspond to the $x_3$ axis and the $\theta=0$ plane to the $x_2=0$ plane. Consider a point force $\vec{F}=(F_r,F_{\theta},F_z)$ located at position $P(r=\rho, \theta=\beta, z=0)$ in cylindrical co-ordinates. The two plane no-slip walls are located at $\theta=\pm \alpha$ where $0<2\alpha<\pi$. 
The governing equations and boundary conditions are
\begin{align}
0&=-\nabla p+\mu \nabla^2\vec{u}+\vec{F}\delta(\vec{x}-\vec{x}_P),
\\\nabla \cdot \vec{u}&=0,~
 \vec{u}=0 \text{ at infinity and on the walls}~\theta=\pm \alpha, 
\end{align}
where 
$\vec{F}=(F_r,F_{\theta},F_z)$ is located at $\vec{x}_P=P(r=\rho, \theta=\beta, z=0)$.

The general solution can be written in terms of four harmonic functions $\phi_n~ (n=1,...,4)$ \cite{tran1982general} satisfying
\begin{align}
\vec{u}&=\nabla (\vec{x}\cdot\bm{\phi}+\phi_4)-2\bm{\phi},~p=2\mu \nabla \cdot \bm{\phi},\\
\nabla^2 \phi_n&=0, ~(n=1,2,3,4),~\bm{\phi}=(\phi_1,\phi_2,\phi_3),~\vec{x}=(x_1,x_2,x_3).
\end{align}
These equations with appropriate boundary conditions can be solved decomposing harmonic functions $\phi_n=\phi_n^{(1)}+\phi_n^{(2)}$ into two parts. The first part, $\phi_n^{(1)}$, is the harmonic function satisfying the flow due to a point force $\vec{F}$ in an unbounded domain while the second part, $\phi_n^{(2)}$, is the correction such that the no-slip boundary conditions are satisfied by the functions $\phi_n$ on the two plane walls, i.e $\vec{u}(\theta=\pm\alpha)=\vec{0}$, written in components as
\begin{align}
u_r&=\left(r\frac{\partial}{\partial r}-1\right)(\phi_1 \cos{\theta}+\phi_2 \sin{\theta})+z\frac{\partial}{\partial r}\phi_3+\frac{\partial}{\partial r}\phi_4=0,\\
u_{\theta}&=\cos{\theta}\frac{\partial}{\partial \theta}\phi_1+\sin{\theta}\frac{\partial}{\partial \theta}\phi_2+\phi_1\sin{\theta}-\phi_2\cos{\theta}+\frac{z}{r}\frac{\partial}{\partial \theta}\phi_3+\frac{1}{r}\frac{\partial}{\partial \theta}\phi_4=0,\\
u_z&=r\frac{\partial}{\partial z}(\phi_1 \cos{\theta}+\phi_2 \sin{\theta})+\left(z\frac{\partial}{\partial z}-1\right)\phi_3+\frac{\partial}{\partial z}\phi_4=0.
\end{align}

Since the solution for  a Stokeslet in an unbounded domain is known, the function  $\phi_n^{(1)}$ is known. In order to find $\phi_n^{(2)}$ we Fourier transform the problem with respect to $z$
\begin{align}
\Phi_n^{(2)}(r,\theta,k)=\int_{-\infty}^{\infty} e^{ikz} \phi_n^{(2)}(r,\theta,z)~dz,
\end{align}
followed by the Kontorovich-Lebedev transform \cite{jones1980kontorovich, jones1986acoustic, sneddon1972use} with respect to r
\begin{align}
\tilde{\Phi}_n^{(2)}(\nu,\theta,k)=\int_{0}^{\infty} r^{-1} K_{i\nu}(|k|r) \Phi_n^{(2)}(r,\theta,k)~dr,
\end{align}
where $K_{i\nu}(|k|r)$ is the modified Bessel function of imaginary order.
This leads to solving
\begin{align}
\left(\frac{d^2}{d\theta^2}-\nu^2\right)\tilde{\Phi}_n^{(2)}=0, ~~(n=1,2,3,4)
\end{align}
with no-slip boundary conditions at $\theta=\pm\alpha$. The solution is
\begin{align}
 \tilde{\Phi}_n^{(2)}=A_n\sinh{\theta\nu}+B_n\cosh{\theta\nu},
\end{align}
where the expressions for $A_n,~B_n$ are given in the Sano \& Hasimoto paper  for the acute and obtuse corners, i.e.~$0<2\alpha\leq\pi$ \cite{sano1978effect}.  %The salient corner case, i.e.~$\pi<\alpha<2\pi$, was solved by Kim \cite{kim1983effect}, but we are not analysing salient corners in this paper. 
From this, the final solution is written as the superposition $\phi_n=\phi_n^{(1)}+\phi_n^{(2)}$, where we find $\phi_n^{(2)}$ formally by applying the inverse transforms
\begin{align}
\phi_n^{(2)}(r,\theta,z)=\frac{1}{\pi^3}\int_{-\infty}^{\infty}  e^{-ikz}~dk \int_{0}^{\infty}  \nu \sinh{(\pi\nu)} K_{i\nu}(|k|r) \tilde{\Phi}_n^{(2)}(\nu,\theta,k) ~d\nu.
\end{align}

By linearity of the Stokes equations, the solutions due to point forces in radial, azimuthal and axial directions can be analysed separately. In what follows, we will address  each case by first evaluating the inverse Fourier transform and then by writing the inverse Kontorovich-Lebedev transform \cite{jones1980kontorovich, jones1986acoustic, sneddon1972use} as the contour integral, as illustrated schematically in Fig.~\ref{fig:1}b, the value of which will be  evaluated using the Residue theorem. Thus, the full flow field will be decomposed into the sum of the flow fields due to individual poles, each satisfying the  incompressibility and no-slip boundary conditions, and leading to a  series expansion of the flow and pressure fields.

%%%%%%%%%%%%%%%%%%%%%%%%%%%%%%%
%%%%%%%%%%%%%%%%%%%%%%%%%%%%%%%
%%%%%%%%%%%%%%%%%%%%%%%%%%%%%%%
\section{Point force parallel to both walls}
We first consider a point force parallel to both walls, i.e.~along the $z$ direction, the only configuration in which a net flux of fluid could be generated along the corner. 
\subsection{Exact integral solution}
Consider a point force in the $z$ direction $\vec{F}=(0,0,F_z)$ located at $P(r=\rho, \theta=\beta, z=0)$ in cylindrical co-ordinates. The solution is given by $\phi_n=\phi_n^{(1)}+\phi_n^{(2)}$ where we find $\phi_n^{(2)}$ by applying the inverse transform for $\tilde{\Phi}_n^{(2)}$
\begin{align}
\phi_n^{(2)}(r,\theta,z)=\frac{1}{\pi^3}\int_{-\infty}^{\infty}  e^{-ikz}~dk \int_{0}^{\infty}  \nu \sinh{(\pi\nu)} K_{i\nu}(|k|r) \tilde{\Phi}_n^{(2)}(\nu,\theta,k) ~d\nu.
\end{align}
The solution in an unbounded domain, $\phi_n^{(1)}$, is given by
\begin{align}
\phi_1^{(1)}&=\phi_2^{(1)}=\phi_4^{(1)}=0,~
\phi_3^{(1)}=c_z/R,\\
R^2&=\rho^2+r^2-2\rho r \cos(\theta-\beta)+z^2,~
c_z=-F_z/(8\pi\mu).
\end{align}
We analytically evaluate the inverse Fourier Transform using the recurrence relation \cite{gradshteyn2007table} for the modified Bessel function
\begin{align}
z\frac{d}{dz}K_{\nu}(z)+\nu K_{\nu}(z)&=-zK_{\nu-1}(z),
\end{align}
to obtain
\begin{align}
\label{eq:rf1}
\phi_1^{(2)}=&\frac{16}{\pi^2}\int_{0}^{\infty} S_{i\nu}f_1(\alpha,\beta,\theta,\nu)\sin{\alpha}\sinh{(\pi \nu)}\, d\nu,\\
\label{eq:rf2}
\phi_2^{(2)}=&-\frac{16}{\pi^2}\int_{0}^{\infty}S_{i\nu}
f_2(\alpha,\beta,\theta,\nu)\cos{\alpha}\sinh{(\pi \nu)}\, d\nu,\\
\label{eq:rf3}
\phi_3^{(2)}=&-\frac{4}{\pi^2}\int_{0}^{\infty}I_{i\nu}
f_3(\alpha,\beta,\theta,\nu)\, d\nu,\\
\phi_4^{(2)}=&0,
\end{align}
where the functions $f_n$ are given by
\begin{align}
\label{eq:f1}
f_1=&\frac{\sinh{(\theta \nu)\cosh{(\alpha \nu)q^A}}}{\Delta^+(\nu;\alpha) D^-(\nu;\alpha)}+\frac{\cosh{(\theta \nu)\sinh{(\alpha \nu)q^S}}}{\Delta^-(\nu;\alpha) D^+(\nu;\alpha)},\\ 
\label{eq:f2}
f_2=&\frac{\sinh{(\theta \nu)\cosh{(\alpha \nu)q^S}}}{\Delta^-(\nu;\alpha) D^+(\nu;\alpha)}+\frac{\cosh{(\theta \nu)\sinh{(\alpha \nu)q^A}}}{\Delta^+(\nu;\alpha) D^-(\nu;\alpha)},\\ 
\label{eq:f3}
f_3=&\frac{\sinh{(\theta \nu)\sinh{(\beta \nu)\sinh{[(\pi-\alpha)\nu]}}}}{\sinh{(\alpha \nu)}}+\frac{\cosh{(\theta \nu)\cosh{(\beta \nu)\cosh{[(\pi-\alpha)\nu]}}}}{\cosh{(\alpha \nu)}},
\end{align}
with
\begin{align}
\label{eq:app1}
\xi&=\frac{\hat{r}^2+\hat{z}^2+1}{2\hat{r}}, \hat{r}=r/\rho, \hat{z}=z/\rho,\\
I_{i\nu}&=\rho\int_0^{\infty}  \cos{(kz)} K_{i\nu}(k\rho)K_{i\nu}(kr) dk=\frac{\pi^2}{4}\hat{r}^{-1/2}\sech{(\pi \nu)}P_{i \nu -1/2}(\xi),\\
S_{i\nu}&=\rho^2\int_0^{\infty}  k \sin{(kz)} K_{i\nu}(k\rho)K_{i\nu}(kr) dk=-\rho\frac{\partial}{\partial \hat{z}}I_{i\nu},\\
S_{i\nu}&=\frac{\pi}{4} \hat{r}^{-3/2}\hat{z}(\xi^2-1)^{-1/2}\Gamma{\left(\frac{3}{2}+i\nu\right)}\Gamma{\left(\frac{3}{2}-i\nu\right)}P_{i\nu-1/2}^{-1}(\xi),\\
q^A&=\sin{\alpha}\cosh{\alpha \nu}\cos{\beta}\sinh{\beta\nu}-\cos{\alpha}\sinh{\alpha \nu}\sin{\beta}\cosh{\beta\nu},\\
q^S&=\sin{\alpha}\sinh{\alpha \nu}\cos{\beta}\cosh{\beta\nu}-\cos{\alpha}\cosh{\alpha \nu}\sin{\beta}\sinh{\beta\nu},\\
\Delta^{\pm}(\nu;\alpha)&=\sinh{2\alpha \nu}\pm \nu\sin{2\alpha},~D^{\pm}(\nu;\alpha)=\cosh{2\alpha \nu}\pm \cos{2\alpha}.
\end{align}
In the equations above, $K_{i\nu}(|k|r)$ is the modified Bessel function of imaginary order, $P_m^n(z)$ is the associated Legendre function, $\Gamma{(z)}$ is the Gamma function. The integrals $I_{i\nu},S_{i\nu}$ can be found in the classic book of tables of Gradshteyn \& Ryzhik \cite{gradshteyn2007table}. 
Both associated Legendre functions $P_{i\nu-1/2}(\xi)$ and $P_{i\nu-1/2}^{-1}(\xi)$ have arguments in $\xi=(\hat{r}^2+\hat{z}^2+1)/(2\hat{r})$. Notice that if $\hat{r}=1$ and $\hat{z}=0$, i.e.~we are on the circle which goes through the location of the point force, then $\xi=1$, but otherwise $\xi>1$. Writing the Legendre functions in terms of hypergeometric series $F(a,b;c;\xi)$ for $\xi>1$ we have \cite{gradshteyn2007table}
\begin{align}
\label{eq:lege1}
P_{i \nu-1/2}^{-1}(\xi)&=\frac{\Gamma{\left(i \nu\right)}(\xi^2-1)^{-1/2}(2\xi)^{i\nu+1/2}}{2\sqrt{\pi}\Gamma{\left(\frac{3}{2}+i \nu\right)}}F\left(-\frac{i\nu}{2}+\frac{1}{4},-\frac{i\nu}{2}-\frac{1}{4},1-i\nu,\xi^{-2}\right)+\text{c.c.},\\
\label{eq:lege2}
P_{i \nu-1/2}(\xi)&=\frac{\Gamma{\left(i \nu\right)}}{\sqrt{\pi}\Gamma{\left(\frac{1}{2}+i \nu\right)}}(2\xi)^{i\nu-1/2}F\left(-\frac{i\nu}{2}+\frac{3}{4},-\frac{i\nu}{2}+\frac{1}{4},1-i\nu,\xi^{-2}\right)+\text{c.c.},\\
F(a,b;c;z)&=1+\frac{a\cdot b}{c \cdot 1}z+\frac{a(a+1)b(b+1)}{c \cdot 1\cdot 2}z^2+\frac{a(a+1)(a+2)b(b+1)(b+2)}{c \cdot 1\cdot 2\cdot 3}z^3+...
\end{align}
where the abbreviation $\text{c.c.}$ stands for `complex conjugate'. These expressions will be useful in order to find the asymptotic behaviour when $\xi\gg1$. Far away from the point force we have $L=(\hat{r}^2+\hat{z}^2)^{1/2}\gg1$, while very close to the corner, i.e.~$\hat{r}\ll1$; in both cases we thus obtain $\xi\gg1$. The structure of $\xi$ suggests that there will be similarities between the nature of the flow very close to the corner and that far away from it because expansions have to match in the case when $\hat{z}\gg1$, $\hat{r}\ll1$, as was already observed for   2D corner flows by Moffatt~\cite{moffatt1964viscous}.

In addition, we note that the flow velocity has to satisfy the incompressibility condition
\begin{align}
\nabla\cdot\vec{u}=\frac{1}{\hat{r}}\frac{\partial (\hat{r}u_r)}{\partial \hat{r}}+\frac{1}{\hat{r}}\frac{\partial u_{\theta}}{\partial \theta}+\frac{\partial u_z}{\partial \hat{z}}=0.
\end{align}
This then leads to  scalings for the flow components far away from the singularity $u_r/L\sim u_{\theta}/L\sim u_{z}/L$ while close to the corner $u_r/r\sim u_{\theta}/r\sim u_{z}$. The flow close to the corner in axial direction is thus expected to be dominant in all cases where it is not identically equal to zero.

In order to compute the contributions of the residues, we may simplify the  form of the integral expressions. Setting $\rho=1, c_z=-F_z/(8\pi\mu)=1$ we have formally
\begin{align}
\label{eq:pz1}
\phi_1^{(2)}=&\Re\left\{\Res\left[S(\hat{r},\hat{z};\nu)f_1(\alpha,\beta,\theta,\nu)\sin{\alpha};\nu^{*}\right]\right\},\\
\label{eq:pz2}
\phi_2^{(2)}=&-\Re\left\{\Res\left[S(\hat{r},\hat{z};\nu)f_2(\alpha,\beta,\theta,\nu)
\cos{\alpha};\nu^{*}\right]\right\},\\
\label{eq:pz3}
\phi_3^{(2)}=&\Re\left\{\Res\left[\frac{I(\hat{r},\hat{z};\nu)}{\sinh{(\pi\nu)}}f_3(\alpha,\beta,\theta,\nu);\nu^{*}\right]\right\},\\
\phi_4^{(2)}=&0,
\end{align}
where
\begin{align}
S(\hat{r},\hat{z};\nu)&=8 \pi^{1/2}\hat{r}^{-3/2}\hat{z}\frac{(2\xi)^{i\nu+1/2}F\left(-\frac{i\nu}{2}+\frac{1}{4},-\frac{i\nu}{2}-\frac{1}{4};1-i\nu;\xi^{-2}\right)}{(\xi^2-1)}\frac{\Gamma{\left(\frac{3}{2}-i \nu\right)}}{\Gamma{\left(1-i \nu\right)}},\\
I(\hat{r},\hat{z};\nu)&=-4\pi^{1/2}\hat{r}^{-1/2}(2\xi)^{i\nu-1/2}F\left(-\frac{i\nu}{2}+\frac{3}{4},-\frac{i\nu}{2}+\frac{1}{4};1-i\nu;\xi^{-2}\right)\frac{\Gamma{\left(\frac{1}{2}-i \nu\right)}}{\Gamma{\left(1-i \nu\right)}},
\end{align}
where the functions $f_1,f_2,f_3$ are defined in equations \eqref{eq:f1}~--~\eqref{eq:f3}, and the the poles, $\nu^{*}$, are taken in the first quadrant of the complex plane (see Fig.~\ref{fig:1}b). If $\Re(\nu^{*})=0,~ \Im(\nu^{*})>0$ then take the half value of the residue by the Indentation lemma.

\subsection{Understanding the residues of the integrands}
The flow field far away from the point force, $L^2=\hat{r}^2+\hat{z}^2\gg1$, can be decomposed into flows decaying spatially with  increasing power laws $\sim 1/L^n$, with the  flow at each  order   satisfying both the no-slip boundary conditions and incompressibility. The flow field near the corner can similarly be written in the power series of $\hat{r}^n\ll1$. 

\begin{figure}[t!]
	\includegraphics[height=0.344\textwidth]{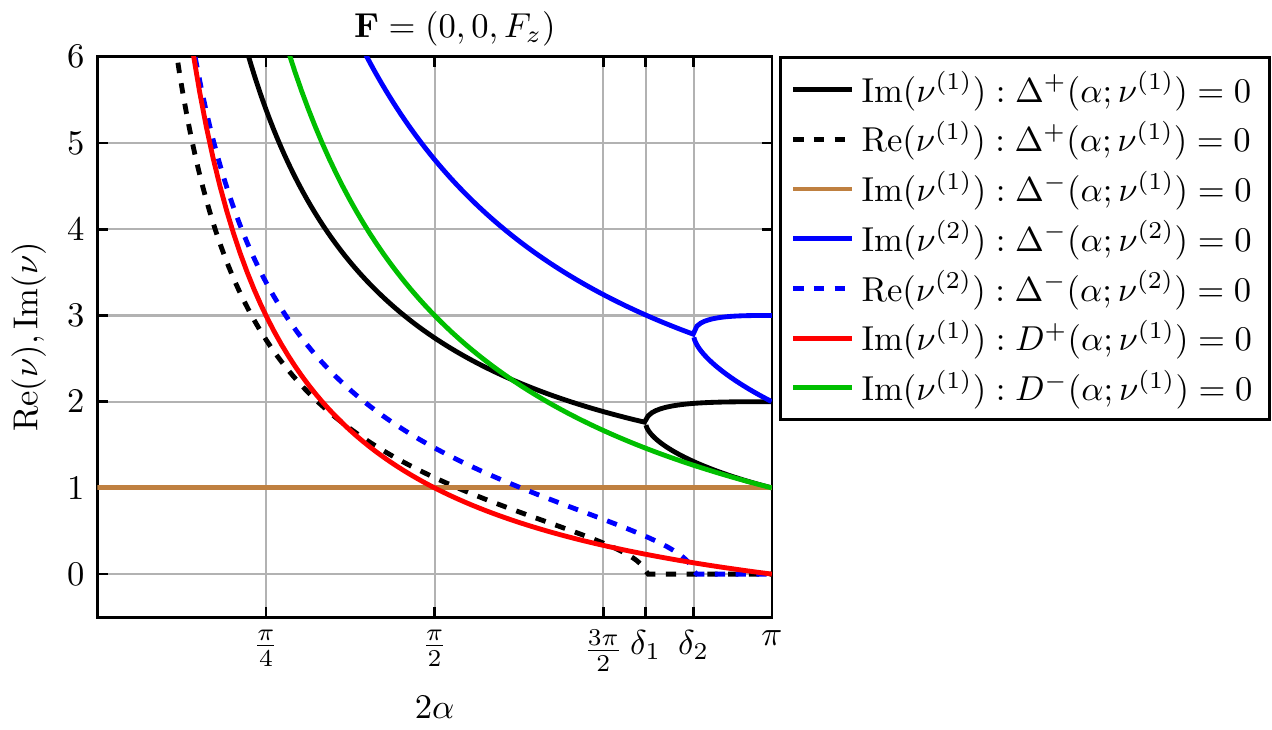}
	\caption{Real and imaginary parts of leading-order poles $\nu^*$ satisfying $\Delta^{\pm}(\nu^*;\alpha)=0$ and $D^{\pm}(\nu^*;\alpha)=0$ with minimal positive imaginary part. The real part of $\nu^*$, such that $\Delta^{\pm}(\nu^*;\alpha)=0$, becomes zero when $2\alpha=\delta_1=146.31^{\circ}$ and $2\alpha=\delta_2=159.11^{\circ}$, which corresponds to critical angles for antisymmetric and symmetric flows about the bisectoral plane $\theta=0$. If the real part of $\nu^*$ is not shown, it is equal to zero.}
	\label{fig:111}
\end{figure}

In order to compute the flow, we need to compute the residues of the integrands representing the harmonic functions $\phi_1^{(2)}, \phi_2^{(2)}, \phi_3^{(2)}$. 
Poles for $\phi_1^{(2)}, \phi_2^{(2)}$ are located at $\nu^*:\Re(\nu^*)\geq 0,~\Im(\nu^*)\geq 0$ such that
\begin{align}
\Delta^{\pm}(\nu^*;\alpha)&=\sinh{2\alpha \nu^*}\pm \nu^*\sin{2\alpha}=0,\\
D^{\pm}(\nu^*,\alpha)&=\cosh{2\alpha \nu^*}\pm \cos{2\alpha}=0.
\end{align}
Write $\nu^*$ as a sum of real $\nu_r$ and imaginary $\nu_i$ parts, i.e.~$\nu^*=\nu_r + i \nu_i$ then
\begin{align}
\Delta^{\pm}(\nu^*;\alpha)&=(\sinh{2\alpha \nu_r}\cos{2\alpha \nu_i}\pm\nu_r \sin{2\alpha})+i(\cosh{2\alpha \nu_r}\sin{2\alpha \nu_i}\pm\nu_i \sin{2\alpha}),\\
D^{\pm}(\nu^*;\alpha)&=(\cosh{2\alpha \nu_r}\cos{2\alpha \nu_i}\pm \cos{2\alpha})+i(\sinh{2\alpha \nu_r}\sin{2\alpha \nu_i}).
\end{align}
Assuming  that $0<2\alpha<\pi$ then
\begin{align}
D^{+}(\nu^*;\alpha)=0 \text{ when } \nu_r&=0, ~\nu_i^{(M)}=\frac{(2M-1)\pi}{2\alpha}\pm1:\nu_i\geq0 ~ (M=1,2,3,...),\\
D^{-}(\nu^*;\alpha)=0 \text{ when } \nu_r&=0, ~\nu_i^{(M)}=1,~\frac{M\pi}{\alpha}\pm1:\nu_i\geq0 ~ (M=1,2,3,...).
\end{align}
The flow associated with these poles will be non-oscillatory because $\nu_r=0$.
We cannot find analytically solutions of $\Delta^{\pm}(\nu^*;\alpha)=0$, except the obvious one, $\Delta^{-}(\alpha;\nu^*)=0$, when $\nu_r=0,~\nu_i=1$, but we can find them numerically, as shown in Fig. \ref{fig:111}. 
We call the flows arising from these poles eddy-type flows, because these flows can have an infinite number of eddies if the real part of the pole is non-zero, i.e.~$\nu_r\neq0$.

\begin{figure}[t!]  
  \includegraphics[height=0.344\textwidth]{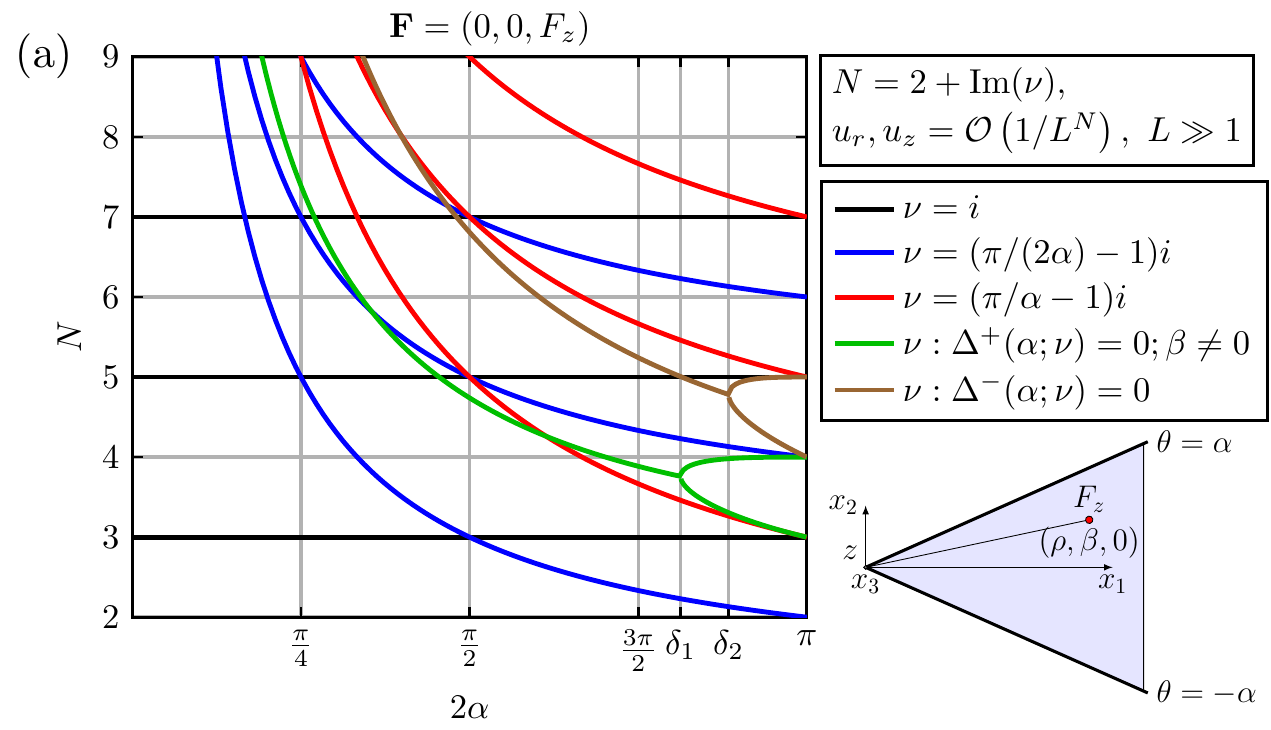}
   \includegraphics[height=0.344\textwidth]{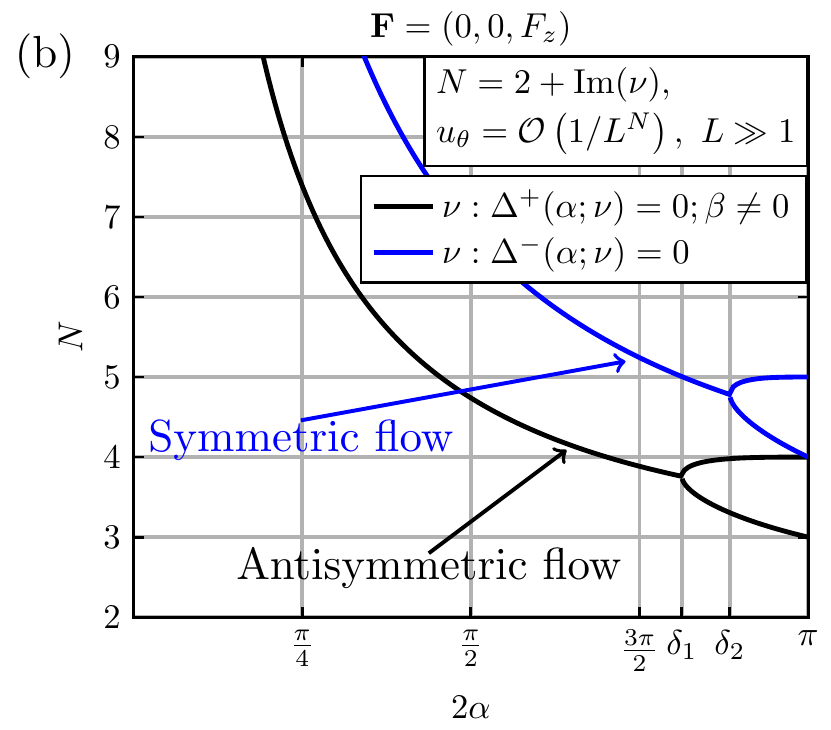}
     \includegraphics[height=0.344\textwidth]{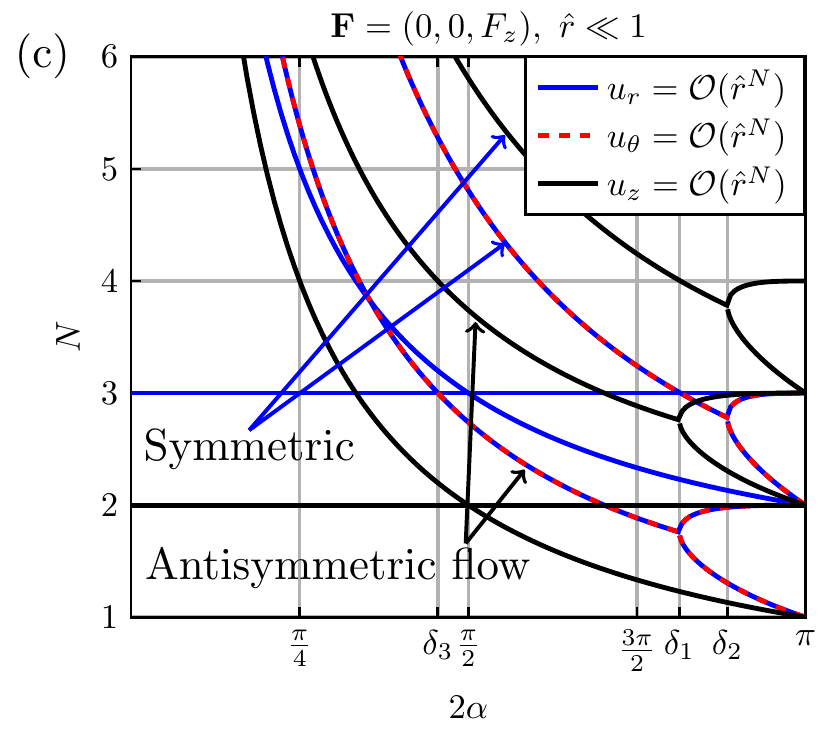}
       \caption{Decomposition of the full flow due to a point force in the axial direction, $\vec{F}=(0,0,F_z)$; (a) Flow in radial and axial directions, $u_r,u_z$; (b) Flow in the azimuthal direction $u_{\theta}$. The flows are at large distance away from the point force, i.e.~$L\gg1$. They decay as $\mathcal{O}(1/L^{2+\Im(\nu)})+\mathcal{O}(1/L^{4+\Im(\nu)})+...$ where   $\Im(\nu)$ is the imaginary part of the pole $\nu$.        The decay power $N$ is plotted against the full corner angle, $2\alpha$; (c) The flow decay close to the corner $\hat{r}\ll1$. The critical angles for antisymmetric and symmetric flows are $\delta_1=146.31^{\circ}$ and $\delta_2=159.11^{\circ}$. The angle $\delta_3=81.87^{\circ}$ indicates the point at which the leading-order radial flow changes.}
   \label{fig:2}
\end{figure}

Poles for $\phi_3^{(2)}$ are at $\nu^*:\Re(\nu^*)\geq 0,~\Im(\nu^*)\geq 0$ with
\begin{align}
\sinh{\pi\nu^*}&=0 \implies  \nu_r=0,~\nu_i^{(M)}=M,~(M=0,1,2,3,...),\\
\sinh{\alpha \nu^*}&=0 \implies \nu_r=0,~\nu_i^{(M)}=\frac{M\pi}{\alpha},~ (M=1,2,3,...),\\
\cosh{\alpha \nu^*}&=0 \implies \nu_r=0,~\nu_i^{(M)}=i\frac{(2M-1)\pi}{2\alpha},~ (M=1,2,3,...).
\end{align}
In order to visualise the decomposition of the full flow into flows arising from different poles as a function of the corner angle, $2\alpha$, see Fig.~\ref{fig:2} for $L\gg1$ and $\hat{r}\ll1$. From this figure one can read off the decay rate $N$, where $u_r,u_{\theta},u_{z}=\mathcal{O}(1/L^N)$, for a given angle and a given flow component in the far field Fig.~\ref{fig:2}(a-b) and near field Fig.~\ref{fig:2}c, where $u_r,u_{\theta},u_{z}=\mathcal{O}(\hat{r}^N)$.

\subsection{Leading-order flow for the acute corner, $0<2\alpha<\pi/2$}
Using the expressions obtained in \eqref{eq:pz1}-\eqref{eq:pz3}
 for the harmonic functions $\phi_1, \phi_2,\phi_3,\phi_4$, we can next evaluate the leading-order flow for the acute angle case, i.e.~when $0<2\alpha<\pi/2$. In this case the pole at $\nu=i$ contributes to the $\phi_1 =\mathcal{O}(1/L^3),~\phi_2=\mathcal{O}(1/L^3)$. The poles at $\nu=0,~i,~2i$ for $\phi_3^{(2)}$ exactly cancel the free-space Stokeslet given by $\phi_3^{(1)}$, i.e.~we have $\phi_3=\phi_3^{(1)}+\phi_3^{(2)}=\phi_4=0$ at leading order. The point force $\vec{F}=(0,0,F_z)$ is located at $(r,\theta,z)=(\rho,\beta,0)$. At large distances away from the point force, i.e.~for $L=(\hat{r}^2+\hat{z}^2)^{1/2}\gg1$, the flow is given by
\begin{align}
u_r&=\frac{5f(\alpha,\beta,\theta)\hat{r}^3\hat{z}}{(\hat{r}^2+\hat{z}^2)^{7/2}},\\
u_{\theta}&=\mathcal{O}\left(\frac{1}{L^{2+\Im(\nu^{*})}}\right) :\Delta^{+}(\alpha;\nu^{*}=0), \beta\neq0 \text{ and } \Delta^{-}(\alpha;\nu^{*}=0), \beta=0, \\
u_z&=-\frac{f(\alpha,\beta,\theta)\hat{r}^2(\hat{r}^2-4\hat{z}^2)}{(\hat{r}^2+\hat{z}^2)^{7/2}},\\
\label{eq:acutez1}
f(\alpha,\beta,\theta)&=\frac{F_z}{8\pi\mu\rho}\frac{3\pi(1-\cos{2\beta}\sec{2\alpha})(1-\cos{2\theta}\sec{2\alpha})}{2(\tan{2\alpha}-2\alpha)}\cdot
\end{align}
The leading-order flow satisfies the  incompressibility condition, $\nabla \cdot \vec{u}=0$, and the no-slip boundary conditions at $\theta=\pm\alpha$. It may be written in terms of the streamfunction $\Psi$ defined by $\vec{u}=\nabla \times (\Psi \vec{e}_{\theta})$ and the streamlines will be given by $\hat{r}\Psi=C$ for some constant $C$. In this case
\begin{align}
\label{eq:a1}
u_r&=-\frac{\partial\Psi}{\partial z}, \quad u_{\theta}=0,\quad u_z=\frac{1}{r}\frac{\partial(r\Psi)}{\partial r},\\
\Psi&=\frac{\rho f(\alpha,\beta,\theta)\hat{r}^3}{(\hat{r}^2+\hat{z}^2)^{5/2}}.
\end{align}

 \begin{figure}
 \includegraphics[width=0.49\textwidth]{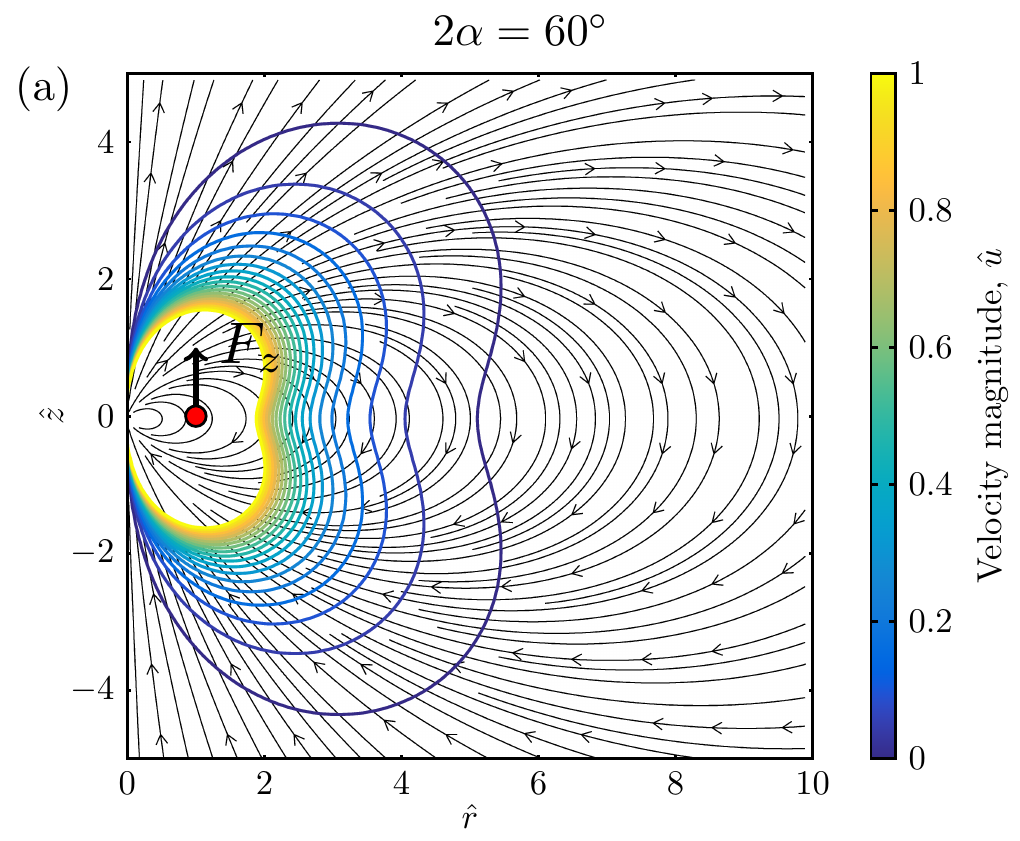}
 \includegraphics[width=0.49\textwidth]{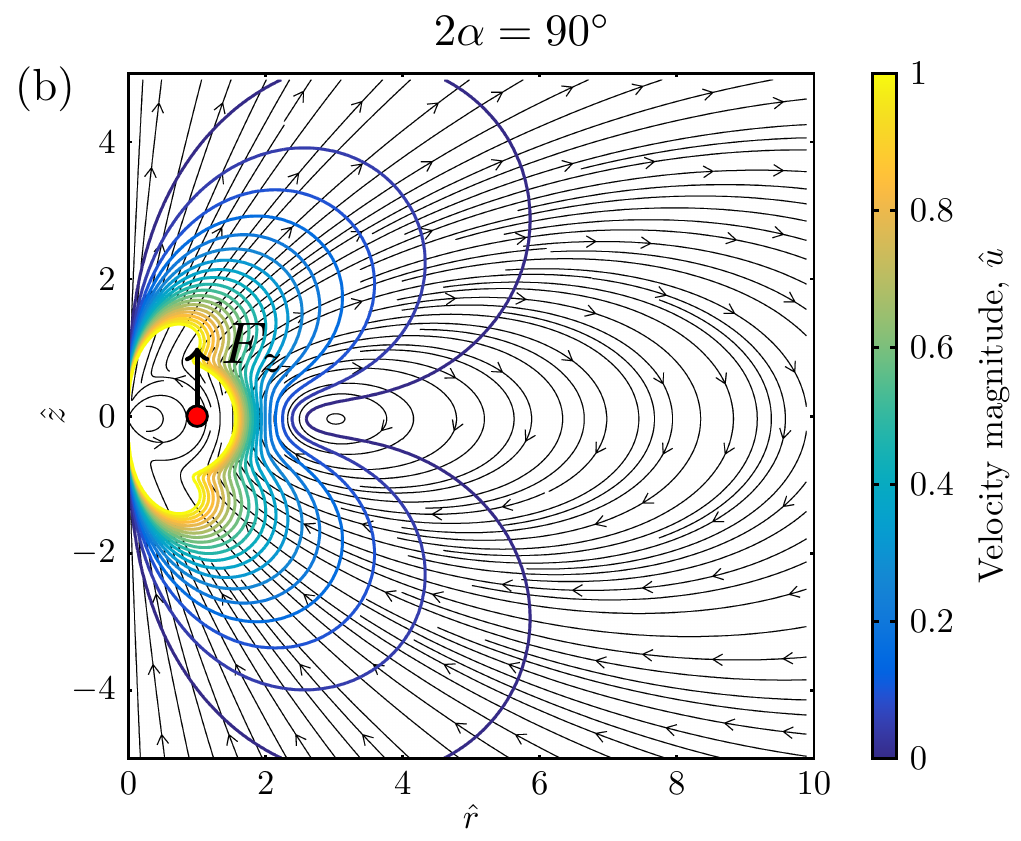}  
 \includegraphics[width=0.49\textwidth]{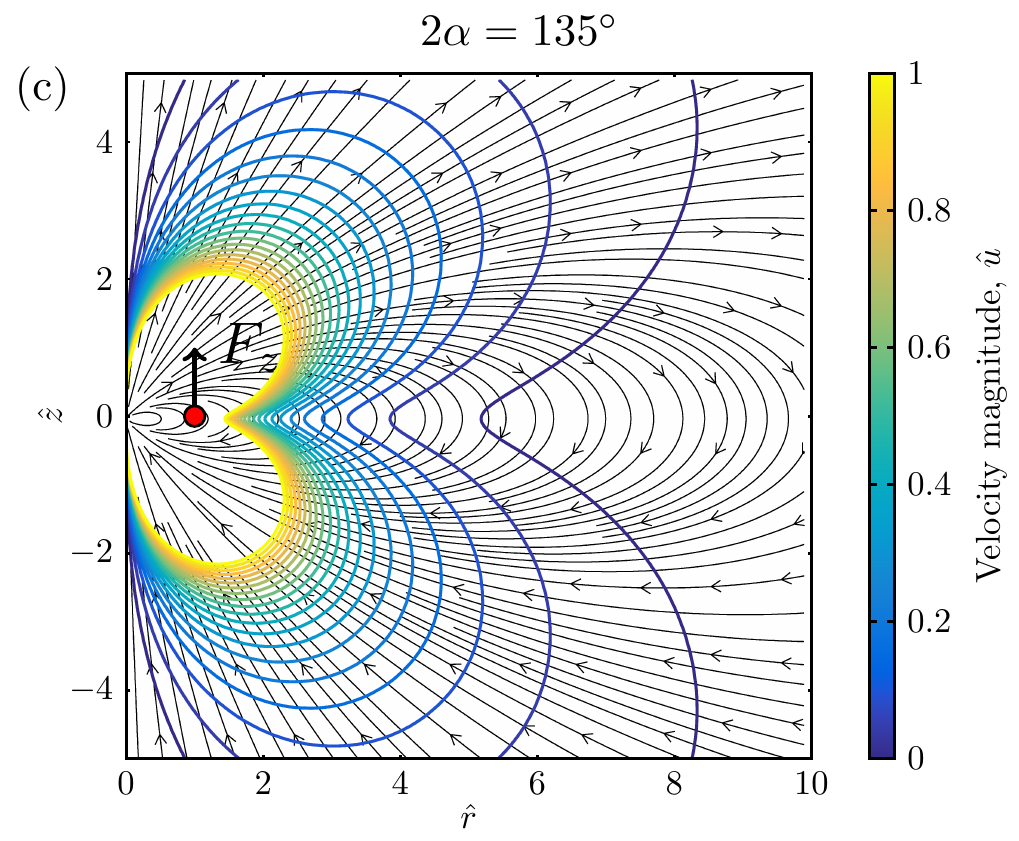}
 \includegraphics[width=0.49\textwidth]{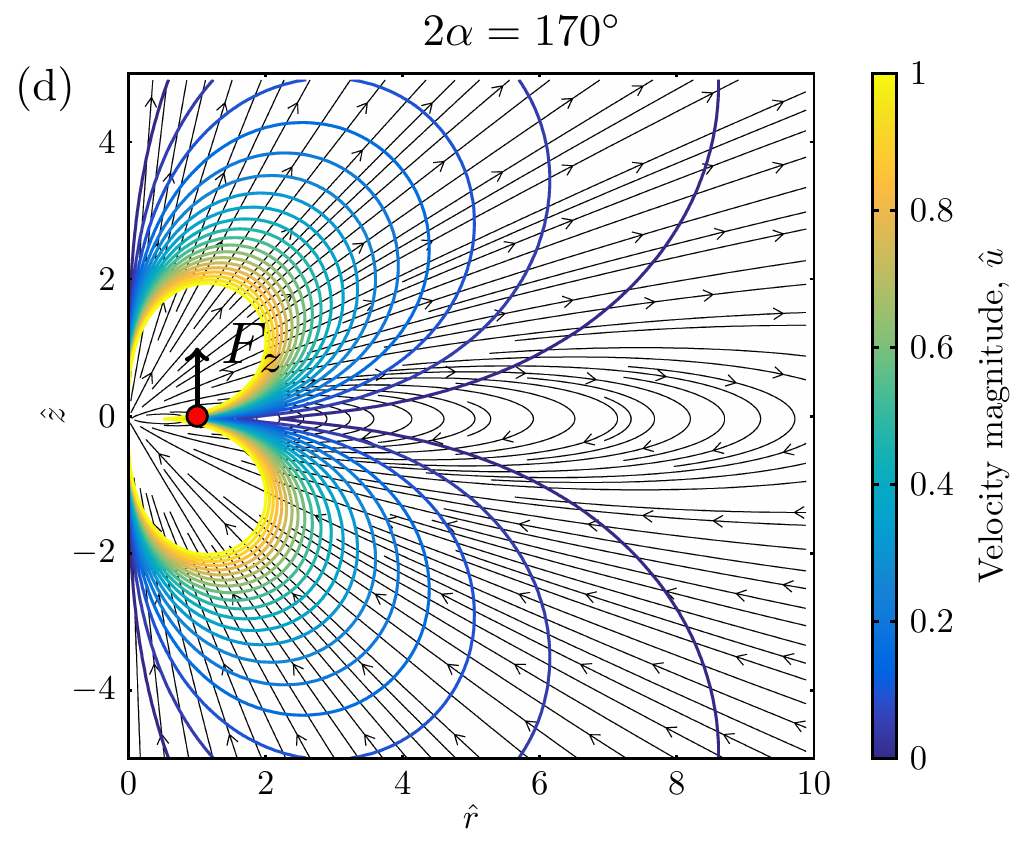}
              \caption{
              Illustration of the far-field flow. Streamlines and lines of constant velocity magnitude, $\hat{u}=8\pi\mu\rho u/F_z=(\hat{u}_r^2+\hat{u}_z^2)^{1/2}$, for the leading-order  flow with an axial point force,  $\vec{F}=(0,0,F_z)$: (a) $2\alpha=60^{\circ}$; (b) $2\alpha=90^{\circ}$; (c) $2\alpha=135^{\circ}$; (d) $2\alpha=170^{\circ}$. In all cases $\theta=\beta=0$, and the solution is valid in the far field, i.e.~for $\hat{r}^2+\hat{z}^2\gg1$.
              }
   \label{fig:3}
\end{figure}

The streamlines and iso-contours of velocity magnitude  are shown in  Fig.~\ref{fig:3}a for the case when $\theta=0$, $\beta=0$, $2\alpha=60^{\circ}$. Note that  the structure of the  streamlines  is independent of angles $\theta, \beta, \alpha$ in this case and only the flow  magnitude depends on them. The flow is two dimensional, because $u_r,u_z\gg u_{\theta}$, and it decays algebraically as $\mathcal{O}(1/L^3)$.

The flux generated along the axial direction is given by the formula
\begin{align}
Q&=\int_{-\alpha}^{\alpha} \int_0^{\infty}u_z r dr d\theta,
\end{align}
and its value is independent of $z$ by mass conservation.  
We can evaluate this flux exactly by looking at the flow far away from the point force, i.e.~in the limit $\hat{z}\gg1$, and use for this the leading-order flow obtained above \eqref{eq:a1}. We obtain that  the zero flux, $Q=0$, and the flow created by the point force is seen to  recirculate. Looking at the scaling the flow decays as $\mathcal{O}(1/L^3)$, the surface area for the flux integration scales as $\mathcal{O}(L^2)$. This gives zero flux as $L$ tends to infinity. The flux would be finite if the flow decayed as $\mathcal{O}(L^2)$, e.g., for  a point force above a single no-slip wall. Note that both a point force between two parallel plates and in a pipe give rise to recirculating flows and zero net flux with an algebraic flow decay in the case of two walls and an exponential decay in the pipe  \cite{liron1976stokes,liron1978stokes}.

The leading-order pressure field may be deduced from the results above and we get
\begin{align}
p=-\frac{F_z}{8\pi\rho}\frac{3\pi(1-\cos{2\beta}\sec{2\alpha})(-6\hat{r}^2+4\hat{z}^2+10\hat{r}^2\cos{2\theta}\sec{2\alpha})\hat{z}}{2(\tan{2\alpha}-2\alpha)(\hat{r}^2+\hat{z}^2)^{7/2}}\cdot
\end{align}
It decays algebraically to zero at large distance away from the point force, $p=\mathcal{O}(1/L^4)$. As a difference with the case of a pipe, there is no net pressure jump induced by the point force  near an acute corner \cite{liron1978stokes}.

Next, we consider the flow field near  the corner. In that case we also obtain power-law decays with results illustrated numerically  in Fig.~\ref{fig:2}c. If $r\ll1$  we still have $\xi\gg1$, and can therefore use the same expansion for the Legendre associated function $P_m^n$ as above, except that this time  the total flow will be written in powers of $\hat{r}$. In the  case of an acute angle, $0<2\alpha<\pi/2$, the leading-order flow is characterised  by a streamfunction
\begin{equation}
\label{eq:z_acute_c}
\Psi=\frac{\rho f(\alpha,\beta,\theta)\hat{r}^3(4\hat{z}^2-1)}{4(\hat{z}^2+1)^{7/2}},
\end{equation}
for some constant $C$ and radial and axial flow components decay as
\begin{align}
u_r=-\frac{\partial \Psi}{\partial z}=\mathcal{O}(\hat{r}^3),\quad
u_z=\frac{1}{r}\frac{\partial (r\Psi)}{\partial r}=\mathcal{O}(\hat{r}^2).
\end{align}
In this case, exactly as for the far field calculation above, the leading-order potentials $\phi_3=\phi_4=0$ and the pole at $\nu=i$ gives rise to $\phi_1=\phi_2=\mathcal{O}(\hat{r})+\mathcal{O}(\hat{r}^3)+...$. The leading-order pressure field is given by
\begin{align}
p=\frac{6\pi \hat{z}(1-\cos 2\beta \sec 2\alpha)}{(1+\hat{z}^2)^{5/2}(\tan 2\alpha-2\alpha)}+\frac{15\pi\hat{r}^2\hat{z}(3-4\hat{z}^2)(1-\cos 2\beta \sec 2\alpha)(2-\cos 2\theta \sec 2\alpha)}{4(1+\hat{z}^2)^{9/2}(\tan{2\alpha}-2\alpha)}\cdot
\end{align}
\begin{figure}
 \includegraphics[width=0.49\textwidth]{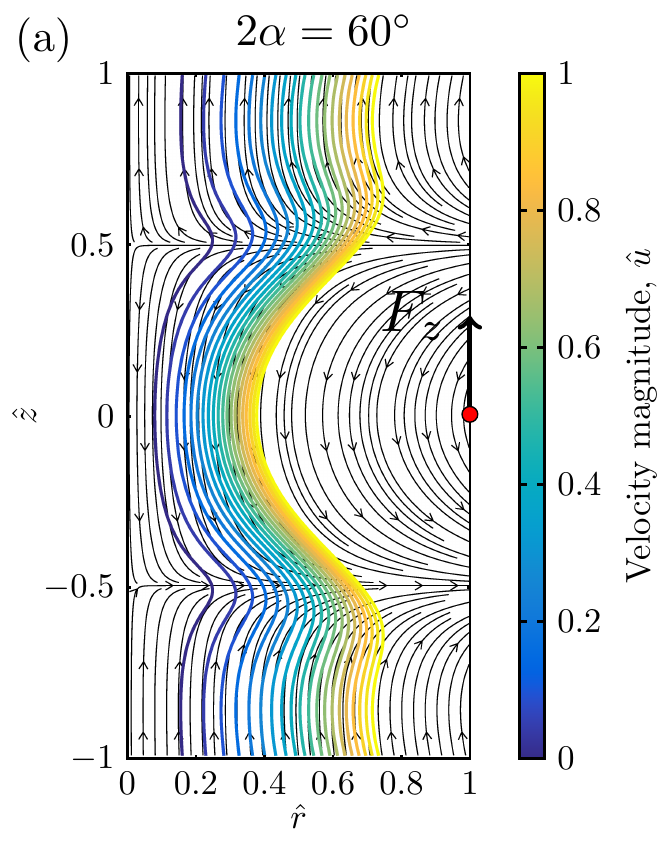}
 \includegraphics[width=0.49\textwidth]{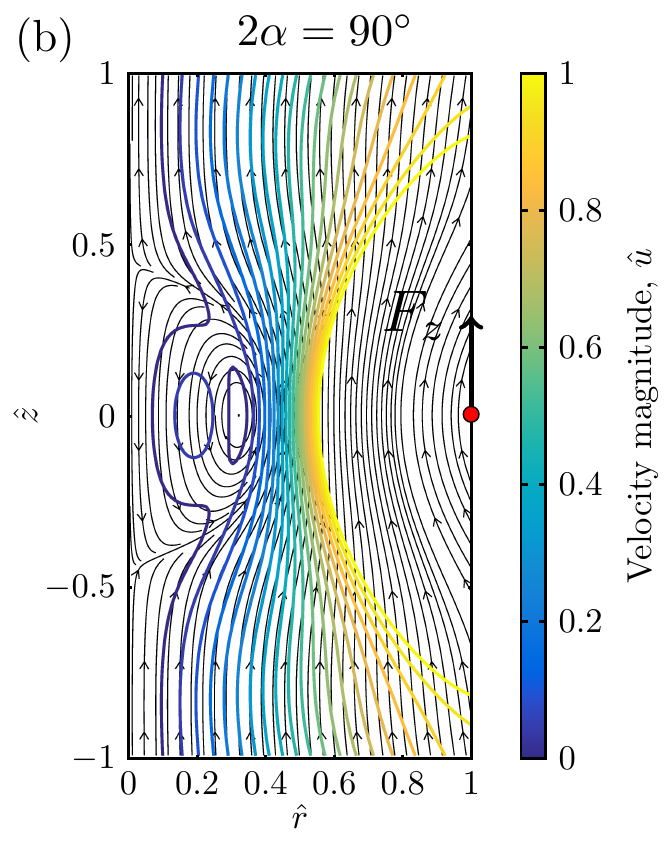}\\
 \includegraphics[width=0.49\textwidth]{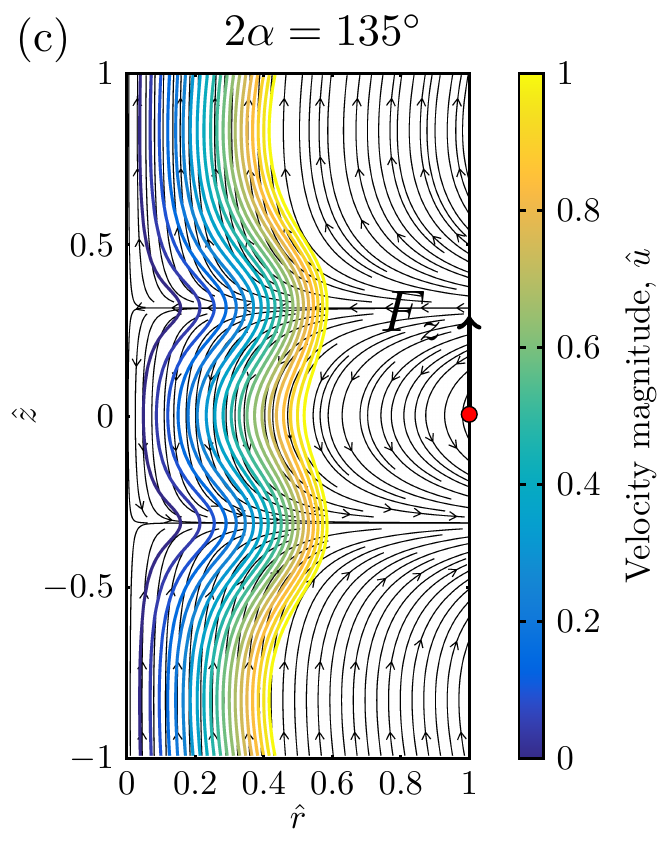}
 \includegraphics[width=0.49\textwidth]{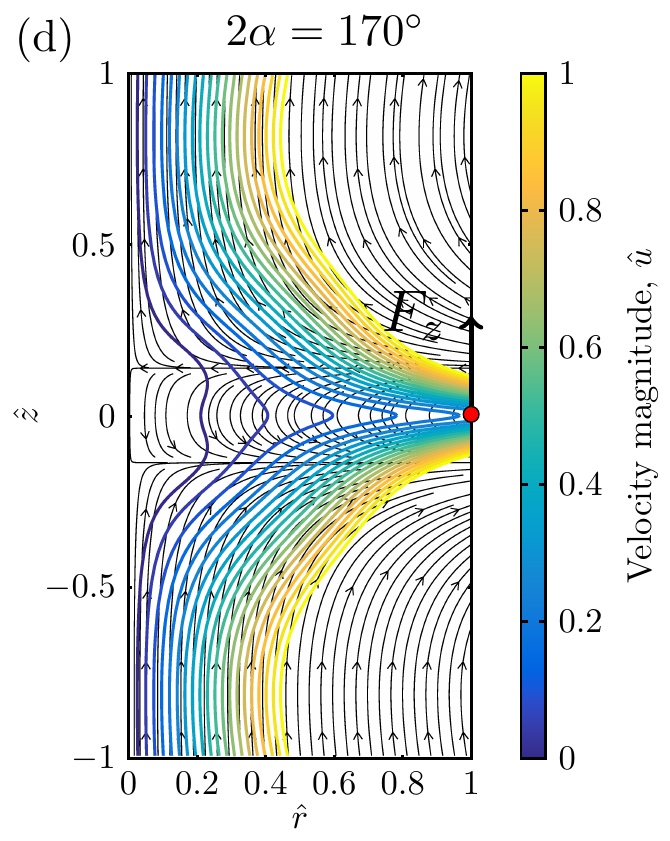}
              \caption{Illustration of the  flow near the corners. 
              Streamlines and lines of constant velocity magnitude, $\hat{u}=8\pi\mu\rho u/F_z=(\hat{u}_r^2+\hat{u}_z^2)^{1/2}$, for the corner  flow with an axial point force,  
              $\vec{F}=(0,0,F_z)$; (a) $2\alpha=60^{\circ}$; (b) $2\alpha=90^{\circ}$; (c) $2\alpha=135^{\circ}$; (d) $2\alpha=170^{\circ}$. 
In all cases $\theta=\beta=0$, and the solution is valid near the corner, i.e.~for $\hat{r}\ll1$.
              }
   \label{fig:4}
\end{figure}

The streamlines and contour lines are shown in Fig.~\ref{fig:4}a. As above, the flow structure  is independent of the corner angle, $2\alpha$, and only the overall magnitude of the flow depends on it. Notice that $\hat{r}\Psi=0$ at $\hat{z}=\pm1/2$, and therefore there is a fluid cell of width $\hat{d}=d/\rho$  in which there is a counter flow. The axial extent of this cell is independent of the corner angle for acute angles, i.e.~$0<2\alpha<\pi/2$, see Fig.~\ref{fig:5}. In the azimuthal direction the flow is given by
\begin{align}
u_{\theta}&=\mathcal{O}\left(\hat{r}^{\Im(\nu^{*})}\right) :\Delta^{+}(\alpha;\nu^{*}=0), \beta\neq0 \text{ and } \Delta^{-}(\alpha;\nu^{*}=0), \beta=0.
\end{align}
It is always smaller than the axial flow, i.e.~$u_{\theta}\ll u_{z}$, see Fig.~\ref{fig:2}c for the decay rate $N$ as a function of the corner angle $2\alpha$, where $u_r,u_{\theta},u_z=\mathcal{O}(\hat{r}^N)$. Finally, we note that the leading-order flow at large distances, $L\gg1$, and the flow near the corner, $\hat{r}\ll1$, match each other in the overlap region $\hat{r}\ll1$ and $\hat{z}\gg1$, as expected.

%%%%
\subsection{Leading-order flow when $2\alpha=\pi/2$}
When the corner is exactly a right angle, we obtain as a special case in which the leading-order flow is given by a double pole at $\nu=i$ for $\phi_1=\phi_2=\mathcal{O}(\ln L/L^3+1/L^3)$ and this time $\phi_3=\mathcal{O}(1/L^3)$. The flow in the radial and axial directions,  as above,  can be written in terms of a streamfunction $\Psi$ defined by $\vec{u}=\nabla \times (\Psi \vec{e}_{\theta})$ where
\begin{align}
\label{eq:z_right_f}
\frac{8\pi\mu}{F_z}\Psi&=-\frac{3 [\pi -4 \beta \sin (2 \beta)] \cos (2
   \theta)}{2}\frac{\hat{r}^3}{\left(\hat{r}^2+\hat{z}^2\right)^{5/2}}+\\
   \nonumber
   &-\frac{\cos (2 \beta) \left[\cos (2 \theta) \left(-24 \ln \left(\frac{4 \left(\hat{r}^2+\hat{z}^2\right)}{\hat{r}}\right)-3 \pi ^2+28\right)+6 [\pi -4 \theta \sin (2 \theta)]\right]\hat{r}^3}{4 \left(\hat{r}^2+\hat{z}^2\right)^{5/2}}\cdot
\end{align}
This  radial and axial components of this flow decay spatially as $\mathcal{O}(\ln{L}/L^{3})+\mathcal{O}(1/L^{3})$. The azimuthal component of the flow has the same behaviour as above, namely
\begin{align}
u_{\theta}&=\mathcal{O}\left(\frac{1}{L^{2+\Im(\nu^{*})}}\right) :\Delta^{+}(\alpha;\nu^{*}=0), \beta\neq0 \text{ and } \Delta^{-}(\alpha;\nu^{*}=0), \beta=0.
\end{align} 
The streamlines and contour lines for this flow are shown in Fig.~\ref{fig:3}b in the case
 $\theta=\beta=0$. We observe the formation of  a single eddy  with closed streamlines. The flow is nearly two-dimensional, because $u_r,u_z\gg u_{\theta}$ in the limit $L=(\hat{r}^2+\hat{z}^2)^{1/2}\gg1$. The net flux generated along the axial direction, $Q$, is again exactly zero.

If $r\ll1$ then we get flow due to a double pole at $\nu=i$ for $\phi_1=\phi_2=\mathcal{O}(\hat{r}\ln\hat{r}+\hat{r})+\mathcal{O}(\hat{r}^3\ln\hat{r}+\hat{r}^3)$ and this time $\phi_3=\mathcal{O}(\hat{r}^2+\hat{r}^4)$. The leading-order flow near the corner is given by the streamfunction
\begin{align}
\label{eq:z_right_c}
\frac{8\pi\mu}{F_z}\Psi&=-\frac{3 [\pi -4 \beta \sin (2 \beta)] \cos (2
   \theta)}{2}\frac{\hat{r}^3(\hat{z}^2-1/4)}{\left(1+\hat{z}^2\right)^{7/2}}+\\\nonumber
   &-\frac{\cos (2 \beta) \left[\cos (2 \theta) \left(-24 \ln \left(\frac{4 \left(1+\hat{z}^2\right)}{\hat{r}}\right)\right)\right]\hat{r}^3(\hat{z}^2-1/4)}{4 \left(1+\hat{z}^2\right)^{7/2}}\\\nonumber
     &-\frac{\cos (2 \beta) \cos (2 \theta) \left[(-3 \pi ^2+28)(\hat{z}^2-1/4)-9/2\right]\hat{r}^3}{4 \left(1+\hat{z}^2\right)^{7/2}}\\\nonumber
      &-\frac{3\cos (2 \beta) [\pi -4 \theta \sin (2 \theta)]\hat{r}^3(\hat{z}^2-1/4)}{2 \left(1+\hat{z}^2\right)^{7/2}},\\
    u_r&=\mathcal{O}(\hat{r}^3\ln{\hat{r}})+\mathcal{O}(\hat{r}^3),\\
    u_z&=\mathcal{O}(\hat{r}^2\ln{\hat{r}})+\mathcal{O}(\hat{r}^2),
\end{align}
with streamlines and contour lines  shown in Fig.~\ref{fig:4}b.
The leading-order flow in the azimuthal direction is
\begin{align}
u_{\theta}&=\mathcal{O}\left(\hat{r}^{\Im(\nu^{*})}\right) :\Delta^{+}(\alpha;\nu^{*}=0), \beta\neq0 \text{ and } \Delta^{-}(\alpha;\nu^{*}=0), \beta=0.
\end{align}
with $u_{\theta}\ll u_z$, as shown in Fig.~\ref{fig:2}c.

%%%%%%
\subsection{Leading-order flow for the obtuse corner, $\pi/2<2\alpha<\pi$}
The leading-order flow in the case where the corner is obtuse, i.e.~$\pi/2<\alpha<\pi$, is given by the pole at $\nu=i[\pi/(2\alpha)-1]$ for $\phi_1=\phi_2=\mathcal{O}(1/L^{1+\pi/(2\alpha)})$ and by the pole at $\nu=i\pi/(2\alpha)$ for $\phi_3=\mathcal{O}(1/L^{1+\pi/(2\alpha)})$. The flow is written as
\begin{align}
u_{r}&=\frac{(\alpha+\pi)f(\alpha,\beta,\theta)\hat{r}^{1+\pi/(2\alpha)}\hat{z}}{\alpha(\hat{r}^2+\hat{z}^2)^{3/2+\pi/(2\alpha)}},\\
u_{\theta}&=\mathcal{O}\left(\frac{1}{L^{2+\Im(\nu^{*})}}\right) :\Delta^{+}(\alpha;\nu^{*}=0), \beta\neq0 \text{ and } \Delta^{-}(\alpha;\nu^{*}=0), \beta=0, \\
u_{z}&=\frac{f(\alpha,\beta,\theta)[(2\alpha-\pi)\hat{r}^{2}+(4\alpha+\pi)\hat{z}^2]\hat{r}^{\pi/(2\alpha)}}{2\alpha(\hat{r}^2+\hat{z}^2)^{3/2+\pi/(2\alpha)}},
\end{align}
with
\begin{equation}
f(\alpha,\beta,\theta)=\frac{F_z}{8\pi\mu\rho}\frac{16\pi^{1/2}\alpha\Gamma \left(\frac{3}{2}+\frac{\pi}{2\alpha}\right)\cos \left(\frac{\pi  \beta}{2 \alpha}\right) \cos\left(\frac{\pi  \theta}{2 \alpha}\right)}{(\alpha+\pi)(4 \alpha-\pi)\Gamma \left(1+\frac{\pi}{2\alpha}\right)}\cdot
\end{equation}

This flow is incompressible, it satisfies the no-slip boundary condition at $\theta=\pm \alpha$ and it can be written in terms of a streamfunction, $\Psi$,  given by 
\begin{align}
\label{eq:z_obtuse_f}
\Psi&=\frac{\rho f(\alpha,\beta,\theta)\hat{r}^{1+\pi/(2\alpha)}}{(\hat{r}^2+\hat{z}^2)^{1/2+\pi/(2\alpha)}}\cdot
\end{align}
The streamlines and velocity magnitude contours are shown in  Fig.~\ref{fig:3}c-d in the case where $\theta=0$, $\beta=0$. 
Similarly to  the acute-angle case, the structure of the streamlines  is independent of the angles $\theta, \beta$ and it is  only the magnitude of the flow which depends on the value of  these angles. However, as a difference with above, this time the flow structure  depends on the corner angle, $2\alpha$. The flow is quasi-two-dimensional, because $u_r,u_z\gg u_{\theta}$, and it decays algebraically as $\mathcal{O}(1/L^{1+\pi/(2\alpha)})$ for $L=(\hat{r}^2+\hat{z}^2)^{1/2}\gg1$. Notably, the  power decay rate of the flow, $N=1+\pi/(2\alpha)$, is in general not an integer. The flux, $Q$, generated along the axial direction is again exactly zero because $N>2$.

The leading-order pressure field for this flow is given by
\begin{align}
p&=\frac{F_z}{4\pi\rho}\frac{16\pi^{1/2}\Gamma \left(\frac{3}{2}+\frac{\pi}{2\alpha}\right)\cos \left(\frac{\pi  \beta}{2 \alpha}\right) \cos\left(\frac{\pi  \theta}{2 \alpha}\right)}{(4 \alpha-\pi)\Gamma \left(1+\frac{\pi}{2\alpha}\right)}\frac{\hat{r}^{\pi/(2\alpha)}\hat{z}}{(\hat{r}^2+\hat{z}^2)^{3/2+\pi/(2\alpha)}},
\end{align}
and it algebraically decays to zero at large distance away from the point force, as $p=\mathcal{O}(1/L^{2+\pi/(2\alpha)})$, for $L=(\hat{r}^2+\hat{z}^2)^{1/2}\gg1$. 

\begin{figure}
 \includegraphics[width=0.85\textwidth]{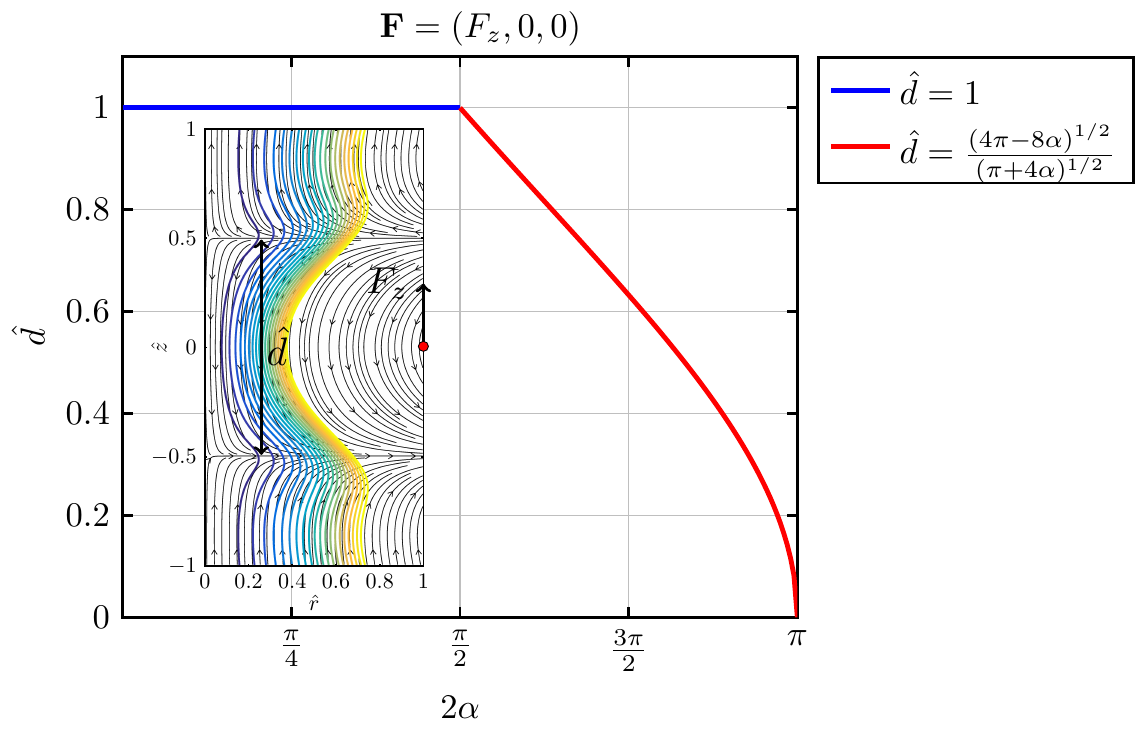}
              \caption{Flow recirculation near the corner due to the point force in the axial direction, $\vec{F}=(0,0,F_z)$. The width of the recirculation is $\hat{d}=d/\rho$ which is constant for the acute angles $0<2\alpha<\pi/2$ and the recirculation cell gets smaller for obtuse angles $\pi/2<2\alpha<\pi$.}
   \label{fig:5}
\end{figure}

If $\hat{r}\ll1$ then the leading-order flow near the corner is given poles $\nu=i[\pi/(2\alpha)-1];~i[\pi/(2\alpha)+1]$ for $\phi_1=\phi_2=\mathcal{O}(\hat{r}^{-1+\pi/(2\alpha)})+\mathcal{O(}\hat{r}^{1+\pi/(2\alpha)})$ and by the pole at $\nu=i\pi/(2\alpha)$ for $\phi_3=\mathcal{O}(\hat{r}^{\pi/(2\alpha)})+\mathcal{O}(\hat{r}^{2+\pi/(2\alpha)})$. The flow is described  by the streamfunction
\begin{align}
\label{eq:z_obtuse_c}
\Psi&=\frac{\rho f(\alpha,\beta,\theta)\hat{r}^{1+\pi/(2\alpha)}[(4\alpha+\pi)\hat{z}^2+(2\alpha-\pi)]}{(4\alpha+\pi)(1+\hat{z}^2)^{3/2+\pi/(2\alpha)}},
\end{align}
while
\begin{align}
u_r=\mathcal{O}(\hat{r}^{1+\pi/(2\alpha)}),~u_z=\mathcal{O}(\hat{r}^{\pi/(2\alpha)}).
\end{align}
The streamlines and contour lines in that case are shown in Fig. \ref{fig:4}c-d. This time $\Psi=0$ at the lines
\begin{equation}
\hat{z}=\pm(\pi-2\alpha)^{1/2}/(\pi+4\alpha)^{1/2},
\end{equation}
 i.e.~there is a fluid cell of width $\hat{d}=(4\pi-8\alpha)^{1/2}/(\pi+4\alpha)^{1/2}$  in which one observes a counter flow, see Fig.~\ref{fig:5}. The leading-order flow in the azimuthal direction is given by
\begin{align}
u_{\theta}&=\mathcal{O}\left(\hat{r}^{\Im(\nu^{*})}\right) :\Delta^{+}(\alpha;\nu^{*}=0), \beta\neq0 \text{ and } \Delta^{-}(\alpha;\nu^{*}=0), \beta=0,
\end{align}
where $u_{\theta}\ll u_z$ for $\hat{r}\ll1$ (see Fig.~\ref{fig:2}c).

Note that exactly the same leading-order solutions will be dominant in the case of salient corners with $\pi<2\alpha<2\pi$, see details in Appendix \ref{salient}.
\subsection{Recovering Blake's solution when $2\alpha=\pi$}
In the limit when $2\alpha=\pi$, the corner becomes a flat surface. 
If we calculate residues for $\phi_n$, in equations \eqref{eq:rf1}~--~\eqref{eq:rf3}, we  recover in this case exactly  Blake's solution near a wall \cite{blake1974fundamental}. The leading-order solution in the obtuse angle case tends in the limit as $2\alpha\to\pi$ to
\begin{align}
u_{r}&=\frac{3f(\pi/2,\beta,\theta)\hat{r}^{2}\hat{z}}{(\hat{r}^2+\hat{z}^2)^{5/2}},\\
u_{z}&=\frac{3f(\pi/2,\beta,\theta)\hat{r}\hat{z}^{2}}{(\hat{r}^2+\hat{z}^2)^{5/2}},\\
f(\pi/2,\beta,\theta)&=\frac{F_z}{2\pi\mu\rho}\cos \left(\beta\right) \cos\left(\theta\right),\\
u_{\theta}&=\mathcal{O}\left(\frac{1}{L^{2+\Im(\nu^{*})}}\right) :\nu^{*}=i, \beta\neq0 \text{ and } \nu^{*}=2i, \beta=0.
\end{align}
The radial and axial leading-order flow components, $u_r, u_z$, continuously tend to the Blake's result whereas the azimuthal component $u_{\theta}$ jumps discontinuously from $1/L^{3}$ (antisymmetric) and $1/L^{4}$ (symmetric) to $1/L^{2}$. 

%%%%%%
\subsection{Leading-order eddy-type flow}
The leading-order flows calculated for the acute, obtuse and right angle cases in the previous sections were non-oscillatory because the poles $\nu$ had zero real part. In this section, we characterise the flows due to the poles when $\Delta^{\pm}(\alpha;\nu)=\sinh{2\alpha\nu}\pm\nu\sin{2\alpha}$, which we  refer to as eddy-type flows.

 The antisymmetric flow component is given by poles $\Delta^{+}(\alpha;\nu)=0$ and the symmetric one by $\Delta^{-}(\alpha;\nu)=0$.
The flow due to the leading-order pole when $\beta\neq0$ (one can similarly calculate for the case when $\beta=0$) is given by
\begin{align}
u_r=-\frac{F_{z}}{\pi^{1/2}\mu\rho}\Re\left\{\frac{\Gamma{(3/2-i\nu)}}{\Gamma{(1-i\nu)}D^{-}(\alpha;\nu)\Delta^{+'}(\alpha;\nu)}N_r(\alpha,\beta,\theta;\nu)G_r(\hat{r},\hat{z};\nu)\right\},\\
u_{\theta}=-\frac{F_{z}}{\pi^{1/2}\mu\rho}\Re\left\{\frac{\Gamma{(3/2-i\nu)}}{\Gamma{(1-i\nu)}D^{-}(\alpha;\nu)\Delta^{+'}(\alpha;\nu)}N_{\theta}(\alpha,\beta,\theta;\nu)G_{\theta}(\hat{r},\hat{z};\nu)\right\},\\
u_z=-\frac{F_{z}}{\pi^{1/2}\mu\rho}\Re\left\{\frac{\Gamma{(3/2-i\nu)}}{\Gamma{(1-i\nu)}D^{-}(\alpha;\nu)\Delta^{+'}(\alpha;\nu)}N_z(\alpha,\beta,\theta;\nu)G_z(\hat{r},\hat{z};\nu)\right\},
\end{align}
where
\begin{align}
N_r=&N_z=q^{A}(\beta)q^{A}(\theta),\\
N_{\theta}=&q^A(\beta)\cosh{\alpha\nu}\sin{\alpha}(\nu\cos{\theta}\cosh{\theta\nu}+\sin{\theta}\sinh{\theta\nu})\\
&+q^A(\beta)\sinh{\alpha\nu}\cos{\alpha}(\cos{\theta}\cosh{\theta\nu}-\nu\sin{\theta}\sinh{\theta\nu}),\nonumber\\
q^A(t)=&\sin{\alpha}\cosh{\alpha \nu}\cos{t}\sinh{t\nu}-\cos{\alpha}\sinh{\alpha \nu}\sin{t}\cosh{t\nu},
\end{align}
and
\begin{align}
\Delta^{+'}(\alpha;\nu)=&2\alpha\cosh{2\alpha\nu}+\sin{2\alpha},~D^{-}(\alpha;\nu)=\cosh{2\alpha \nu}- \cos{2\alpha},\\
(G_r,G_{\theta},G_z)=&\left(\hat{r}\frac{\partial}{\partial\hat{r}}-1,1,\hat{r}\frac{\partial}{\partial\hat{z}}\right)G_0(\hat{r},\hat{z};\nu),\\
G_0(\hat{r},\hat{z};\nu)=&\hat{r}^{-3/2}\hat{z}\frac{(2\xi)^{1/2+i\nu}F\left(-\frac{i\nu}{2}+\frac{1}{4},-\frac{i\nu}{2}-\frac{1}{4};1-i\nu;\xi^{-2}\right)}{\xi^2-1}\cdot
\end{align}

If $L\gg1$ then the leading-order flow is given by
\begin{align}
G_r&=4i\frac{\hat{r}^{-i\nu}\hat{z}(\hat{r}^2(4i+\nu)-(-i+\nu)\hat{z}^2)}
{(\hat{r}^2+\hat{z}^2)^{5/2-i\nu}},~
G_{\theta}=\frac{4\hat{r}^{-i\nu}\hat{z}}
{(\hat{r}^2+\hat{z}^2)^{3/2-i\nu}},\\
G_{z}&=4\frac{\hat{r}^{1-i\nu}(\hat{r}^2+2i(i+\nu)\hat{z}^2)}
{(\hat{r}^2+\hat{z}^2)^{5/2-i\nu}},
\end{align}
which is fully three-dimensional, but  is always subdominant in the far field. In contrast, in the limit where  $\hat{r}\ll1$ then the leading-order flow is
\begin{align}
G_r&=4i(i-\nu)\frac{\hat{r}^{-i\nu}\hat{z}}
{(1+\hat{z}^2)^{3/2-i\nu}},~G_{\theta}=\frac{4\hat{r}^{-i\nu}\hat{z}}
{(1+\hat{z}^2)^{3/2-i\nu}},~G_{z}=0,
\end{align}
which in this case is two-dimensional and  has an infinite number of eddies for $2\alpha<146.31^{\circ}$, similarly to the classical corner flow problem of Moffatt \cite{moffatt1964viscous}. 
The streamfunction, $\Psi$, defined as $\vec{u}=\nabla\cross (\Psi \vec{e}_z)$ is given then by
\begin{align}
\Psi=\frac{4F_{z}}{\pi^{1/2}\mu}\Re\left\{\frac{\Gamma{(3/2-i\nu)}}{(1-i\nu)\Gamma{(1-i\nu)}D^{-}(\alpha;\nu)\Delta^{+'}(\alpha;\nu)}N_{\theta}(\alpha,\beta,\theta;\nu)\frac{\hat{r}^{-i\nu+1}\hat{z}}
{(1+\hat{z}^2)^{3/2-i\nu}}\right\},
\end{align}
where $\nu$ is the pole of $\Delta^{+}(\alpha;\nu)=0$.

%%%%%%%%%%%%%%%%%%%%%%%%%%%%%%%
%%%%%%%%%%%%%%%%%%%%%%%%%%%%%%%
%%%%%%%%%%%%%%%%%%%%%%%%%%%%%%%
\section{Point force perpendicular to the corner axis}

We next  consider the case where the  point force is perpendicular to the corner axis, i.e.~the force is aligned along either the $r$ or the $\theta$ directions. We use the solution calculated in integral form by Sano \& Hasimoto \cite{sano1978effect} for the acute and obtuse corners.

\subsection{Exact solution}
Consider a point force $\vec{F}=(F_r, F_{\theta},0)$ which is perpendicular to the intersection of the walls \cite{sano1978effect}. Denote the relative strength of these forces by $\tan{\gamma}=F_{\theta}/F_{r}$.
The first-order harmonic functions are then given by
\begin{align}
\phi_1^{(1)}&= -\frac{1}{R}\cos{(\beta+\gamma)},~\phi_2^{(1)}= -\frac{1}{R}\sin{(\beta+\gamma)},~\phi_3^{(1)}= 0,~\phi_4^{(1)}= \frac{1}{R} \cos{\gamma},\\
R&=1+\hat{r}^2-2\hat{r}\cos{(\theta-\beta)}+\hat{z}^2.
\end{align}
The solution to $\phi_n^{(2)}$ is given by computing the inverse integral transforms of $\tilde{\Phi}_n^{(2)}$
\begin{align}
\tilde{\Phi}_n^{(2)}&=A_n \sinh{\theta \nu}+B_n \cosh{\theta\nu} ~~(n=1,2,4), \tilde{\Phi}_3^{(2)}=0,\\
A_1&=\frac{4\pi c}{\nu \sinh{\pi \nu}\Delta^{+}}\left\{a_1^0 K_{i\nu}(|k|\rho)+\Re[a_1^{+}|k|\rho K_{i(\nu+i)}(|k|\rho)]\right\},\\
B_1&=\frac{4\pi c}{\nu \sinh{\pi \nu}\Delta^{-}}\left\{b_1^0 K_{i\nu}(|k|\rho)+\Re[{b_1^{+}|k|\rho K_{i(\nu+i)}(|k|\rho)]}\right\},\\
A_2&=\frac{4\pi c}{\nu \sinh{\pi \nu}\Delta^{-}}\left\{a_2^0 K_{i\nu}(|k|\rho)+\Re[a_2^{+}|k|\rho K_{i(\nu+i)}(|k|\rho)]\right\},\\
B_2&=\frac{4\pi c}{\nu \sinh{\pi \nu}\Delta^{+}}\left\{b_2^0 K_{i\nu}(|k|\rho)+\Re[b_2^{+}|k|\rho K_{i(\nu+i)}(|k|\rho)]\right\},\\
A_4&=\frac{2\pi c\cos{\gamma}}{\nu \sinh{\pi \nu}}\frac{\sinh{[(\pi-\alpha)\nu]}\sinh{\beta\nu}}{\sinh{\alpha\nu}} \rho K_{i\nu}(|k|\rho),\\
B_4&=\frac{2\pi c\cos{\gamma}}{\nu \sinh{\pi \nu}}\frac{\cosh{[(\pi-\alpha)\nu]}\cosh{\beta\nu}}{\cosh{\alpha\nu}} \rho K_{i\nu}(|k|\rho),
\end{align}
where
\begin{align}
a_1^0&=\nu \sin{\alpha}\left[\cos{\alpha}\cos{(\beta+\gamma)}\sinh{\beta\nu}\cosh{\pi \nu}+\sin{\alpha}\sin{(\beta+\gamma)}\cosh{\beta \nu}\sinh{\pi \nu}\right]\\ \nonumber
&-\cos{(\beta+\gamma)}\cosh{\alpha \nu}\sinh{[(\pi-\alpha)\nu]}\sinh{\beta \nu},\\
b_1^0&=\nu \sin{\alpha}\left[\cos{\alpha}\cos{(\beta+\gamma)}\cosh{\beta\nu}\cosh{\pi \nu}+\sin{\alpha}\sin{(\beta+\gamma)}\sinh{\beta \nu}\sinh{\pi \nu}\right]\\ \nonumber
&-\cos{(\beta+\gamma)}\sinh{\alpha \nu}\cosh{[(\pi-\alpha)\nu]}\cosh{\beta \nu},\\
a_2^0&=-\nu \cos{\alpha}\left[\cos{\alpha}\cos{(\beta+\gamma)}\cosh{\beta\nu}\sinh{\pi \nu}+\sin{\alpha}\sin{(\beta+\gamma)}\sinh{\beta \nu}\cosh{\pi \nu}\right]\\ \nonumber
&-\sin{(\beta+\gamma)}\cosh{\alpha \nu}\sinh{[(\pi-\alpha)\nu]}\sinh{\beta \nu},\\
b_2^0&=-\nu \cos{\alpha}\left[\cos{\alpha}\cos{(\beta+\gamma)}\sinh{\beta\nu}\sinh{\pi \nu}+\sin{\alpha}\sin{(\beta+\gamma)}\cosh{\beta \nu}\cosh{\pi \nu}\right]\\ \nonumber
&-\sin{(\beta+\gamma)}\sinh{\alpha \nu}\cosh{[(\pi-\alpha)\nu]}\cosh{\beta \nu},
\end{align}
and 
\begin{align}
a_1^{+}&=-i \sin{\alpha}\cos{\gamma}\cosh{\alpha \nu}\sinh{\pi \nu}\frac{\sinh{[\beta(\nu + i)}]}{\sinh{[\alpha(\nu + i)}]}\\ \nonumber
&=- \frac{2\sin{\alpha}\cos{\gamma}\cosh{\alpha \nu}\sinh{\pi \nu}}{D^{-}}\left(q^A+i p^A\right),\\
b_1^{+}&=-i \sin{\alpha}\cos{\gamma}\sinh{\alpha \nu}\sinh{\pi \nu}\frac{\cosh{\beta(\nu + i)}}{\cosh{\alpha(\nu + i)}}\\ \nonumber
&=- \frac{2\sin{\alpha}\cos{\gamma}\sinh{\alpha \nu}\sinh{\pi \nu}}{D^{+}}\left(q^S+i p^S\right),\\
a_2^{+}&=i \cos{\alpha}\cos{\gamma}\cosh{\alpha \nu}\sinh{\pi \nu}\frac{\cosh{\beta(\nu + i)}}{\cosh{\alpha(\nu + i)}}\\ \nonumber
&= \frac{2\cos{\alpha}\cos{\gamma}\cosh{\alpha \nu}\sinh{\pi \nu}}{D^{+}}\left(q^S+i p^S\right),\\
b_2^{+}&=i \cos{\alpha}\cos{\gamma}\sinh{\alpha \nu}\sinh{\pi \nu}\frac{\sinh{\beta(\nu + i)}}{\sinh{\alpha(\nu + i)}}\\ \nonumber
&= \frac{2\cos{\alpha}\cos{\gamma}\sinh{\alpha \nu}\sinh{\pi \nu}}{D^{-}}\left(q^A+i p^A\right),
\end{align}
with
\begin{align}
q^A&=\sin{\alpha}\cosh{\alpha \nu}\cos{\beta}\sinh{\beta\nu}-\cos{\alpha}\sinh{\alpha \nu}\sin{\beta}\cosh{\beta\nu},\\
q^S&=\sin{\alpha}\sinh{\alpha \nu}\cos{\beta}\cosh{\beta\nu}-\cos{\alpha}\cosh{\alpha \nu}\sin{\beta}\sinh{\beta\nu},\\
p^A&=\cos{\alpha}\sinh{\alpha \nu}\cos{\beta}\sinh{\beta\nu}+\sin{\alpha}\cosh{\alpha \nu}\sin{\beta}\cosh{\beta\nu},\\
p^S&=\cos{\alpha}\cosh{\alpha \nu}\cos{\beta}\cosh{\beta\nu}+\sin{\alpha}\sinh{\alpha \nu}\sin{\beta}\sinh{\beta\nu},\\
\Delta^{\pm}&=\sinh{2\alpha \nu}\pm \nu\sin{2\alpha},\\
D^{\pm}&=\cosh{2\alpha \nu}\pm \cos{2\alpha}.
\end{align}

To evaluate the integral transforms, we need to do some manipulations on Bessel functions. 
We first denote  
\begin{align}
J_{i\nu}&=\rho\int_0^{\infty}  k\rho \cos{(kz)} K_{i\nu-1}(k\rho)K_{i\nu}(kr) dk,\\
I_{i\nu}&=\rho\int_0^{\infty}  \cos{(kz)} K_{i\nu}(k\rho)K_{i\nu}(kr) dk=\frac{\pi^2}{4}\hat{r}^{-1/2}\sech{(\pi \nu)}P_{i \nu -1/2}(\xi),
\end{align}
and use the recurrence relation for the modified Bessel function
\begin{align}
K_{i\nu-1}(z)&=-\frac{d}{dz} K_{i\nu}(z)-\frac{i\nu}{z}K_{i\nu}(z),
\end{align}
to write
\begin{align}
J_{i\nu}&=-\rho\int_0^{\infty} k\rho \cos{(kz)} \left(\frac{d}{d(k\rho)} K_{i\nu}(k\rho)+\frac{i\nu}{k\rho}K_{i\nu}(k\rho)\right) K_{i\nu}(kr) dk\\\nonumber&=-\left(i\nu+\rho \frac{\partial}{\partial \rho}\right) I_{i\nu}.
\end{align}
We can simplify this expression to
\begin{align}
J_{i\nu}&=-\frac{\pi^2}{8}\hat{r}^{-1/2}\sech{(\pi \nu)}\left[(2i\nu-1)P_{i \nu -1/2}(\xi)+\frac{(\xi-2\hat{r}^{-1})}{(\xi^2-1)^{1/2}}P^{-1}_{i \nu -1/2}(\xi)\right],\\ 
\Im\{J_{i\nu}\}&=-\nu I_{i\nu},\\ 
\Re\{J_{i\nu}\}&=-\rho\frac{\partial}{\partial \rho} I_{i\nu}=\frac{1}{2}I_{i\nu}-
\frac{\pi^2}{4\rho}\hat{r}^{-1/2}\sech{(\pi \nu)}\rho\frac{\partial \xi}{\partial \rho}\frac{d}{d \xi}P_{i \nu -1/2}(\xi),\\
\rho \frac{\partial \xi}{\partial \rho}&=\frac{1}{\hat{r}}-\xi,\\
\frac{d}{d\xi} P_{i\nu-1/2}&=-\left(\nu^2+\frac{1}{4}\right)\frac{1}{(\xi^2-1)^{1/2}}P_{i\nu-1/2}^{-1}(\xi),\\
\Re\{J_{i\nu}\}&=\frac{1}{2}I_{i\nu}+
\frac{\pi^2}{4}\sech{(\pi \nu)}\left(\nu^2+\frac{1}{4}\right)\frac{(\hat{r}^{-3/2}-\hat{r}^{-1/2}\xi)}{(\xi^2-1)^{1/2}}P_{i\nu-1/2}^{-1}(\xi).
\end{align}
This then allows us to evaluate the inverse Fourier transform to get
\begin{align}
\phi_1^{(2)}&=\frac{8}{\pi^2}\int_0^{\infty} \left(\frac{\sinh{\theta\nu}}{\Delta^{+}}\left[a_1^0 I_{i\nu}+\Re\{a_1^{+} J_{i\nu}\}\right]+\frac{\cosh{\theta\nu}}{\Delta^{-}}\left[b_1^0 I_{i\nu}+\Re\{b_1^{+} J_{i\nu}\}\right]\right)d\nu,\\
\phi_2^{(2)}&=\frac{8}{\pi^2}\int_0^{\infty} \left(\frac{\sinh{\theta\nu}}{\Delta^{-}}\left[a_2^0 I_{i\nu}+\Re\{a_2^{+} J_{i\nu}\}\right]+\frac{\cosh{\theta\nu}}{\Delta^{+}}\left[b_2^0 I_{i\nu}+\Re\{b_2^{+} J_{i\nu}\}\right]\right)d\nu,\\
\phi_3^{(2)}&=0,\\
\phi_4^{(2)}&=\frac{4\cos{\gamma}}{\pi^2}\int_0^{\infty} I_{i\nu} f(\alpha,\beta,\theta,\nu) d\nu,\\
f&=\left(\frac{\sinh{(\theta\nu)}\sinh{[(\pi-\alpha)\nu]}\sinh{(\beta\nu)}}{\sinh{(\alpha\nu)}}+\frac{\cosh{(\theta\nu)}\cosh{[(\pi-\alpha)\nu]}\cosh{(\beta\nu)}}{\cosh{(\alpha\nu)}}\right).
\end{align}

\subsection{Point force in azimuthal direction}
\begin{figure}      
  \includegraphics[height=0.335\textwidth]{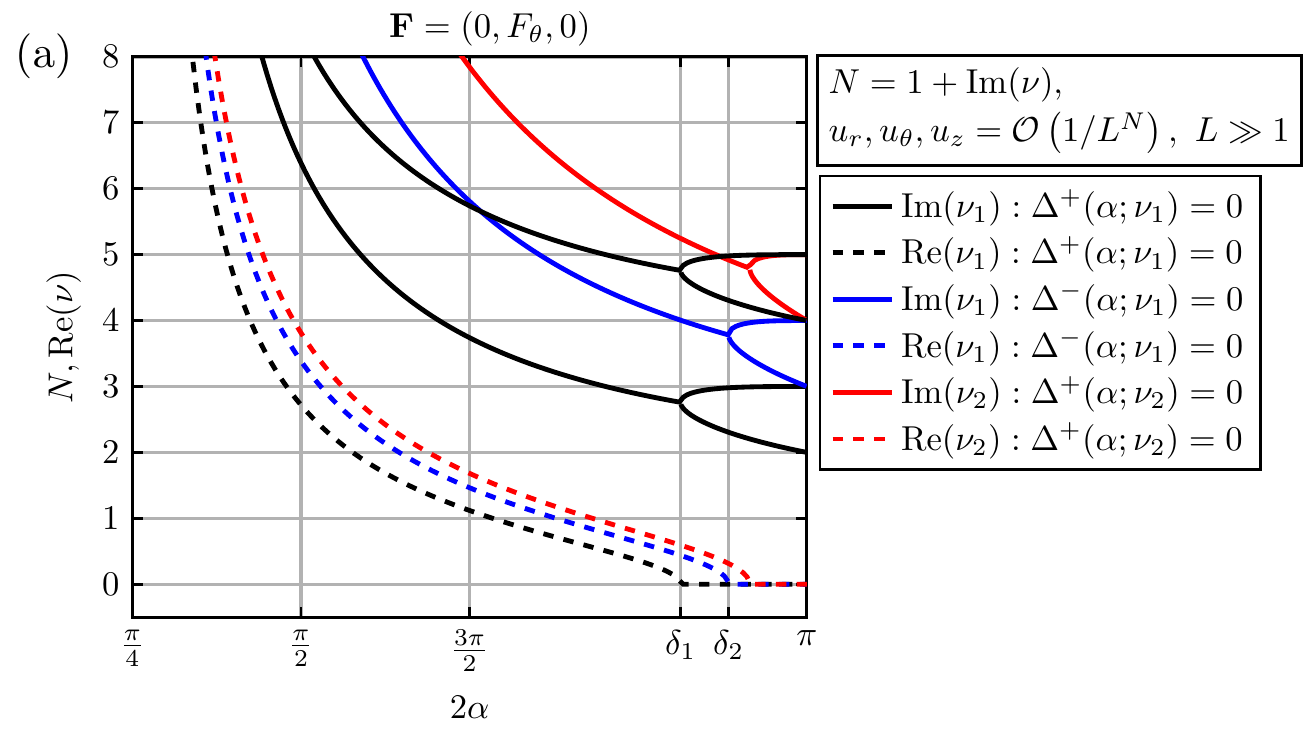}
    \includegraphics[height=0.335\textwidth]{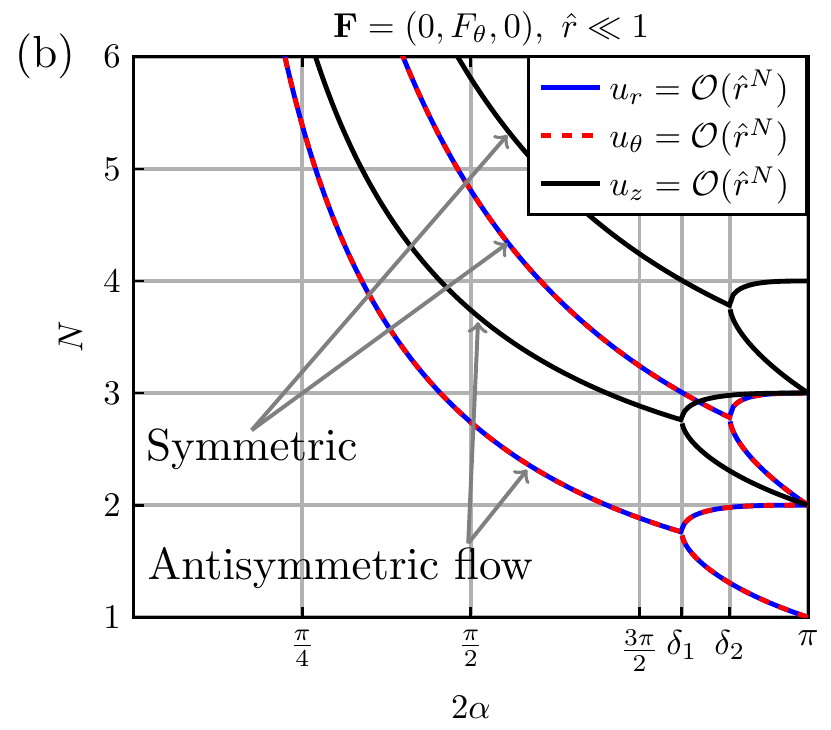}
       \caption{Decomposition of the full flow due to a point force in the azimuthal direction, $\vec{F}=(0,F_{\theta},0)$. The velocity components  in radial, azimuthal and axial directions, $u_r,u_{\theta},u_z$, decay as $\mathcal{O}(1/L^{1+\Im(\nu)})+\mathcal{O}(1/L^{3+\Im(\nu)})+...$ where $\Im(\nu)$ is the imaginary part of the pole $\nu$. The oscillations of the flow are determined by the real part of the pole, $\Re(\nu)$, because the flow is proportional to $\cos(\Re(\nu)\ln{L})$. There is 
 an infinite number of corner eddies if $\Re(\nu)>0$. 
 (a) Real and imaginary parts of poles $\nu$ are shown in the far field $L\gg1$; (b) same in the near field $\hat{r}\ll1$. The critical angles for antisymmetric and symmetric flows are given by $\delta_1=146.31^{\circ}$ and $\delta_2=159.11^{\circ}$.}
   \label{fig:7}
\end{figure}

The ratio between the strength of the force in the  azimuthal and radial directions is given by $\tan{\gamma}=F_{\theta}/F_{r}$. We consider first the case when $\gamma=\pi/2$, i.e.~$F_r=0$. The force is therefore in the azimuthal direction and in this case 
 the expressions simplify to 
\begin{align}
\phi_1^{(1)}&= -\frac{1}{R}\cos{(\beta+\pi/2)},~\phi_1^{(2)}=\frac{8}{\pi^2}\int_0^{\infty} \left(\frac{\sinh{\theta\nu}}{\Delta^{+}}a_1^0 I_{i\nu}+\frac{\cosh{\theta\nu}}{\Delta^{-}}b_1^0 I_{i\nu}\right)d\nu,\\
\phi_2^{(1)}&= -\frac{1}{R}\sin{(\beta+\pi/2)},~\phi_2^{(2)}=\frac{8}{\pi^2}\int_0^{\infty} \left(\frac{\sinh{\theta\nu}}{\Delta^{-}}a_2^0 I_{i\nu}+\frac{\cosh{\theta\nu}}{\Delta^{+}}b_2^0 I_{i\nu}\right)d\nu,\\
\phi_3^{(1)}&=\phi_3^{(2)}=0,~\phi_4^{(1)}=\phi_4^{(2)}=0,~
R=1+\hat{r}^2-2\hat{r}\cos{(\theta-\beta)}+\hat{z}^2.
\end{align}
Writing as before
\begin{align}
I_{i\nu}&=\frac{i\pi}{16}\frac{I(\hat{r},\hat{z};\nu)}{\sinh{(\pi\nu)}}+\text{c.c.},\\
I(\hat{r},\hat{z};\nu)&=-4\pi^{1/2}\hat{r}^{-1/2}(2\xi)^{i\nu-1/2}F\left(-\frac{i\nu}{2}+\frac{3}{4},-\frac{i\nu}{2}+\frac{1}{4};1-i\nu;\xi^{-2}\right)\frac{\Gamma{\left(\frac{1}{2}-i \nu\right)}}{\Gamma{\left(1-i \nu\right)}},
\end{align}
the harmonic functions are then given by
\begin{align}
\phi_1^{(2)}=-\Re\left\{\Res\left[\frac{I(\hat{r},\hat{z};\nu)}{\sinh{(\pi\nu)}}\left(\frac{\sinh{\theta\nu}}{\Delta^{+}(\alpha;\nu)}a_1^0(\alpha;\beta;\nu)+\frac{\cosh{\theta\nu}}{\Delta^{-}(\alpha;\nu)}b_1^0(\alpha,\beta;\nu)\right);\nu^{*}\right]\right\},\\
\phi_2^{(2)}=-\Re\left\{\Res\left[\frac{I(\hat{r},\hat{z};\nu)}{\sinh{(\pi\nu)}}\left(\frac{\sinh{\theta\nu}}{\Delta^{-}(\alpha;\nu)}a_2^0(\alpha;\beta;\nu)+\frac{\cosh{\theta\nu}}{\Delta^{+}(\alpha;\nu)}b_2^0(\alpha,\beta;\nu)\right);\nu^{*}\right]\right\},
\end{align}
where we choose for the contour of integration the quarter of the circle in the first quadrant of the complex plane.  The poles where $\sinh{\pi\nu}=0$ cancel the Stokeslet in the free space, i.e.~$\phi_1^{(1)}$ and $\phi_2^{(1)}$. The harmonic functions are determined by evaluating the contribution from the poles such that $\Delta^{\pm}(\alpha;\nu)=\sinh{2\alpha\nu}\pm\nu\sin{2\alpha}=0$ (see Fig.~\ref{fig:7} for the leading-order poles, $\nu$). 
 The antisymmetric flow component is given by poles such that $\Delta^{+}(\alpha;\nu)=0$ and the symmetric flow component when $\Delta^{-}(\alpha;\nu)=0$. The antisymmetric flow  is dominant and present for all locations, i.e.~for all $\beta$, because the point force being in the azimuthal direction always gives rise to the antisymmetric flow.   There are no non-oscillatory flows present in this case, hence no need to distinguish between acute and obtuse cases.
The flow due to the leading-order pole is given by
\begin{align}
\label{eq:azi1}
u_r=-\frac{F_{\theta}}{2\pi^{1/2}\mu\rho}\Re\left\{\frac{\Gamma{(1/2-i\nu)}}{\Gamma{(1-i\nu)}\Delta^{+'}(\alpha;\nu)}N_r(\alpha,\beta,\theta;\nu)G_r(\hat{r},\hat{z};\nu)\right\},\\
\label{eq:azi2}
u_{\theta}=-\frac{F_{\theta}}{2\pi^{1/2}\mu\rho}\Re\left\{\frac{\Gamma{(1/2-i\nu)}}{\Gamma{(1-i\nu)}\Delta^{+'}(\alpha;\nu)}N_{\theta}(\alpha,\beta,\theta;\nu)G_{\theta}(\hat{r},\hat{z};\nu)\right\},\\
\label{eq:azi3}
u_{z}=-\frac{F_{\theta}}{2\pi^{1/2}\mu\rho}\Re\left\{\frac{\Gamma{(1/2-i\nu)}}{\Gamma{(1-i\nu)}\Delta^{+'}(\alpha;\nu)}N_{z}(\alpha,\beta,\theta;\nu)G_{z}(\hat{r},\hat{z};\nu)\right\},
\end{align}
where
\begin{align}
N_r=&N_z=\sin{\beta}\sinh{\beta\nu}(\nu\cos^2{\alpha}\sin{\theta}\cosh{\theta\nu}+\cosh^2{\alpha\nu}
\cos{\theta}\sinh{\theta\nu})\\\nonumber
&+\cos{\beta}\cosh{\beta\nu}(\sinh^2{\alpha\nu}\sin{\theta}\cosh{\theta\nu}+\nu\sin^2{\alpha}
\cos{\theta}\sinh{\theta\nu}),\\
N_{\theta}=&(\nu\sin^2{\alpha}\cos{\beta}\cosh{\beta\nu}+
\cosh^2{\alpha\nu}\sin{\beta}\sinh{\beta\nu})(\nu\cos{\theta}\cosh{\theta\nu}+
\sin{\theta}\sinh{\theta\nu})\\\nonumber
&+(\sinh^2{\alpha\nu}\cos{\beta}\cosh{\beta\nu}+
\nu\cos^2{\alpha}\sin{\beta}\sinh{\beta\nu})(\nu\sin{\theta}\sinh{\theta\nu}-\cos{\theta}\cosh{\theta\nu}).
\end{align}
and
\begin{align}
\Delta^{+'}(\alpha;\nu)=&2\alpha\cosh{2\alpha\nu}+\sin{2\alpha},\\
(G_r,G_{\theta},G_z)=&\left(\hat{r}\frac{\partial}{\partial\hat{r}}-1,1,\hat{r}\frac{\partial}{\partial\hat{z}}\right)G_0(\hat{r},\hat{z};\nu),\\
G_0(\hat{r},\hat{z};\nu)=&\hat{r}^{-1/2}(2\xi)^{-1/2+i\nu}F\left(-\frac{i\nu}{2}+\frac{3}{4},-\frac{i\nu}{2}+\frac{1}{4};1-i\nu;\xi^{-2}\right).
\end{align}

 \begin{figure}      
  \includegraphics[width=0.621\textwidth]{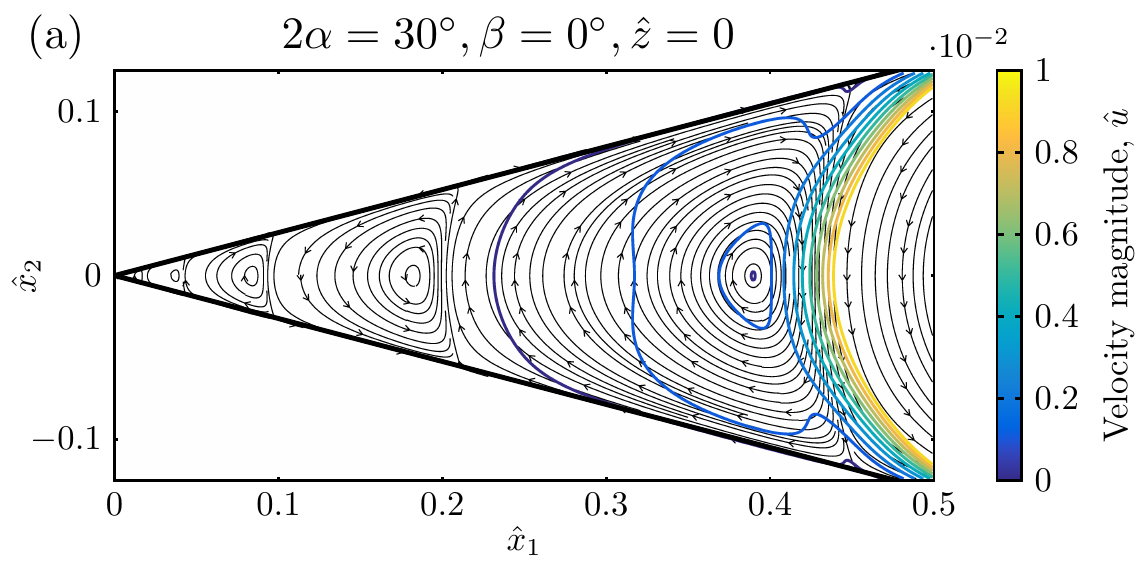}
   \includegraphics[width=0.37\textwidth]{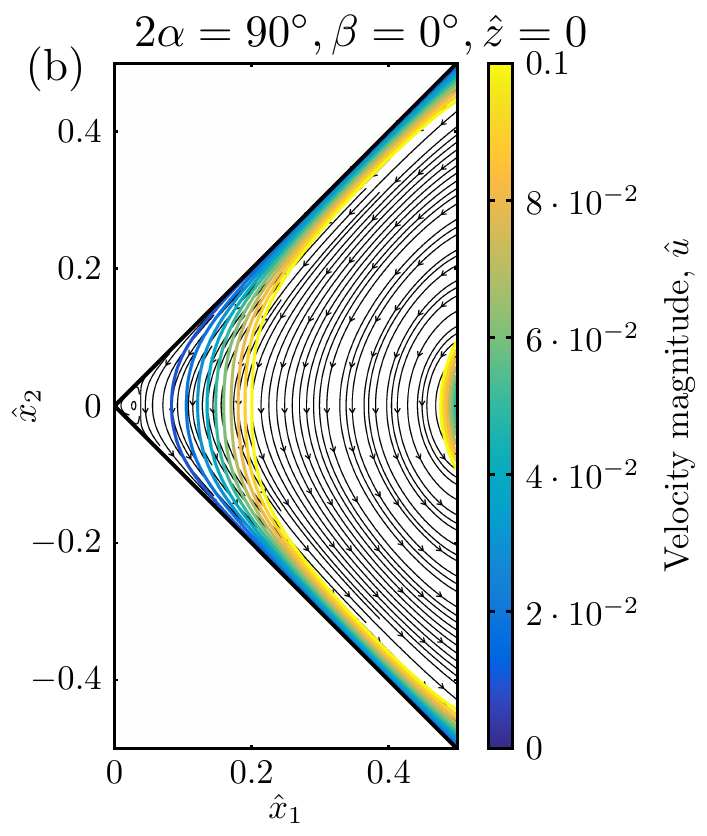}
       \caption{Leading-order flow due to the point force in the azimuthal direction close to the corner. Streamlines and velocity magnitude contour lines are shown  using the scaling $u={F_{\theta}\hat{u}}/{8\pi\mu\rho}$. The flow decays algebraically as $\mathcal{O}(\hat{r}^{-i\nu})$ for $\hat{r}\ll1$: (a) Flow when $2\alpha=30^{\circ}$,~$\beta=0^{\circ}$,~$\hat{z}=0$,~$\nu=4.2+8.1i$, decays rapidly as $\mathcal{O}(\hat{r}^{8.1})$; (b) Flow when $2\alpha=90^{\circ}$,~$\beta=0^{\circ}$,~$\hat{z}=0$,~$\nu=1.1+2.7i$, which decays slower  as $\mathcal{O}(\hat{r}^{2.7})$.}
   \label{fig:8}
\end{figure}

In Ref.~\cite{sano1980three}, an analysis of the fluid structure in the case where $\beta=0$, i.e.~when the flow is purely antisymmetric about the bisectoral plane $\theta=0$, was carried out. It was found that in the $z=0$ plane, or in the region near
the intersections of planes, eddies are similar to 2D ones. If we look near  the corner, $\hat{r}\ll1$, or far away from the corner, $L=(\hat{r}^2+\hat{z}^2)^{1/2}\gg1$, then $\xi\gg1$, i.e.
\begin{align}
F\left(-\frac{i\nu}{2}+\frac{3}{4},-\frac{i\nu}{2}+\frac{1}{4};1-i\nu;\xi^{-2}\right)=&1+\frac{(3-2i\nu)(i+2\nu)}{16(i+\nu)\xi^2}+\mathcal{O}(\xi^{-4}).
\end{align}
If $L\gg1$ then the leading-order flow, $u_r=u_{\theta}=u_z=\mathcal{O}(L^{-1+i\nu})$, is given by
\begin{align}
G_r&=\frac{i[\hat{r}^2(2i+\nu)+\hat{z}^2(i-\nu)]\hat{r}^{-i\nu}}
{(\hat{r}^2+\hat{z}^2)^{3/2-i\nu}},\,G_{\theta}=\frac{\hat{r}^{-i\nu}}
{(\hat{r}^2+\hat{z}^2)^{1/2-i\nu}},\,G_{z}=\frac{(2i\nu-1)\hat{r}^{1-i\nu}\hat{z}}
{(\hat{r}^2+\hat{z}^2)^{3/2-i\nu}},
\end{align}
while if $\hat{r}\ll1$ then the leading-order flow, i.e.~$u_r=u_{\theta}=\mathcal{O}(\hat{r}^{-i\nu})$, is characterised by
\begin{align}
\label{eq:p_c}
G_r&=i(i-\nu)\frac{\hat{r}^{-i\nu}}
{(1+\hat{z}^2)^{1/2-i\nu}},\,G_{\theta}=\frac{\hat{r}^{-i\nu}}
{(1+\hat{z}^2)^{1/2-i\nu}},\,G_{z}=0.
\end{align}
These leading-order flows are in agreement with the results of Ref.~\cite{sano1980three}, but we disagree with their conclusion that  the flow decays exponentially.  
Here we obtain that the flow decays algebraically with oscillations as $\mathcal{O}(L^{-1-\Im(\nu)}\cos[A\ln(L)+B])$  away from the Stokeslet if the eddy flow is present ($2\alpha<146.31^{\circ}$) where $A,~B$ are constants which depend on the corner angle;  similarly, close to the corner, the flow will decay as $\mathcal{O}(\hat{r}^{\Im(\nu)}\cos[C\ln(\hat{r})+D])$ for some constants $C,~D$ which depend on the corner angle. The flow will approach the exponential decay as the corner angle $2\alpha$ decreases asymptotically to zero, because $N_r=N_z$ and $N_{\theta}$ will all tend to zero  and therefore the flow at all algebraic orders will be  small. This matches nicely to the  case when the point force is perpendicular to two parallel plates giving, classically,  an exponentially decaying flow \cite{liron1976stokes}. The same phenomenon happens in the 2D case as described by Moffat \cite{moffatt1964viscous}.

The decay is faster for smaller corner angles $\alpha$, as illustrated in  Fig.~\ref{fig:7}. The streamlines and the velocity magnitude contours are shown in Fig.~\ref{fig:8} in the cases $2\alpha=30^{\circ}$ and   $2\alpha=90^{\circ}$. The flow magnitude decays faster for smaller angle corners while for larger angles the distance between successive eddies increases. This  might explain why it is difficult experimentally to see many eddies even if the corner is sharp \cite{taneda1979visualization}.

As a final note, we may  check our calculation by considering the  limit $2\alpha \to \pi$ in which the corner becomes a flat wall. As shown in Appendix \ref{B}, we recover in that case the leading-order behaviour of  Blake's  classical solution \cite{blake1974fundamental}.

%%%%%%%%
\subsection{Point force in radial direction}
\begin{figure}[t!]
  \includegraphics[height=0.344\textwidth]{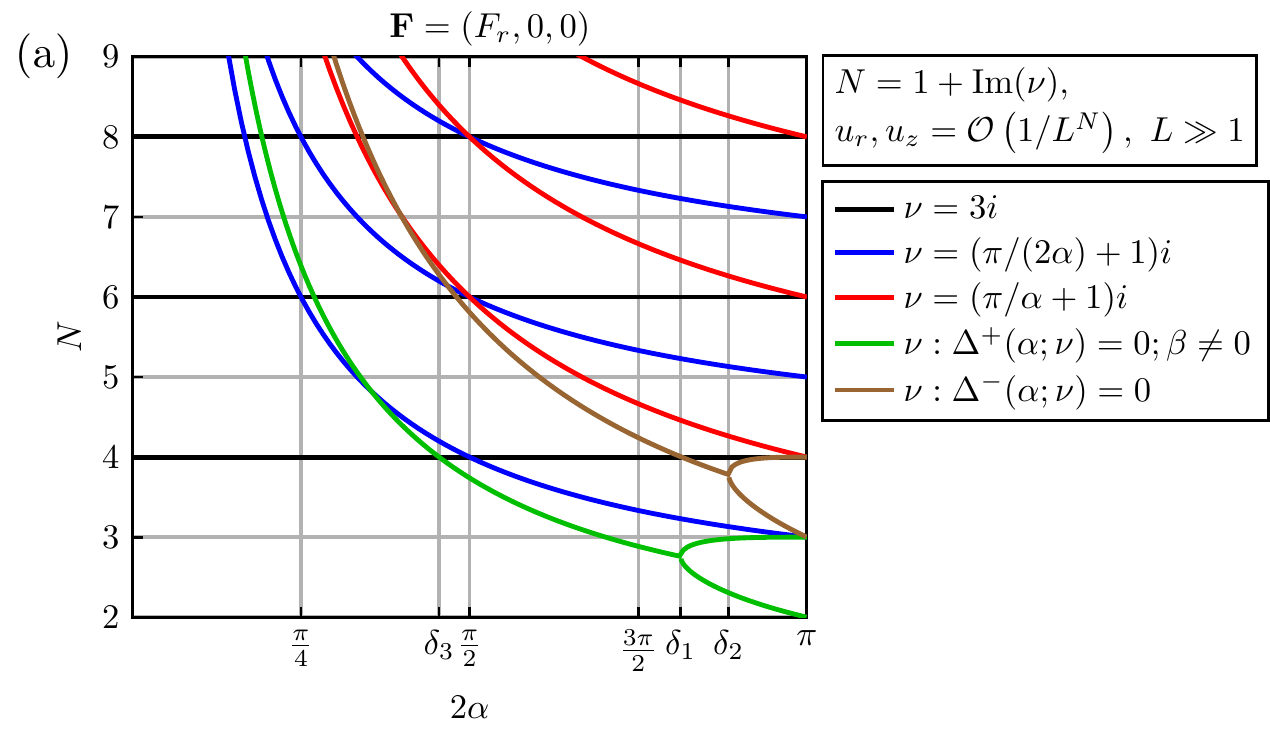}
   \includegraphics[height=0.344\textwidth]{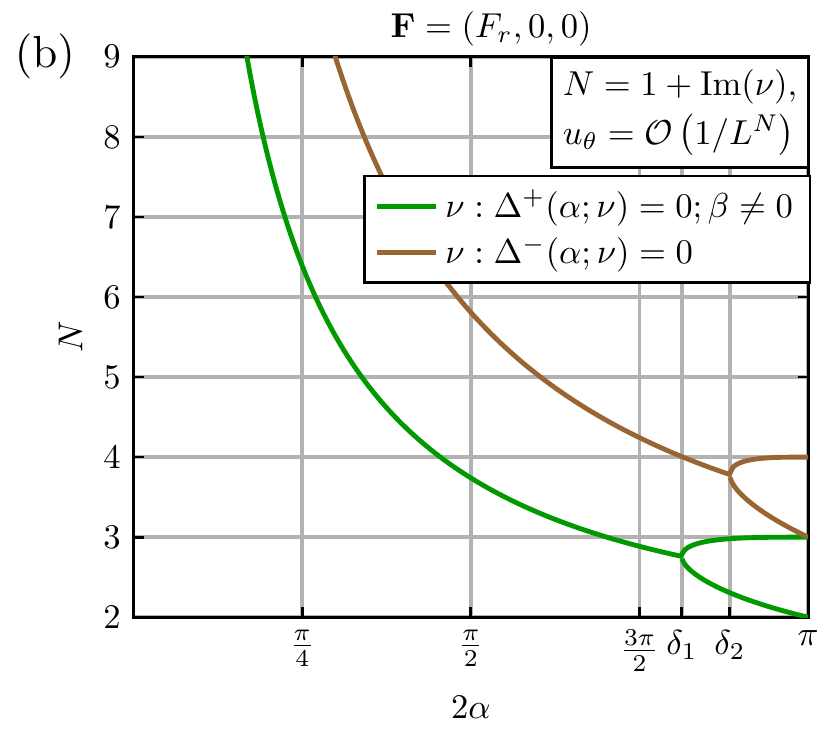}
      \includegraphics[height=0.344\textwidth]{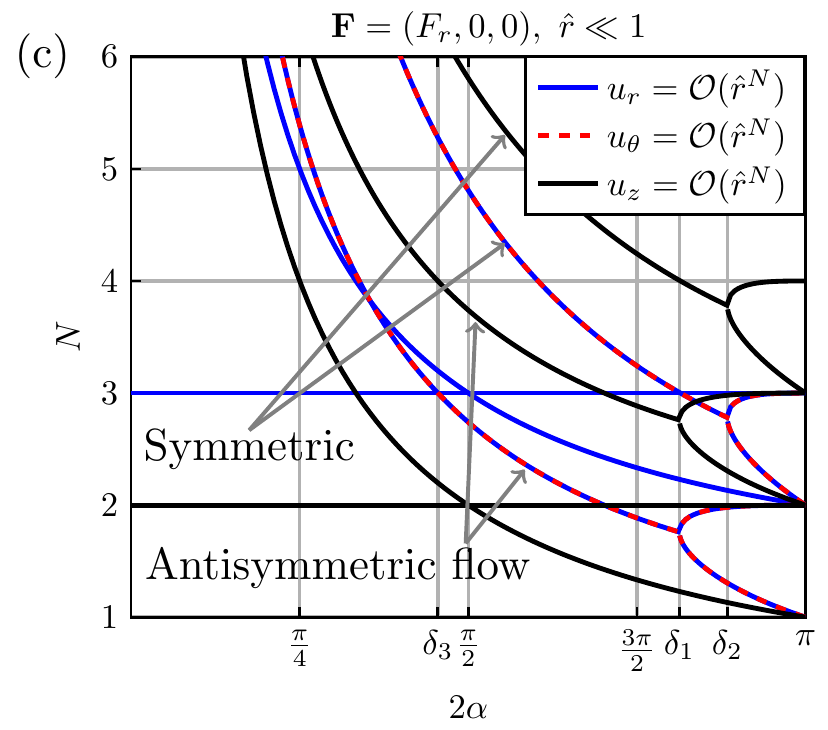}
       \caption{Point force in the radial direction. The flow components in the radial, azimuthal and axial directions, $u_r,u_{\theta},u_z$, decay as $\mathcal{O}(1/L^{1+\Im(\nu)})+\mathcal{O}(1/L^{3+\Im(\nu)})+...$ for $L\gg1$ where $\Im(\nu)$ is the imaginary part of the pole $\nu$. The decay power, $N$, is plotted against the full corner angle, $2\alpha$. (a) Far-field radial and axial velocity components; (b) the flow in the azimuthal direction; (c) flow close to the corner $\hat{r}\ll1$. The critical angles are $\delta_1=146.31^{\circ}$, $\delta_2=159.11^{\circ}$ and $\delta_3=81.87^{\circ}$.}
   \label{fig:9}
\end{figure}

We address now the case where $\gamma=0$, i.e.~$F_{\theta}=0$. The force is in the radial direction. In this case, we have
\begin{align}
\phi_1^{(1)}&= -\frac{1}{R}\cos{(\beta)},~\phi_2^{(1)}= -\frac{1}{R}\sin{(\beta)},~\phi_3^{(1)}= 0,~\phi_4^{(1)}= \frac{1}{R},\\
R&=1+\hat{r}^2-2\hat{r}\cos{(\theta-\beta)}+\hat{z}^2.
\end{align}
Considering one of the harmonic functions
\begin{align}
\phi_1^{(2)}&=\frac{8}{\pi^2}\int_0^{\infty} \left(\frac{\sinh{\theta\nu}}{\Delta^{+}}\left[a_1^0 I_{i\nu}+\Re\{a_1^{+} J_{i\nu}\}\right]+\frac{\cosh{\theta\nu}}{\Delta^{-}}\left[b_1^0 I_{i\nu}+\Re\{b_1^{+} J_{i\nu}\}\right]\right)d\nu
\end{align}
we split  it into two parts as
\begin{align}
\phi_1^{(2)}&=\phi_{11}^{(2)}+\phi_{12}^{(2)},\\
\phi_{11}^{(2)}&=\frac{8}{\pi^2}\int_{0}^{\infty} \left(\frac{\sinh{\theta\nu}}{\Delta^{+}}a_1^0 I_{i\nu}+\frac{\cosh{\theta\nu}}{\Delta^{-}}b_1^0 I_{i\nu}\right)d\nu,\\
\phi_{12}^{(2)}&=\frac{8}{\pi^2}\int_{0}^{\infty} \left(\frac{\sinh{\theta\nu}}{\Delta^{+}}\Re\{a_1^{+} J_{i\nu}\}+\frac{\cosh{\theta\nu}}{\Delta^{-}}\Re\{b_1^{+} J_{i\nu}\}\right)d\nu.
\end{align}
The decomposition of the full flow into leading-order flows is illustrated in Fig.~\ref{fig:9}.

\subsubsection{Leading-order flow for the acute corner, $0<2\alpha<\pi/2$}
 \begin{figure}
 \includegraphics[width=0.49\textwidth]{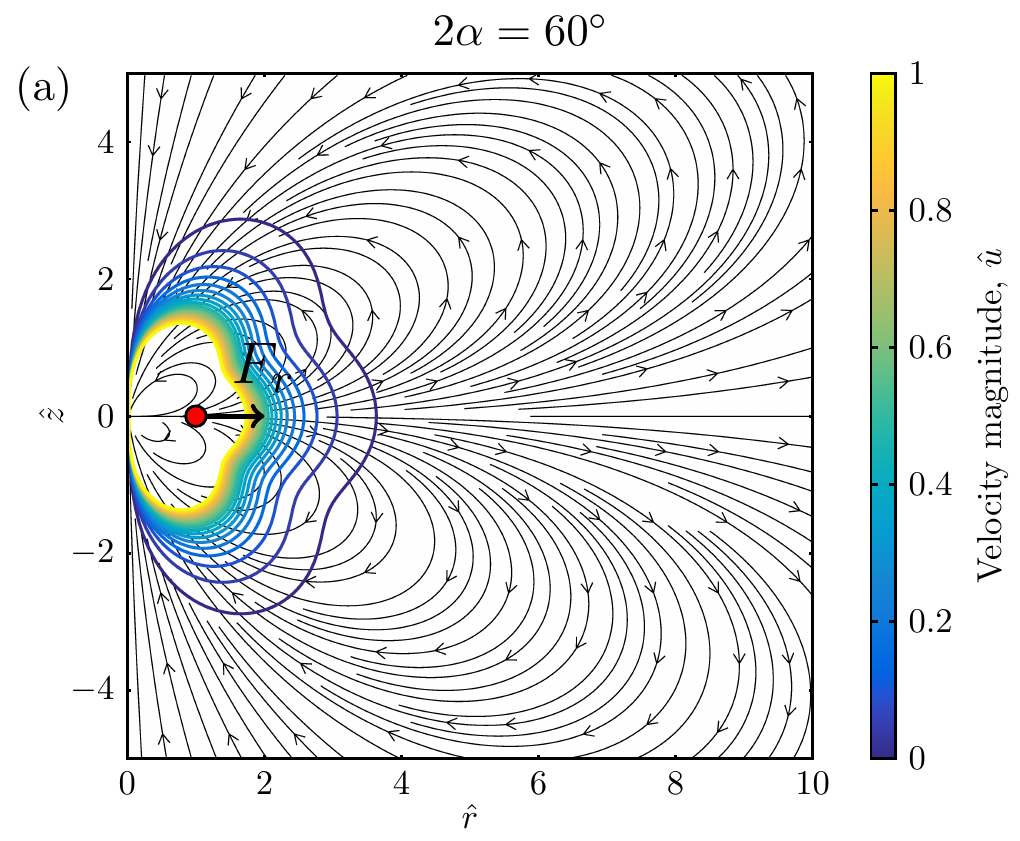}
 \includegraphics[width=0.49\textwidth]{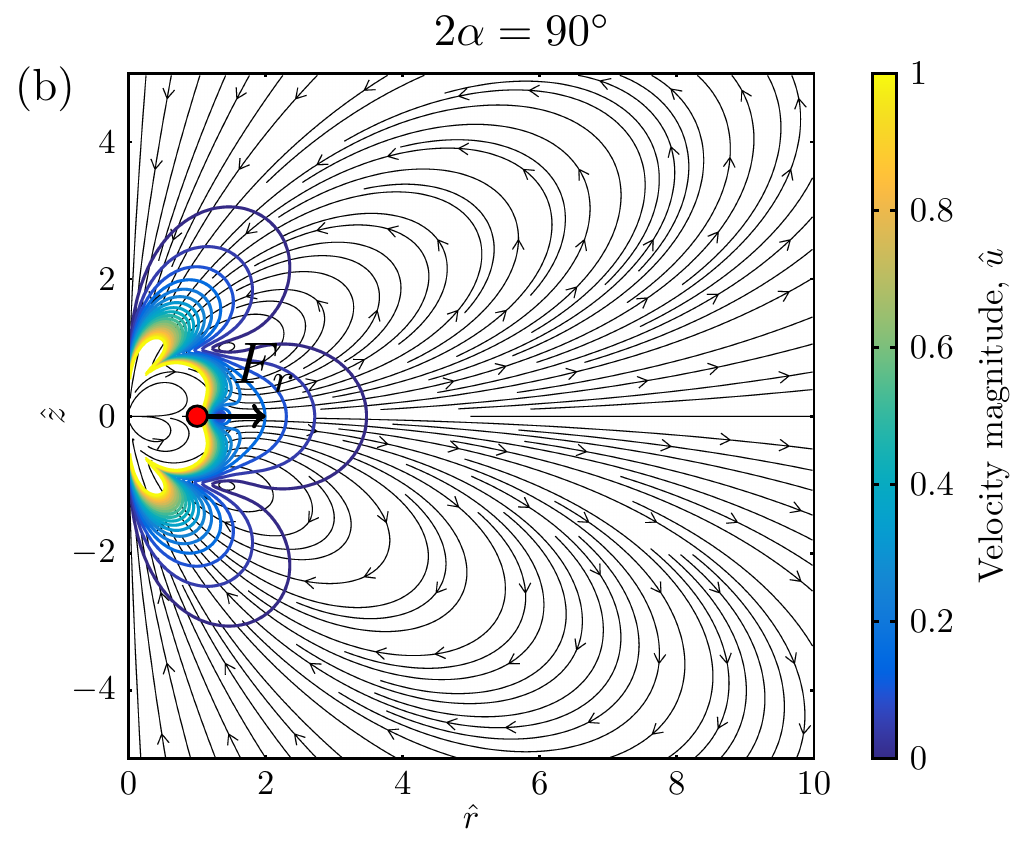}  
 \includegraphics[width=0.49\textwidth]{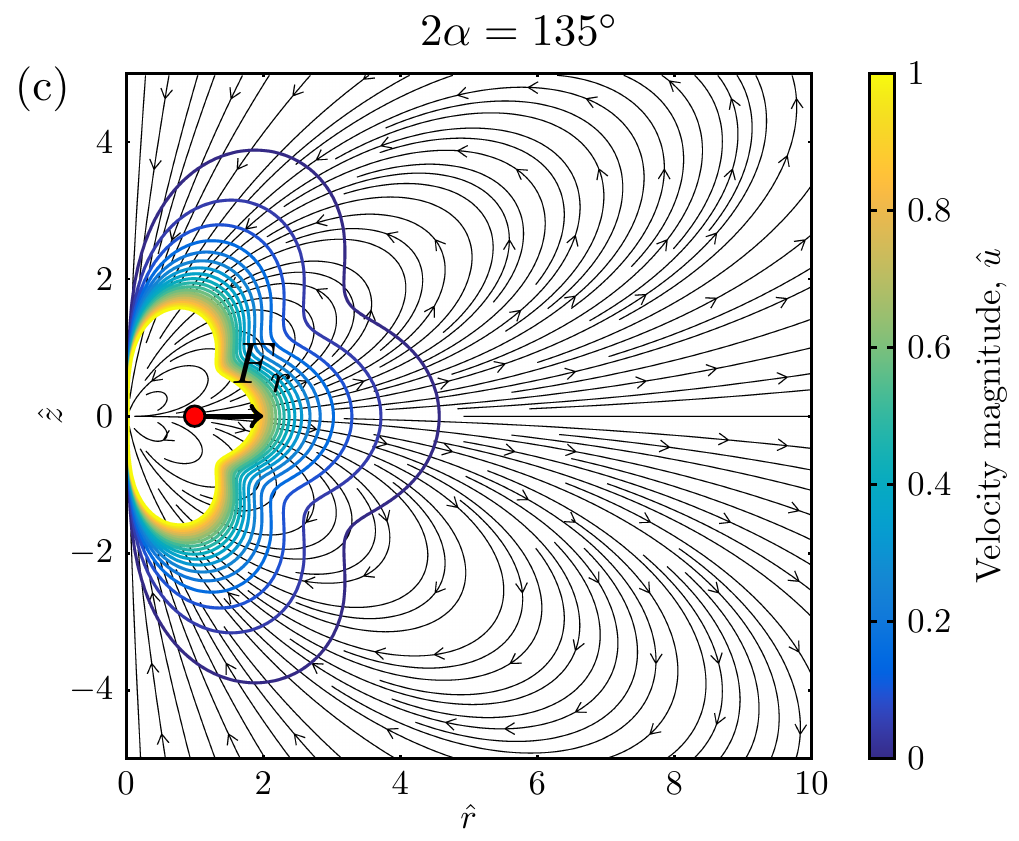}
 \includegraphics[width=0.49\textwidth]{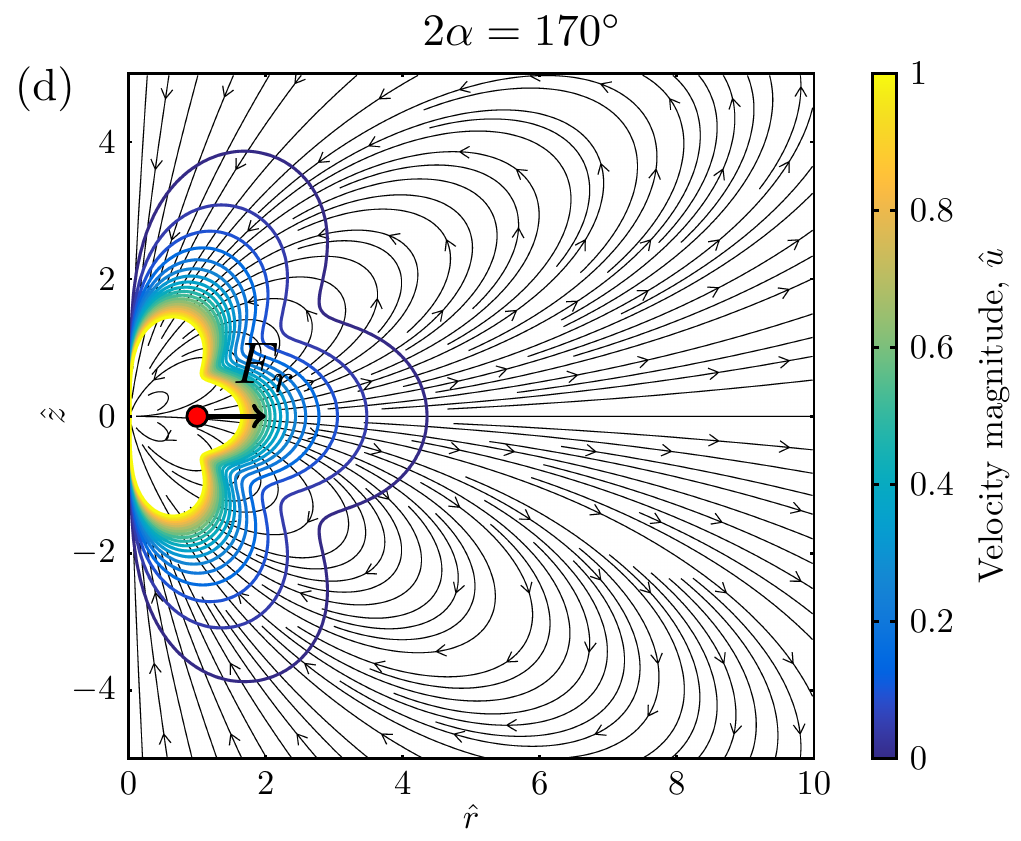}
              \caption{Streamlines and contours of velocity magnitude, $\hat{u}=8\pi\mu\rho u/F_r=\hat{u}_r^2+\hat{u}_z^2$,  for the leading-order 2D flow and a radial point force. (a) $2\alpha=60^{\circ}$, (b) $2\alpha=90^{\circ}$, (c) $2\alpha=135^{\circ}$, (d) $2\alpha=170^{\circ}$. For all cases $\theta=\beta=0$. The flow is valid in the far field, i.e.~for $\hat{r}^2+\hat{z}^2\gg1$.}
   \label{fig:10}
\end{figure}
We can calculate the leading-order flow in the cases $0<2\alpha<\pi/2$ as before. The leading-order harmonic functions decay as $\phi_1=\phi_2=\mathcal{O}(1/L^4),~\phi_3=\phi_4=0$ and the flow is given by
\begin{align}
u_r&=-\frac{f(\alpha,\beta,\theta)\hat{r}^3(\hat{r}^2-6\hat{z}^2)}{(\hat{r}^2+\hat{z}^2)^{9/2}},\\
u_{\theta}&=\mathcal{O}\left(\frac{1}{L^{1+\Im(\nu^{*})}}\right) :\Delta^{+}(\alpha;\nu^{*}=0), \beta\neq0 \text{ and } \Delta^{-}(\alpha;\nu^{*}=0), \beta=0, \\
u_z&=-\frac{f(\alpha,\beta,\theta)\hat{r}^2 \hat{z} (3\hat{r}^2-4\hat{z}^2)}{(\hat{r}^2+\hat{z}^2)^{9/2}},
\end{align}
\begin{align}
f(\alpha,\beta,\theta)&= \frac{F_r}{8\pi\mu\rho}\frac{15\pi(1-\cos{2\beta}\sec{2\alpha})(1-\cos{2\theta}\sec{2\alpha})}{8(2\alpha-\tan{2\alpha})}\cdot
\end{align}
The streamfunction for the flow defined as 
\begin{align}
\vec{u}&=\nabla \times (\Psi \vec{e}_{\theta}),~u_r=-\frac{\partial \Psi}{\partial z},~u_{\theta}=0,~u_z=\frac{1}{r}\frac{\partial (r\Psi)}{\partial r},
\end{align}
is thus given by
\begin{align}
\label{eq:r_acute_f}
\Psi&=  \frac{\rho f(\alpha,\beta,\theta)\hat{r}^3 \hat{z}}{(\hat{r}^2+\hat{z}^2)^{7/2}},
\end{align}
where $\hat{r}\Psi=C$ for some constant $C$ defines streamlines. The streamlines and velocity magnitude contours are shown in  Fig.~\ref{fig:10}a for the case where $\theta=0$, $\beta=0$, $2\alpha=60^\circ$.  We note that the  structure of the streamlines is independent of angles $\theta, \beta, \alpha$ and  it is only the magnitude of the flow which depends on these angles. The flow is essentially two dimensional since $u_r,u_z\gg u_{\theta}$, and it decays algebraically as $\mathcal{O}(1/L^4)$ for $L=(\hat{r}^2+\hat{z}^2)^{1/2}\gg1$. Recall that the flow due to the point force in the axial direction, $(0,0,F_z)$, decays slower, $\mathcal{O}(1/L^3)$, for the acute corner case. The pressure field decays to zero at large distance away from the point force, $p=\mathcal{O}(1/L^5)$. It can be obtained from the velocity field.

 \begin{figure}
 \includegraphics[width=0.4\textwidth]{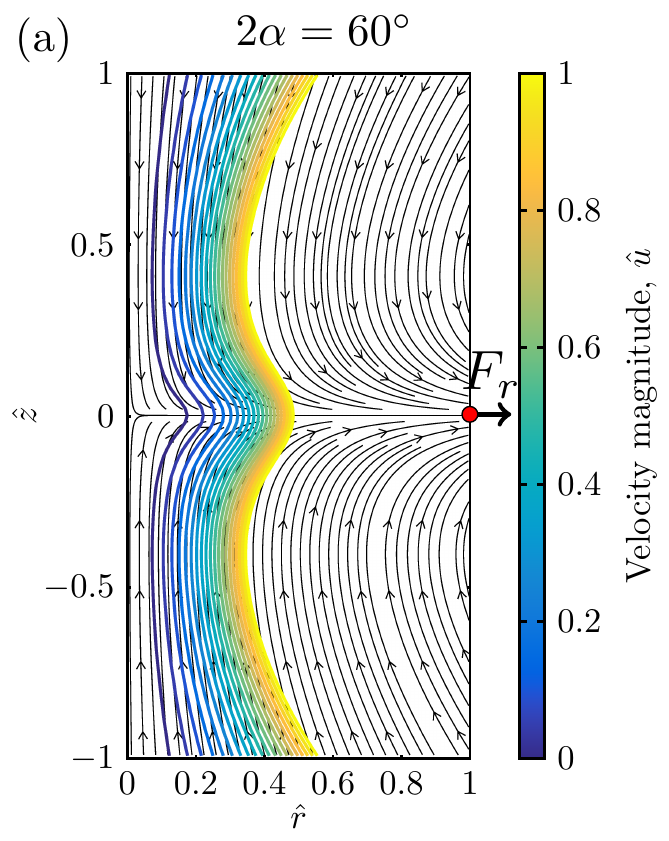}
 \includegraphics[width=0.4\textwidth]{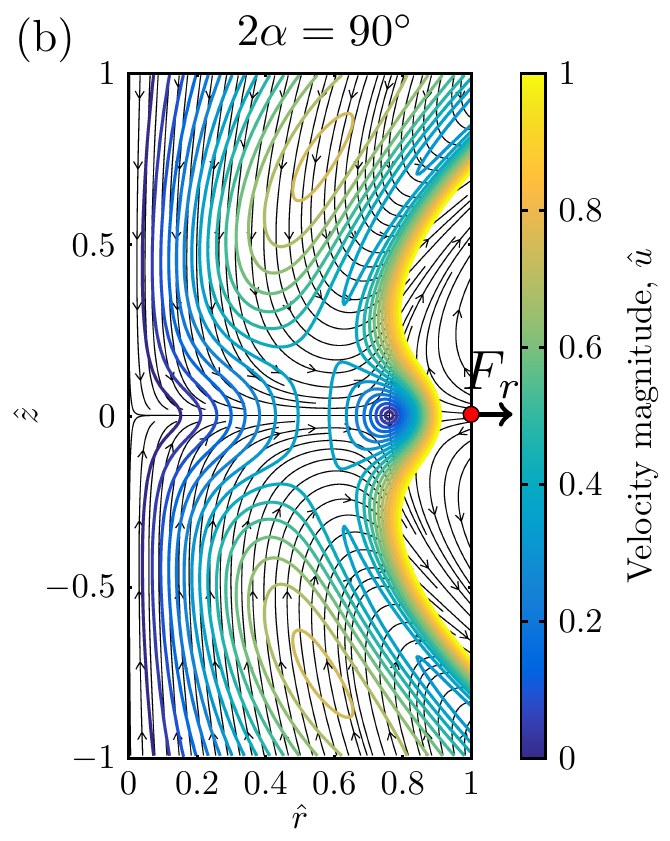}  
 \includegraphics[width=0.4\textwidth]{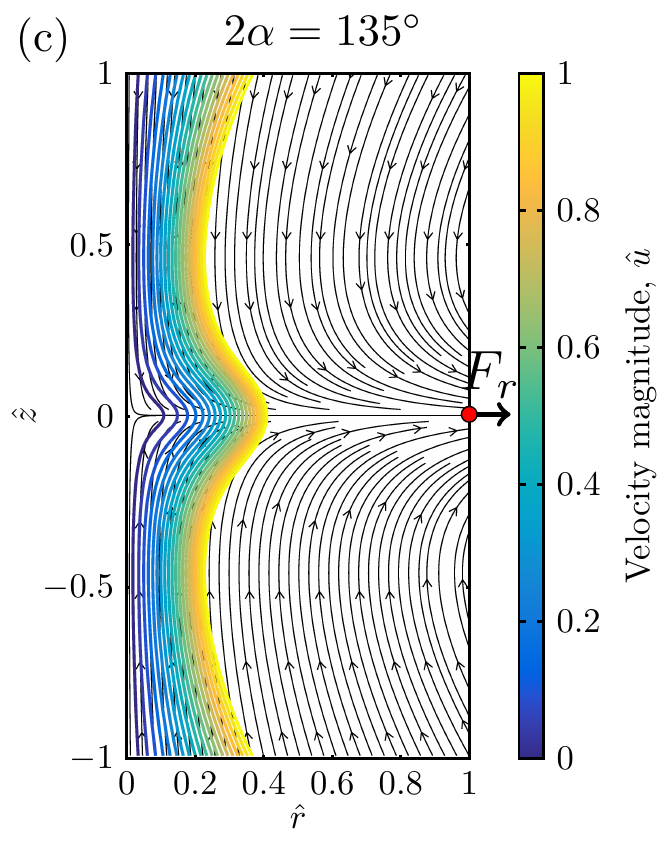}
 \includegraphics[width=0.4\textwidth]{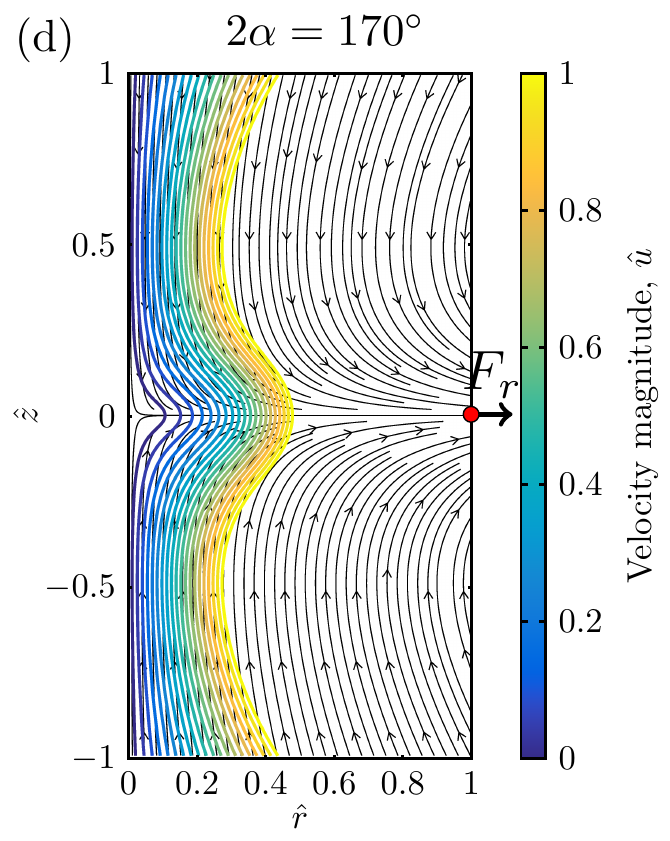}
              \caption{Streamlines and lines of velocity magnitude, $\hat{u}=8\pi\mu\rho u/F_r=\hat{u}_r^2+\hat{u}_z^2$,  for the leading-order 2D flow when $\vec{F}=(0,F_r,0)$. (a) $2\alpha=60^{\circ}$, (b) $2\alpha=90^{\circ}$, (c) $2\alpha=135^{\circ}$, (d) $2\alpha=170^{\circ}$. For all cases $\theta=\beta=0$. The flow is valid near the corner i.e.~in the limit $\hat{r}\ll1$.}
   \label{fig:11}
\end{figure}

Near the corner, $r\ll1$,   the leading-order flow  is given by
\begin{align}
\label{eq:r_acute_c}
\Psi&= \frac{\rho f(\alpha,\beta,\theta)\hat{r}^3\hat{z}}{(1+\hat{z}^2)^{7/2}},~u_r=\mathcal{O}(\hat{r}^3),~u_z=\mathcal{O}(\hat{r}^2).
\end{align}
The streamlines and contour lines are shown in this case in Fig. \ref{fig:11}a. This flow decays at the same rate as the flow near a corner due to the point force in the axial direction.

\subsubsection{Leading-order flow when $2\alpha=\pi/2$}

In the case where the corner has a right angle,  the leading-order flow decays as $\mathcal{O}(L^{-4}\ln{L})+\mathcal{O}(L^{-4})$, with a streamfunction given by
\begin{align}
\label{eq:r_right_f}
\Psi&=\frac{F_r}{8\pi\mu} \Bigg\{30\cos{(2\theta)}\left[\pi-4\beta\sin{(2\beta)}\right]
+30\cos{(2\beta)}\left[\pi-4\theta\sin{(2\theta)}\right]\\
&-\cos{(2\beta)}\cos{(2\theta)\left[-158+15\pi^2+120\ln{\frac{4(\hat{r}^2+\hat{z}^2)}{\hat{r}}}\right]}\Bigg\}\frac{\hat{r}^3\hat{z}}{16(\hat{r}^2+\hat{z}^2)^{7/2}}\cdot\nonumber
\end{align}
The streamfunctions and contour lines are shown in Fig.\ref{fig:10}b. 
The flow in the azimuthal direction decays 
\begin{align}
u_{\theta}&=\mathcal{O}\left(\frac{1}{L^{1+\Im(\nu^{*})}}\right) :\Delta^{+}(\alpha;\nu^{*}=0), \beta\neq0 \text{ and } \Delta^{-}(\alpha;\nu^{*}=0), \beta=0.
\end{align}
If $\hat{r}\ll1$ then we now obtain
\begin{align}
\label{eq:r_right_c}
\Psi&=\frac{F_r}{8\pi\mu} \Bigg\{30\cos{(2\theta)}\left[\pi-4\beta\sin{(2\beta)}\right]
+30\cos{(2\beta)}\left[\pi-4\theta\sin{(2\theta)}\right]\\
&-\cos{(2\beta)}\cos{(2\theta)\left[-158+15\pi^2+120\ln{\frac{4(1+\hat{z}^2)}{\hat{r}}}\right]}\Bigg\}\frac{\hat{r}^3\hat{z}}{16(1+\hat{z}^2)^{7/2}},\nonumber\\
    u_r&=\mathcal{O}(\hat{r}^3\ln{\hat{r}})+\mathcal{O}(\hat{r}^3),~u_z=\mathcal{O}(\hat{r}^2\ln{\hat{r}})+\mathcal{O}(\hat{r}^2).
\end{align}
with streamlines  shown in Fig.~\ref{fig:11}b.

\subsubsection{Leading-order flow for the obtuse corner, $\pi/2<2\alpha<\pi$}
The flow field for the case when $\pi/2<2\alpha<\pi$ is given by
\begin{align}
u_r&=\frac{f(\alpha,\beta,\theta)}{\alpha}\frac{\hat{r}^{1+\pi/(2\alpha)}[-\alpha \hat{r}^2+(2\alpha+\pi)\hat{z}^2]}{(\hat{r}^2+\hat{z}^2)^{5/2+\pi/(2\alpha)}},\\
u_{\theta}&=\mathcal{O}\left(\frac{1}{L^{1+\Im(\nu^{*})}}\right) :\Delta^{+}(\alpha;\nu^{*}=0), \beta\neq0 \text{ and } \Delta^{-}(\alpha;\nu^{*}=0), \beta=0,\\
u_z&=\frac{f(\alpha,\beta,\theta)}{2\alpha}\frac{\hat{r}^{\pi/(2\alpha)}\hat{z}[-(2\alpha+\pi)\hat{r}^2+(4\alpha+\pi)\hat{z}^2]}{(\hat{r}^2+\hat{z}^2)^{5/2+\pi/(2\alpha)}},\\
f(\alpha,\beta,\theta)&=\frac{F_r}{8\pi\mu\rho}\frac{32\alpha\pi^{1/2} \Gamma{(\frac{3}{2}+\frac{\pi}{2\alpha})}\cos \left(\frac{\pi  \beta}{2 \alpha}\right) \cos\left(\frac{\pi  \theta}{2 \alpha}\right)}{(\pi^2-16\alpha^2)\Gamma{(1+\frac{\pi}{2\alpha})}},
\end{align}
with a streamfunction
\begin{align}
\label{eq:r_obtuse_f}
\Psi&=\frac{\rho f(\alpha,\beta,\theta)\hat{r}^{1+\pi/(2\alpha)}\hat{z}}{(\hat{r}^2+\hat{z}^2)^{3/2+\pi/(2\alpha)}}\cdot
\end{align}
This flow is incompressible and it satisfies no-slip boundary conditions at $\theta=\pm\alpha$. 
The streamlines and velocity magnitude contours are shown in   Fig.~\ref{fig:10}c,d for the case when $\theta=0$, $\beta=0$ for two obtuse angles ($135^\circ$ and $170^\circ$). 
Similarly to the acute angle case, the streamline structure is independent of angles $\theta, \beta$ and only the magnitude of the flow depends on these angles. But this time the spatial power decay of the  flow  depends on the value of the corner angle, $2\alpha$.   The flow is nearly two dimensional, because $u_r,u_z\gg u_{\theta}$, and it decays algebraically as $\mathcal{O}(1/L^{2+\pi/(2\alpha)})$ for $L=(\hat{r}^2+\hat{z}^2)^{1/2}\gg1$, one power of $L$ quicker than the flow due to the point force in the axial direction.
In the case when $2\alpha$ tends to $\pi$ we obtain that this solution decays as $\mathcal{O}(1/L^{3})$ which is the expected  decay rate for the leading-order solution for the perpendicular force above  a flat wall.  Adding the eddy-type solution, see Fig.~\ref{fig:9}a-b, we then  recover Blake's classical solution for a point force above a wall  \cite{blake1974fundamental}. 

Now, in the limit $r\ll1$ then we obtain the flow near the corner as
\begin{align}
\label{eq:r_obtuse_c}
\Psi&=\frac{\rho f(\alpha,\beta,\theta)\hat{r}^{1+\pi/(2\alpha)}\hat{z}}{(1+\hat{z}^2)^{3/2+\pi/(2\alpha)}},
\end{align}
with streamlines and contour lines shown in Fig. \ref{fig:11}c,d. This flow decays at the same rate as the flow near a corner due to the point force in the axial direction.

\subsubsection{Leading-order eddy-type flow}
We now calculate the flow due to the poles when $\Delta^{\pm}(\alpha;\nu)=\sinh{2\alpha\nu}\pm\nu\sin{2\alpha}$. The antisymmetric flow component is given by the poles $\Delta^{+}(\alpha;\nu)=0$ while the symmetric one by those for $\Delta^{-}(\alpha;\nu)=0$. The flow due to the leading-order pole which satisfies $\Delta^{+}(\alpha;\nu)=0$ is given by calculating the integrals
\begin{align}
\phi_1^{(2)}&=\frac{8}{\pi^2}\int_0^{\infty} \left(\frac{\sinh{\theta\nu}}{\Delta^{+}}a_1^0 I_{i\nu}\right)d\nu+\frac{8}{\pi^2}\int_0^{\infty} \left(\frac{\sinh{\theta\nu}}{\Delta^{+}}\Re\{a_1^{+} J_{i\nu}\}\right)d\nu,\\
\phi_2^{(2)}&=\frac{8}{\pi^2}\int_0^{\infty} \left(\frac{\cosh{\theta\nu}}{\Delta^{+}}b_2^0 I_{i\nu}\right)d\nu+\frac{8}{\pi^2}\int_0^{\infty} \left(\frac{\cosh{\theta\nu}}{\Delta^{+}}\Re\{b_2^{+} J_{i\nu}\}\right)d\nu.
\end{align}
In the sum the first part is very similar to the azimuthal case except now $\gamma=0$ instead of $\gamma=\pi/2$, so we obtain
%\begin{align}
%N_r=&N_z=-\cos{\beta}\sinh{\beta\nu}(\nu\cos^2{\alpha}\sin{\theta}\cosh{\theta\nu}+\cosh^2{\alpha\nu}
%\cos{\theta}\sinh{\theta\nu})\\\nonumber
%&+\sin{\beta}\cosh{\beta\nu}(\sinh^2{\alpha\nu}\sin{\theta}\cosh{\theta\nu}+\nu\sin^2{\alpha}
%\cos{\theta}\sinh{\theta\nu}),\\
%N_{\theta}=&(\nu\sin^2{\alpha}\sin{\beta}\cosh{\beta\nu}-
%\cosh^2{\alpha\nu}\cos{\beta}\sinh{\beta\nu})(\nu\cos{\theta}\cosh{\theta\nu}+
%\sin{\theta}\sinh{\theta\nu})\\\nonumber
%&+(\sinh^2{\alpha\nu}\sin{\beta}\cosh{\beta\nu}-
%\nu\cos^2{\alpha}\cos{\beta}\sinh{\beta\nu})(\nu\sin{\theta}\sinh{\theta\nu}-\cos{\theta}\cosh{\theta\nu}).
%\end{align}
%otice that if $\beta=0$ then there is no asymmetric flow, i.e.~$N_r=N_z=N_{\theta}=0$. To calculate the second part use the residue theorem.
\begin{align}
\label{eq:red1}
\phi_1^{(2)}=-\Re\left\{\frac{\sinh{\theta\nu}}{\Delta^{+'}(\alpha;\nu)}\left(\frac{\hat{I}}{\sinh{(\pi\nu)}}\left[a_1^{0}+\frac{1}{2}\Re(a_1^{+})-\nu\Im (a_1^{+})\right]+\frac{\hat{J}\Re(a_1^{+})}{\sinh{(\pi\nu)}}\right)\right\},\\
\label{eq:red2}
\phi_2^{(2)}=-\Re\left\{\frac{\cosh{\theta\nu}}{\Delta^{+'}(\alpha;\nu)}\left(\frac{\hat{I}}{\sinh{(\pi\nu)}}\left[b_2^{0}+\frac{1}{2}\Re(b_2^{+})-\nu\Im (b_2^{+})\right]+\frac{\hat{J}\Re(b_2^{+})}{\sinh{(\pi\nu)}}\right)\right\},
\end{align}
where
\begin{align}
I(\hat{r},\hat{z};\nu)&=-4\pi^{1/2}\hat{r}^{-1/2}(2\xi)^{i\nu-1/2}F\left(-\frac{i\nu}{2}+\frac{3}{4},-\frac{i\nu}{2}+\frac{1}{4};1-i\nu;\xi^{-2}\right)\frac{\Gamma{\left(\frac{1}{2}-i \nu\right)}}{\Gamma{\left(1-i \nu\right)}},\\
J(\hat{r},\hat{z};\nu)&=-2\pi^{1/2}\frac{(\hat{r}^{-3/2}-\hat{r}^{-1/2}\xi)(2\xi)^{i\nu+1/2}F\left(-\frac{i\nu}{2}+\frac{1}{4},-\frac{i\nu}{2}-\frac{1}{4};1-i\nu;\xi^{-2}\right)}{\xi^2-1}\frac{\Gamma{\left(\frac{3}{2}-i \nu\right)}}{\Gamma{\left(1-i \nu\right)}}\cdot
\end{align}
With these,  one can next compute the flow field. The leading-order flow near a corner, $\hat{r}\ll1$, will be essentially two dimensional, $u_r,~u_{\theta}\gg u_z$, whereas in the the far field, $L\gg1$, the flow is  fully three-dimensional (see illustration in Fig.~\ref{fig:9}).

%%%%%%%%%%%%%%%%%%%%%%%%%%%%%%%
%%%%%%%%%%%%%%%%%%%%%%%%%%%%%%%
%%%%%%%%%%%%%%%%%%%%%%%%%%%%%%%
\section{Experiments}

In order to test the applicability of our theoretical work, we carried out experiments. Specifically, we investigated the sedimentation of smalls steel spheres near a wedge placed in the middle of  a tank filled with corn syrup under the action of gravity. We then use small tracer particles  illuminated by a laser sheet set on the bisectoral plane of the wedge and record their motion with a video camera  placed perpendicular to the bisectoral plane and  taking long exposure images of the motion of both the sphere and the tracer particles.  

 \begin{figure}      
    \includegraphics[width=0.4\textwidth]{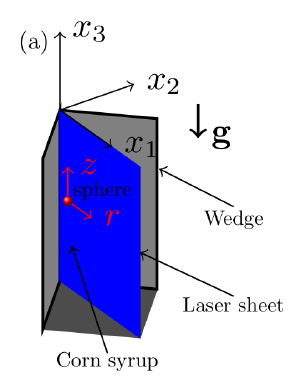}
        \includegraphics[width=0.59\textwidth]{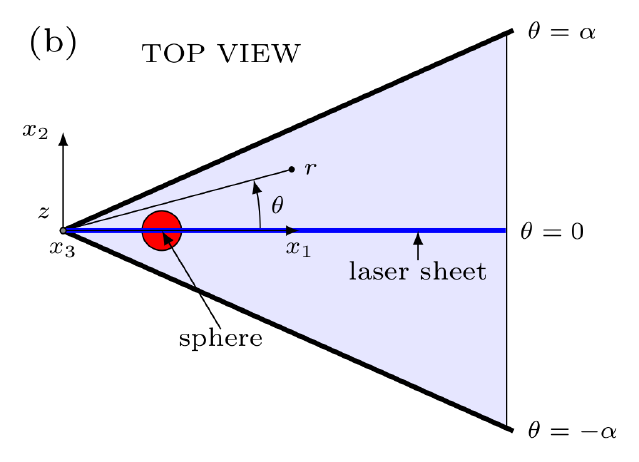}
            \includegraphics[height=0.4\textwidth]{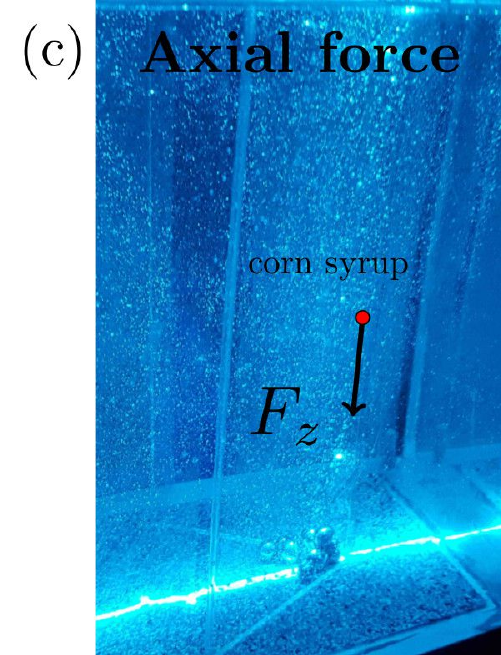}
        \includegraphics[height=0.4\textwidth]{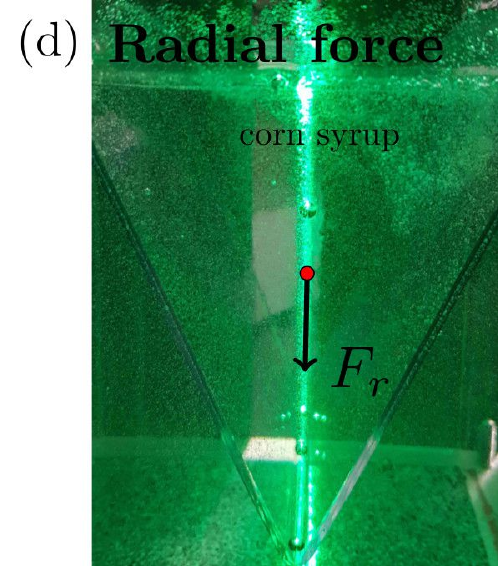}
       \caption{Experimental set-up for the sedimentation of a small sphere  along a rigid wedge. (a)  3D sketch of the wedge inserted into the tank filled with corn syrup mixed with tracer particles; The laser sheet illuminates the tracer particles on the bisectoral plane where the sphere is sedimentating; (b) Top view of the set up; (c) Experimental picture of a  sphere moving along the wedge and (d) towards the edge.}
   \label{fig:88}
\end{figure}

A sketch of the experimental set-up is illustrated in  Fig.~\ref{fig:88}a-b  while the actual experiment is shown in  Fig.~\ref{fig:88}c-d. Specifically, in Fig.~\ref{fig:88}c the sphere is moving in the direction parallel to the corner axis while in Fig.~\ref{fig:88}d it is translating  in the radial direction. The third possibility where the sphere would move in the azimuthal direction proved too difficult to set up experimentally.

 \begin{figure}      
 \includegraphics[height=0.58\textwidth]{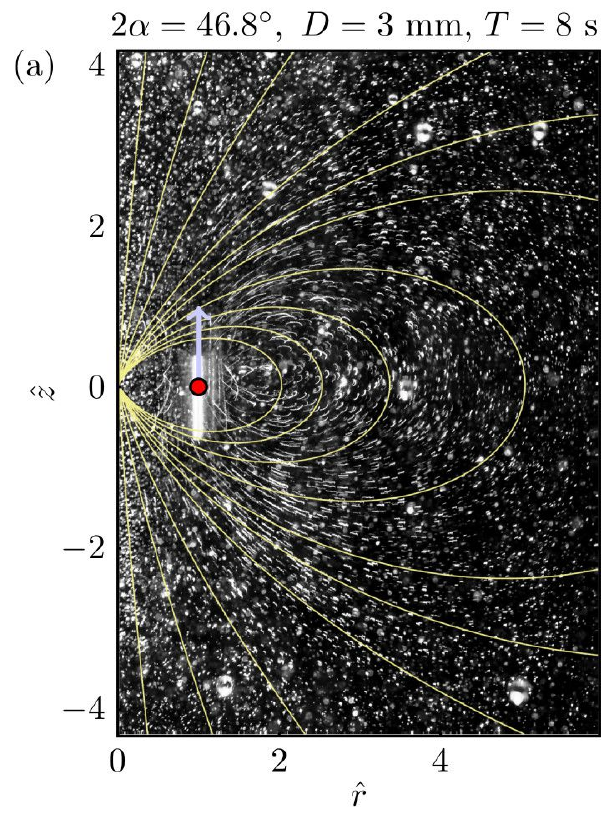}
    \includegraphics[height=0.58\textwidth]{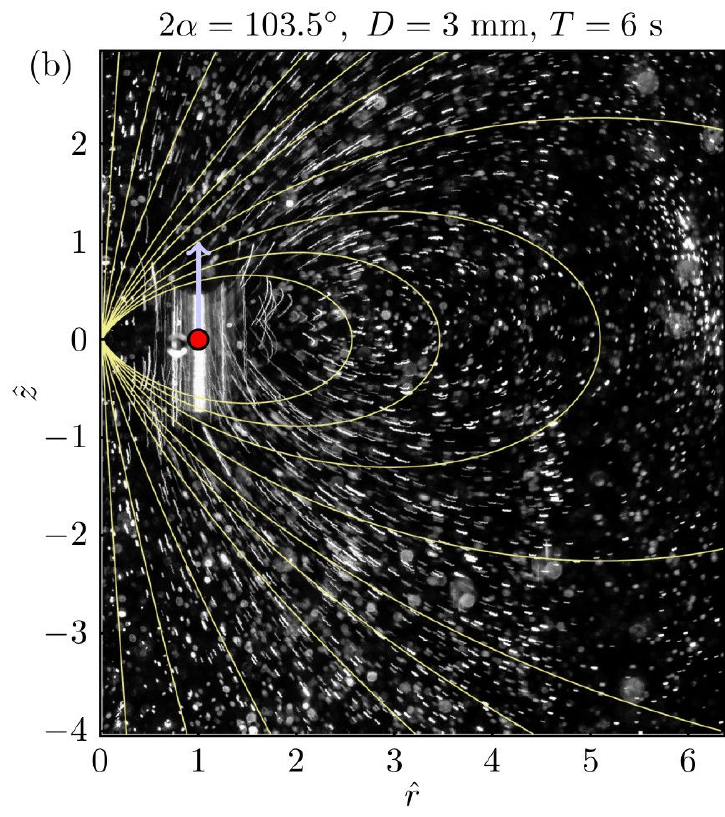}
        \includegraphics[height=0.58\textwidth]{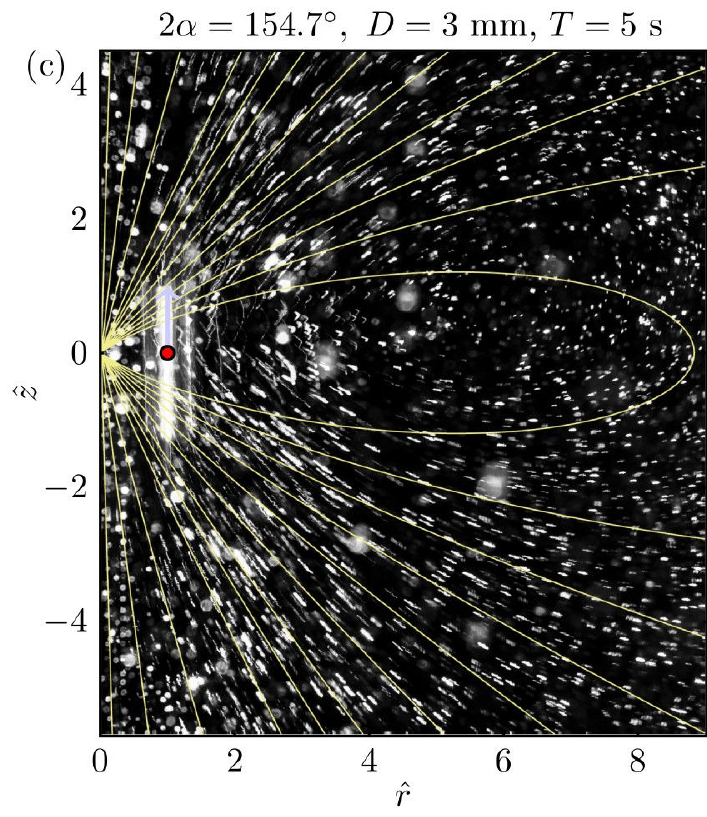}
        \includegraphics[height=0.58\textwidth]{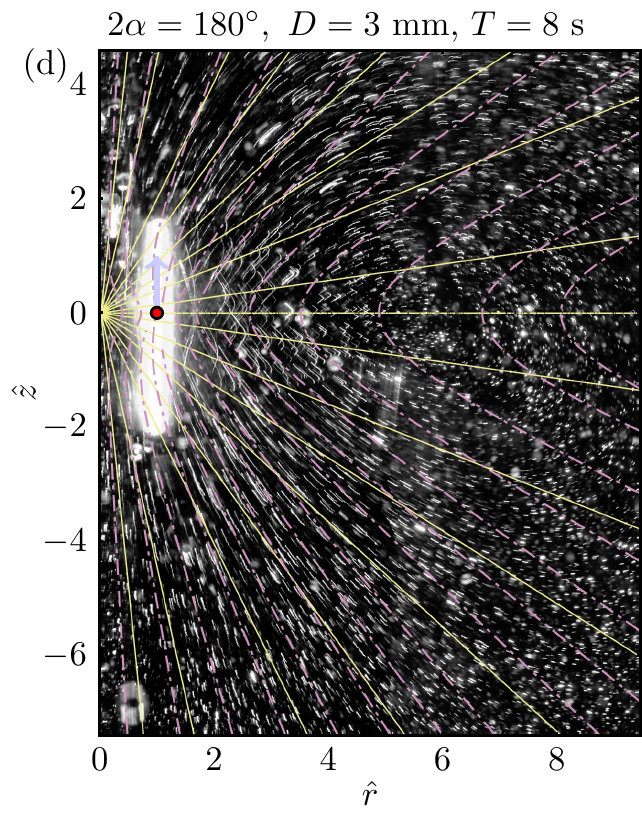}
       \caption{Experiments and theory when the sphere is translating parallel to the corner axis, $\vec{F}=(0,0,F_z)$. For four different values of wedge angle, $2\alpha$, we compare path lines of tracer particles (short white lines) with far-field theoretical predictions of the streamlines (yellow solid lines); the bright white line on the left of each panel indicates the translation of the sphere. The sphere diameter is $D$ and the exposure time $T$. (d) Streamlines for Blake's full solution near a wall   added as  purple dashed lines~\cite{blake1971note} (which agrees with our result in the far field).}
   \label{fig:80}
\end{figure}
 
\begin{figure}      
    \includegraphics[height=0.7\textwidth]{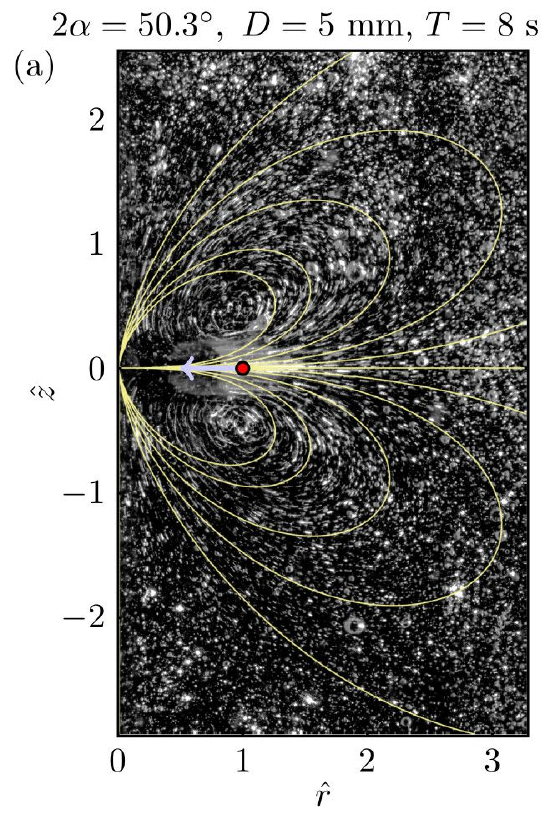}
                    \includegraphics[height=0.7\textwidth]{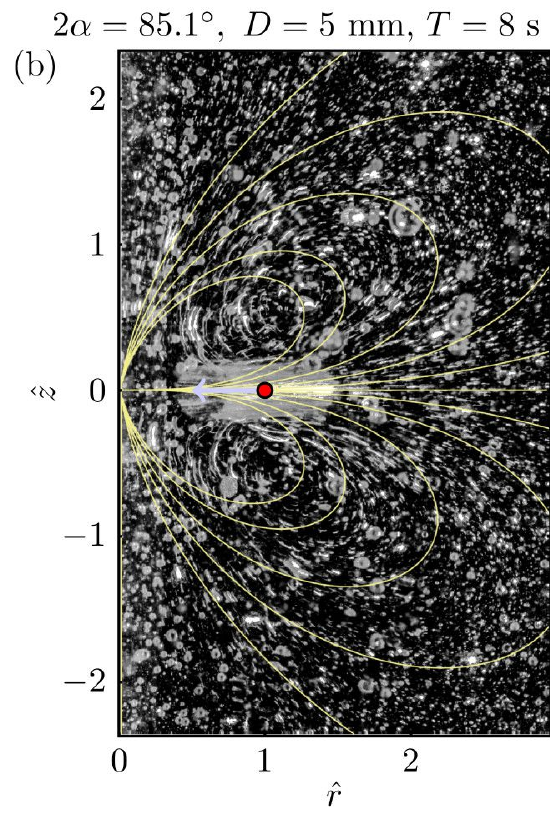}
       \caption{Comparison between experiments and theory in the case where the sphere is translating radially toward the corner axis, $\vec{F}=(F_r,0,0)$ (same details as in Fig.~\ref{fig:80}).}
   \label{fig:81}
\end{figure}

Our experimental results are shown in Fig.~\ref{fig:80} (axial case) and Fig.~\ref{fig:81} (radial case) where we superimpose time-lapse images of the tracer particles (path lines) with our theoretical predictions for the flow far from the sphere. We obtain  qualitative agreement for the far-field streamlines in both cases.  When the sphere is translating along the edge we observe that for the acute corner angle, i.e.~$2\alpha<\pi/2$, the measured streamline structure matches the theoretical prediction, see Fig.~\ref{fig:80}a. For the larger corner angles the streamlines are more in the radial direction, as shown in Fig.~\ref{fig:80}b-d. Note that the path lines recorded in the experiments can match the streamlines in the far field because the  position of the sphere,  which is a function of time, satisfies $z_0(t)\ll (z^2+r^2)^{1/2}$ in the far field and is therefore almost stationary during the exposure time $T$. Theoretically, we also predicted in this case the recirculation regions close to the corner, as shown in   Fig.~\ref{fig:5}, but experimentally it has proven difficult to visualise the flow close to the corner because the sphere is moving.  In the radial case, shown in Fig.~\ref{fig:81}, the additional difficult arises from the slowing down of the sphere (which translated at a steady speed in the axial case) but  we can still observe qualitatively the expected recirculation field.

%%%%%%%%%%%%%%%%%%%%%%%%%%%%%%%
%%%%%%%%%%%%%%%%%%%%%%%%%%%%%%%
%%%%%%%%%%%%%%%%%%%%%%%%%%%%%%%
\section{Discussion}

In this paper, we   computed the leading-order  flows for a point force orientated in an arbitrary direction near a corner both in the near and far field. The flow always recirculates along the axial direction, and therefore the net flux produced in the direction parallel to both walls is zero for all angles with $0<2\alpha<\pi$. The pressure field is not sufficiently high to drive a Poiseuille-type flow; recall that by contrast a point force along the axial direction of a pipe leads to a finite pressure jump \cite{liron1978stokes}. 

\subsection{Leading-order flow near the corner}
\begin{figure}      
    \includegraphics[width=0.45\textwidth]{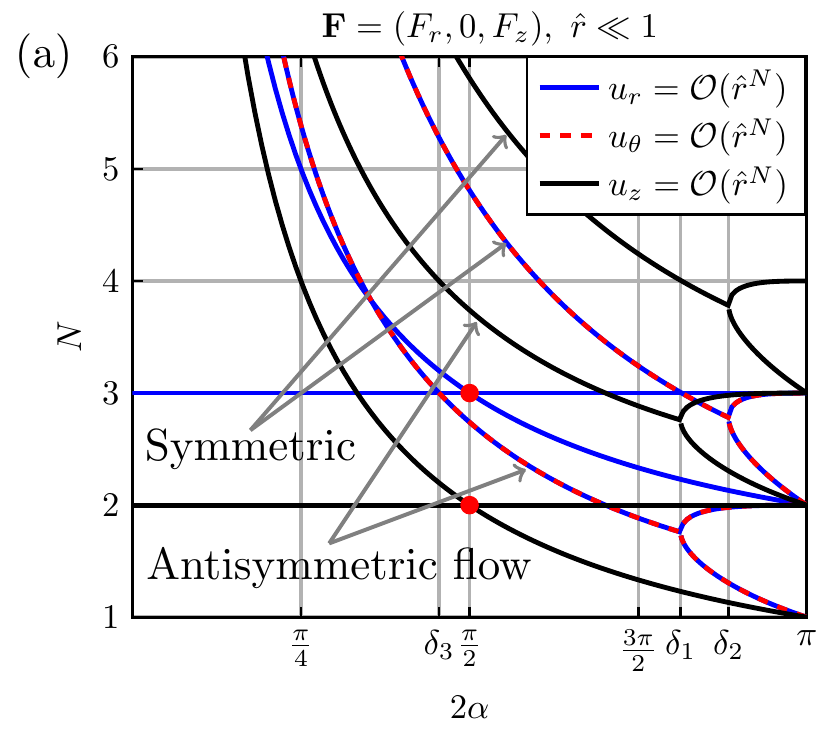}
    \includegraphics[width=0.45\textwidth]{fp_close_all.pdf}
       \caption{Decomposition of the full flow due to a point force, $\vec{F}=(F_r,F_{\theta},F_z)$, close to the corner, $\hat{r}\ll1$. The decay power, $N$, is plotted against the value of the corner angle, $2\alpha$: (a) Flow decay when the point force is in the radial or axial directions; (b)  point force in the azimuthal direction. Gray arrows show symmetric and antisymmetric flow components when they are present. The critical angles are $\delta_1=146.31^{\circ}$, $\delta_2=159.11^{\circ}$, and $\delta_3=81.87^{\circ}$.}
   \label{fig:33}
\end{figure}

The summary of the flow close to the corner, i.e.~$r/\rho=\hat{r}\ll1$, due to the point force $(F_r,F_{\theta},F_z)$ is shown in Fig.~\ref{fig:33}. When the point force is either in the radial or axial direction, $(F_r,0,F_z)$, the flow is dominated by the axial flow component $u_z$, see Fig.~\ref{fig:33}a. When the point force is in azimuthal direction $(0,F_{\theta},0)$, the dominant flow is eddy-type, as shown in Fig.~\ref{fig:33}b. The flow close to the corner has a recirculation cell Fig.~\ref{fig:5} when the point force has a component in the axial direction.

The relative scaling $u_r:u_{\theta}:r u_z$ obtained from the incompressibility equation, $\nabla\cdot\vec{u}=0$, shows that the axial flow component near a corner $\hat{r}\ll1$ is dominant if it is present. For acute angles, $0<2\alpha<\pi/2$, the axial flow decays as $u_z=\mathcal{O}(\hat{r}^2)$, as seen in equations \eqref{eq:z_acute_c} for an axial force  and \eqref{eq:r_acute_c} for a radial one.  In the right angle case, $2\alpha=\pi/2$, the axial flow decays as $u_z=\mathcal{O}(\hat{r}^2\ln\hat{r})$, see equations \eqref{eq:z_right_c} for $(0,0,F_z)$ and \eqref{eq:r_right_c} for $(F_r,0,0)$. For obtuse angles, $\pi/2<2\alpha<\pi$, the decay depends on the value of angle as $u_z=\mathcal{O}(\hat{r}^{\pi/(2\alpha)})$, as shown in equations \eqref{eq:z_obtuse_c}  in the axial case and  \eqref{eq:r_obtuse_c} in the radial case. The radial flow $u_r$ is bigger than the azimuthal flow $u_{\theta}$ when $2\alpha<\delta_3=81.87^{0}$ and smaller when $\delta_3<2\alpha$, see Fig.~\ref{fig:33}a. When the point force in the azimuthal direction $(0,F_{\theta},0)$, the flow  at leading order is in radial and azimuthal directions, see equation \eqref{eq:p_c}.  

Interestingly, our results point to a strong analogy with the local flow driven by a pressure gradient along a wedge \cite{moffatt1980local}. In that case also the leading-order flow is $u_z=\mathcal{O}(\hat{r}^2)$ for $0<2\alpha<\pi/2$, $u_z=\mathcal{O}(\hat{r}^2\ln\hat{r})$ for $2\alpha<\pi/2$ and $u_z=\mathcal{O}(\hat{r}^{\pi/(2\alpha)})$ for $\pi/2<2\alpha<\pi$ close to the corner.

 Finally, there will be an infinite number of eddies close to the corner for symmetric and antisymmetric flows if the leading pole, $\nu$, satisfying equations $\sinh{2\alpha\nu}\pm\nu\sin{2\alpha}=0$, a result consistent with previous work  \cite{shankar1998class}. This eddy-type flow is not dominant if there is the radial or axial component of the point force present in the system. The critical angles for antisymmetric and symmetric flows are $\delta_1=146.31^{\circ}$ and $\delta_2=159.11^{\circ}$.

%%%%%%%%
\subsection{Leading-order flow far from the point force}

 \begin{figure}    
  \includegraphics[width=0.475\textwidth]{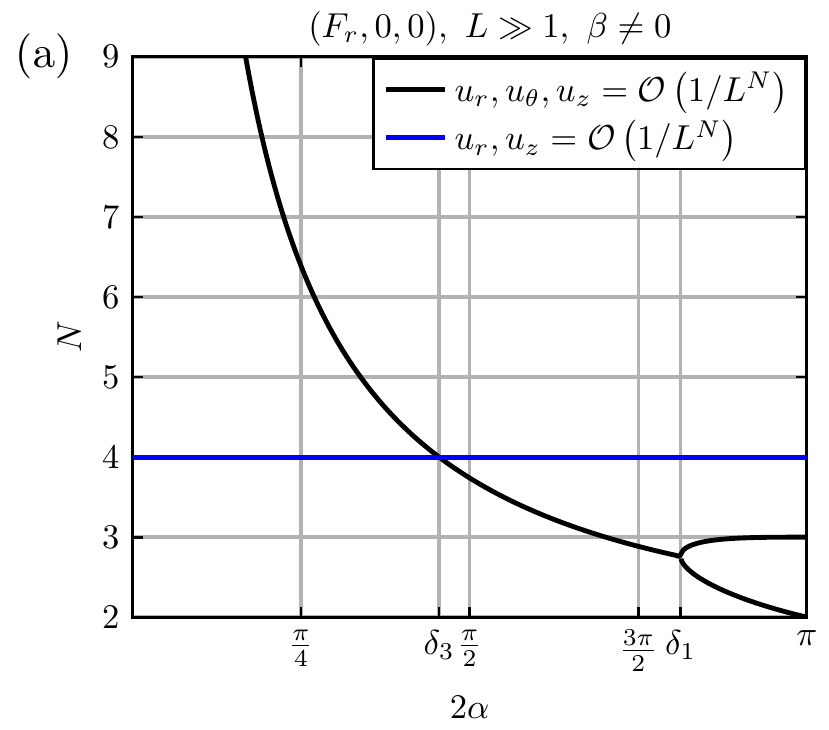}
   \includegraphics[width=0.475\textwidth]{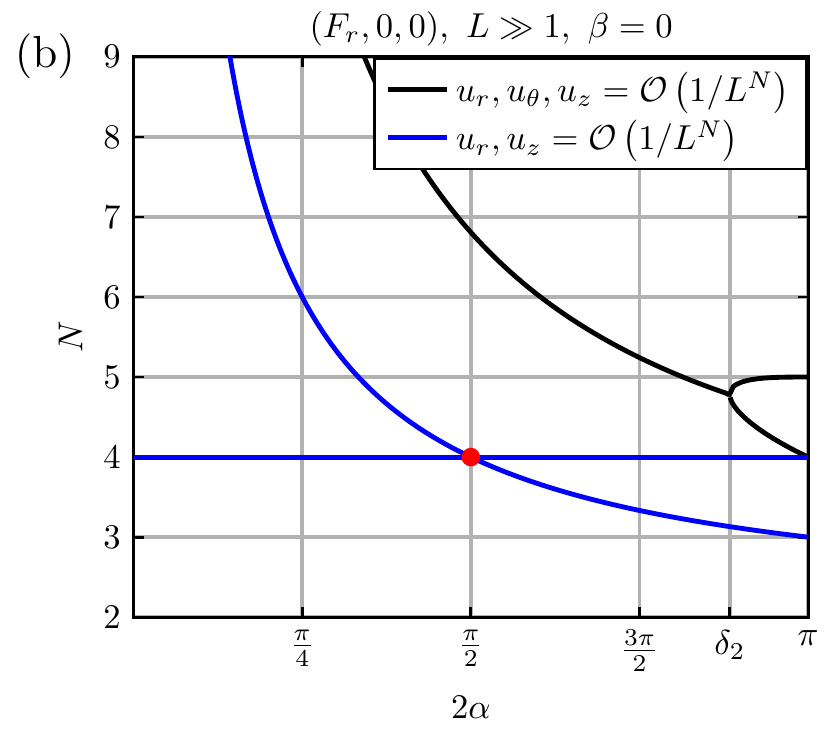}
     \includegraphics[width=0.475\textwidth]{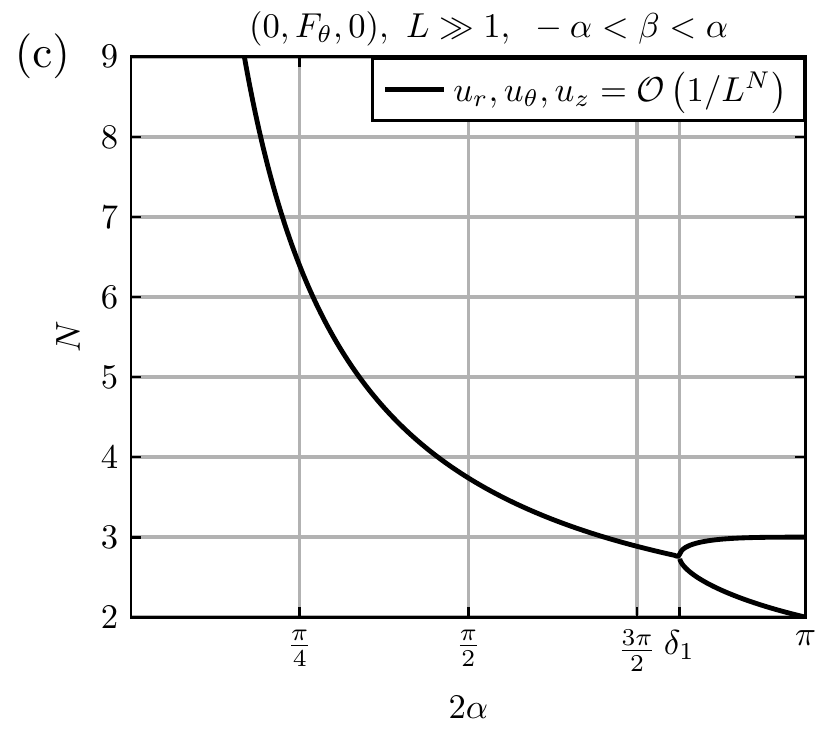}
   \includegraphics[width=0.475\textwidth]{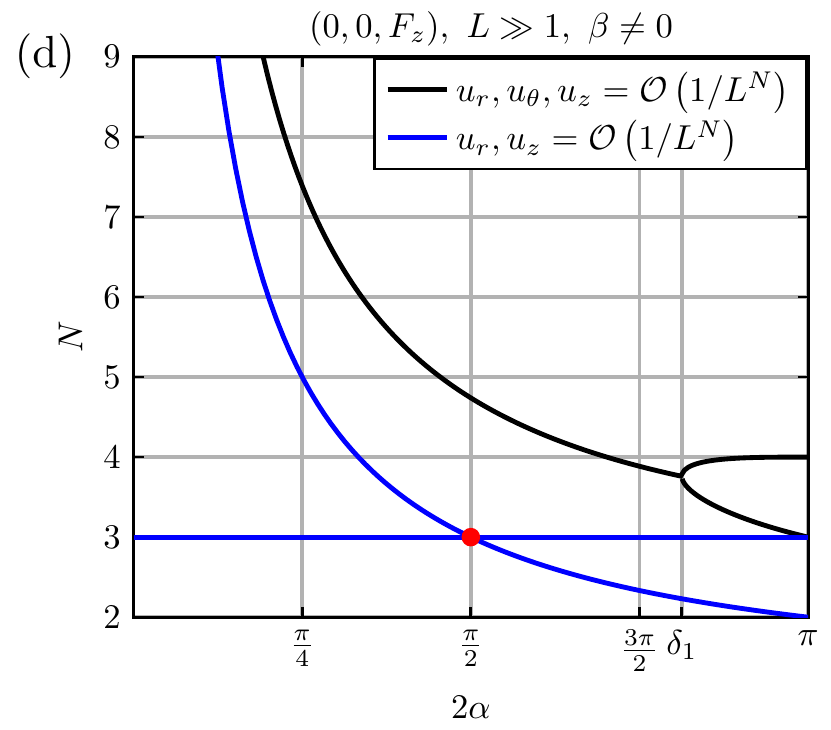}
      \includegraphics[width=0.475\textwidth]{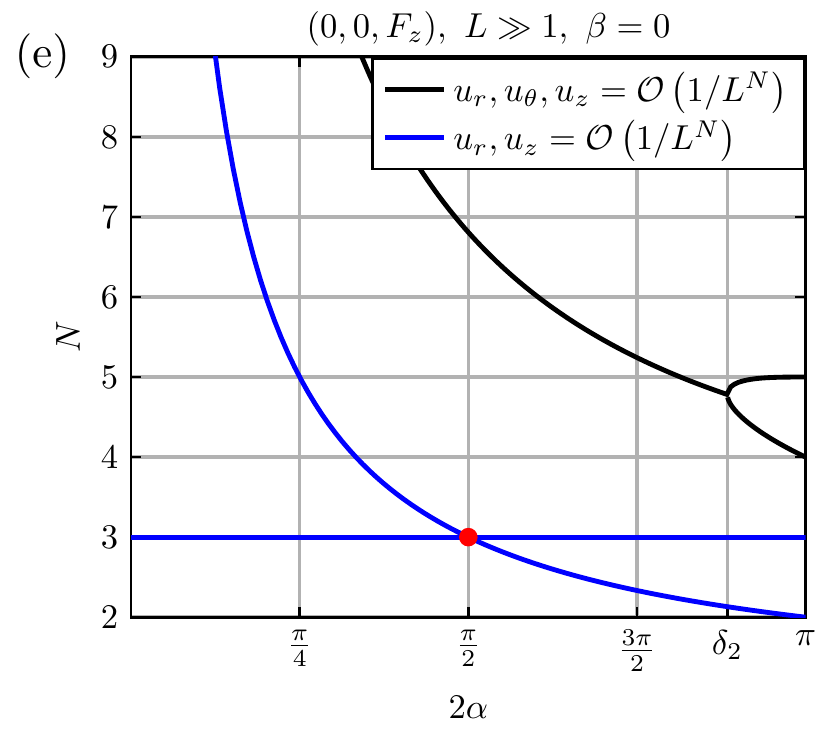}
       \caption{Decay rate, $N$, for the flow $(u_r,u_{\theta},u_z)$   from the point force  $(F_r,F_{\theta},F_z)$ located at $(\rho,\beta,0)$ in the far field limit, $L^2=\hat{r}^2+\hat{z}^2\gg1$, as a function of corner angle $2\alpha$: (a) $(F_r,0,0)$, $\beta\neq0$; (b) $(F_r,0,0)$, $\beta=0$; (c) $(0,F_{\theta},0)$, $-\alpha<\beta<\alpha$; (d) $(0,0,F_z)$, $\beta\neq0$; (e) $(0,0,F_z)$, $\beta=0$. The critical angles are
       $\delta_3=81.87^{\circ}$ when the eddy-type flow overcomes the non-oscillatory flow, see (a). The angles $\delta_1=146.31^{\circ}$ and $\delta_2=159.11^{\circ}$ are the critical angles for antisymmetric and symmetric flows. The red dots denote points when two poles overlap giving the decay $\mathcal{O}(1/L^N \ln{L})$.}
   \label{fig:40}
\end{figure}

We consider  the flow far  from the Stokeslet $(F_r,F_{\theta},F_z)$ situated at $(\rho,\beta,0)$ in cylindrical coordinates, i.e.~address the limit where $L^2=\hat{r}^2+\hat{z}^2\gg1$. The summary for the far-field decay of the flow in this case is given in Fig.~\ref{fig:40}.

When the point force is in the axial direction, $(0,0,F_z)$, the dominant flows are in the planes with constant $\theta$. They can be written in terms of different streamfunctions for acute \eqref{eq:a1}, right-angled \eqref{eq:z_right_f} and obtuse \eqref{eq:z_obtuse_f} cases. These flows decay respectively as $\mathcal{O}(L^{-3})$, $\mathcal{O}(L^{-3}\ln{L})$, $\mathcal{O}(L^{-1-\pi/(2\alpha)})$ in the limit $L \gg1$. In all cases, this spatial decay is too  fast to give any flux along the axial direction, and therefore $Q=0$.

 When the point force is in the azimuthal direction, the flow is dominated by the eddy-type flow which is  asymmetric about the bisectoral plane ($\theta=0$). The flow decays as $\mathcal{O}(L^{-1-\Im(\nu)})$ where $\nu$ satisfies $\sinh{2\alpha\nu}+\nu\sin{2\alpha}=0$ and it has a critical angle at which the eddies disappear of $2\alpha=\delta_1=146.31^{\circ}$, see equations \eqref{eq:azi1}~--~\eqref{eq:azi3}. When the point force is in the radial direction, the dominant flows can be in the planes with constant $\theta$, or fully three dimensional, see Fig.~\ref{fig:40}a-b, and the critical angle is  $2\alpha=\delta_3=81.87^{\circ}$. In the case when the leading flow is two dimensional, it can be written in terms of different streamfunctions for the acute \eqref{eq:r_acute_f}, right-angled \eqref{eq:r_right_f} and obtuse \eqref{eq:r_obtuse_f} cases. The spatial decays are $\mathcal{O}(L^{-4})$, $\mathcal{O}(L^{-4}\ln{L})$, $\mathcal{O}(L^{-2-\pi/(2\alpha)})$ respectively in the limit $L\gg1$. The eddy-type flow is given by the equations \eqref{eq:red1}~--~\eqref{eq:red2}.

%%%%%%%%%%%%%%%%%%%%%%%%%%%%%%%
%%%%%%%%%%%%%%%%%%%%%%%%%%%%%%%
%%%%%%%%%%%%%%%%%%%%%%%%%%%%%%%

 \begin{figure} [t]   
  \includegraphics[width=0.99\textwidth]{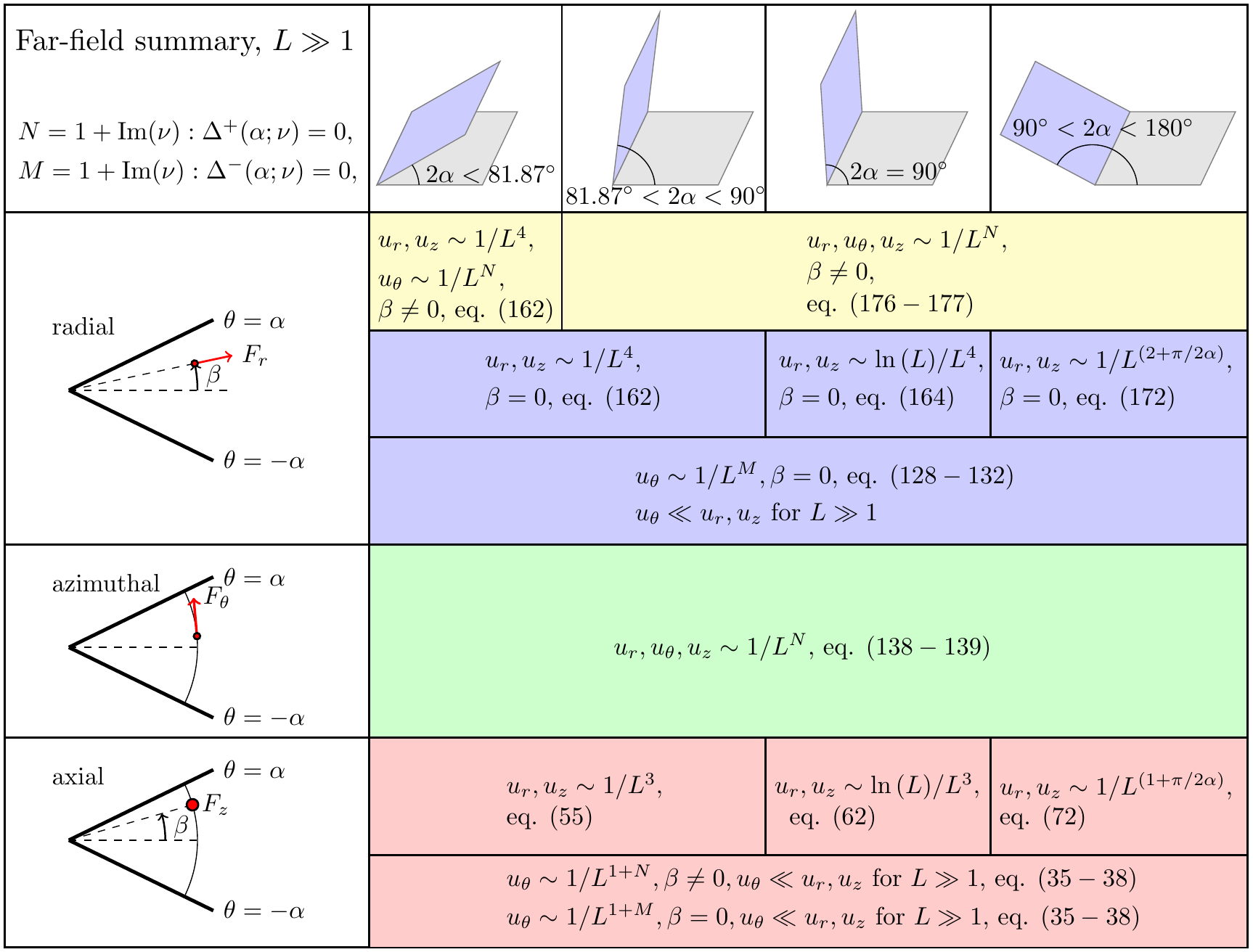}
      \caption{Summary of the flow decay in the far field ($L\gg1$) due to a point force  near a corner.}
  \label{fig:12}
\end{figure}

\section{Summary of leading-order flows}
In this section we present a short summary of the leading-order flows we obtained in the paper. If a point force $(F_r,F_\theta,F_z)$ is situated at a point $(\rho,\beta,0)$ then in a far field $\hat{r}^2+\hat{z}^2\gg1,~\hat{r}=r/\rho,~\hat{z}=r/\rho$ the flow is dominated by $F_z$ (Fig.~\ref{fig:12}) and leading-order flows can be written in terms of streamfunctions. Close to the corner $\hat{r}\ll1$ the flow is largest in the $z$ direction, i.e.~$u_z\gg u_r,u_\theta$, due to $F_z,F_r$.
The leading-order flow is written as a streamfunction $\Psi$ such that the flow field satisfies $\vec{u}=\nabla\cross(\Psi \vec{e_\theta})$ and the streamlines are defined by $r\Psi=\text{constant}$, i.e.
\begin{align}
u_r&=-\frac{\partial\Psi}{\partial z},~ u_{\theta}=0,~ u_z=\frac{1}{r}\frac{\partial(r\Psi)}{\partial r}\cdot
\end{align}
For acute corners $0<2\alpha<\pi/2$
\begin{align}
\Psi&=\frac{f_z(\theta)\hat{r}^3}{(\hat{r}^2+\hat{z}^2)^{5/2}},~\Psi=\frac{f_r(\theta)\hat{r}^3 \hat{z}}{(\hat{r}^2+\hat{z}^2)^{7/2}},~\hat{r}^2+\hat{z}^2\gg1
,\\
\Psi&=\frac{f_z(\theta)\hat{r}^3(4\hat{z}^2-1)}{4(1+\hat{z}^2)^{7/2}},~\Psi=\frac{f_r(\theta)\hat{r}^3\hat{z}}{(1+\hat{z}^2)^{7/2}},~\hat{r}\ll1,\\
f_z(\theta)&=\frac{F_z}{8\pi\mu}\frac{3\pi(1-\cos{2\beta}\sec{2\alpha})(1-\cos{2\theta}\sec{2\alpha})}{2(\tan{2\alpha}-2\alpha)},\\
f_r(\theta)&= \frac{F_r}{8\pi\mu}\frac{15\pi(1-\cos{2\beta}\sec{2\alpha})(1-\cos{2\theta}\sec{2\alpha})}{8(2\alpha-\tan{2\alpha})}\cdot
\end{align}

For obtuse corners $\pi/2<2\alpha<\pi$
\begin{align}
\Psi&=\frac{g_z(\theta)\hat{r}^{1+\pi/(2\alpha)}}{(\hat{r}^2+\hat{z}^2)^{1/2+\pi/(2\alpha)}},~\Psi=\frac{g_r(\theta)\hat{r}^{1+\pi/(2\alpha)}\hat{z}}{(\hat{r}^2+\hat{z}^2)^{3/2+\pi/(2\alpha)}},~\hat{r}^2+\hat{z}^2\gg1
,\\
\Psi&=\frac{g_z(\theta)\hat{r}^{1+\pi/(2\alpha)}[(4\alpha+\pi)\hat{z}^2+(2\alpha-\pi)]}{(4\alpha+\pi)(1+\hat{z}^2)^{3/2+\pi/(2\alpha)}},~\Psi=\frac{g_r(\theta)\hat{r}^{1+\pi/(2\alpha)}\hat{z}}{(1+\hat{z}^2)^{3/2+\pi/(2\alpha)}},~\hat{r}\ll1,\\
g_z(\theta)&=\frac{F_z}{8\pi\mu}\frac{16\pi^{1/2}\alpha\Gamma \left(\frac{3}{2}+\frac{\pi}{2\alpha}\right)\cos \left(\frac{\pi  \beta}{2 \alpha}\right) \cos\left(\frac{\pi  \theta}{2 \alpha}\right)}{(\alpha+\pi)(4 \alpha-\pi)\Gamma \left(1+\frac{\pi}{2\alpha}\right)},\\
g_r(\theta)&=\frac{F_r}{8\pi\mu}\frac{32\alpha\pi^{1/2} \Gamma{(\frac{3}{2}+\frac{\pi}{2\alpha})}\cos \left(\frac{\pi  \beta}{2 \alpha}\right) \cos\left(\frac{\pi  \theta}{2 \alpha}\right)}{(\pi^2-16\alpha^2)\Gamma{(1+\frac{\pi}{2\alpha})}}\cdot
\end{align}
For the salient corners with $\pi<2\alpha<2\pi$, the largest flow will also be due to $F_z$ in the far field and it is given by the same equations as for the obtuse corners, see details in Appendix \ref{salient}.
\section{Conclusion}
In this work, we analysed the flow in a corner geometry induced by a point force. We used complex analysis on an exact solution in the integral form in order to derive both the near- and far-field behaviours of the Stokeslet. 

The flow close to the corner, $\hat{r}\ll1$, is dominantly in the axial direction which was to be expected by the corner geometry since it is easier for the  fluid to move  parallel to the plates with  radial and azimuthal flows required to enforce incompressibility.  When the forcing is in the  azimuthal direction $(0,F_{\theta},0)$ the leading-order flow is two-dimensional in the radial-azimuthal plane. The summary of the flow close to the corner is shown in Fig.~\ref{fig:33}. We also predicted the size of the recirculation eddy  when there is a force acting in the axial direction, see Fig.~\ref{fig:5}.

The  decay rate in the far-field, $L^2=\hat{r}^2+\hat{z}^2\gg1$,  is displayed in Fig.~\ref{fig:40} with a  visual summary of all our results in Fig.~\ref{fig:12}. The flows due to the axial forcing, $(0,0,F_z)$, are dominant. These flows decay as $\mathcal{O}(L^{-3})$ for the angle is acute, $\mathcal{O}(L^{-3}\ln{L})$ when right-angled and $\mathcal{O}(L^{-1-\pi/(2\alpha)})$ when obtuse. They decay too fast to give any flux along the axial direction, and thus we always have $Q=\int_S u_z rdr d\theta=0$. The pressure field decays one order faster than the flow field. For the salient corners the leading-order flow follows the solution for the obtuse corners decaying as $\mathcal{O}(L^{-1-\pi/(2\alpha)})$, see Appendix \ref{salient}.

The flow field in the corner geometry is clearly very complex. However, we saw that it can be written as an asymptotic series in $\xi=(\hat{r}^2+\hat{z}^2+1)/(2\hat{r})$. There are therefore some similarities between the flow close to the corner and far away from the point force because $\xi\gg1$ in both cases. These flows match when $\hat{z}\gg1$, $\hat{r}\ll1$, i.e.~we are far away from the point force ($\hat{z}\gg1$), but close to the corner ($\hat{r}\ll1$). 

There are a number of applications for the asymptotic results derived in this work. The fluid decay far  from the point force can be used in numerical simulations in order to impose far-field boundary conditions. The Stokeslet approximation is also important in studying hydrodynamic coupling of Brownian particles. It was recently demonstrated that a confining surface can influence colloidal dynamics even at large separations, and that this coupling is accurately described by a leading-order Stokeslet approximation \cite{dufresne2000hydrodynamic}.  Our paper could thus be used to extend these results to the hydrodynamic coupling and coupled diffusion of colloidal particles near corners. 
From a biophysical point of view, our results could be exploited to tackle locomotion near corners. 
A recent study addressed  bacteria swimming in the corner region of a large channel modelled as two perpendicular sections of no-slip planes joined with a rounded corner  \cite{shum2015hydrodynamic}. That work could be extended using analytical results from our paper, with applications potentially ranging from 
 reducing biofouling in medical implants to controlling the transport of bacteria in microfluidic devices. 

\section*{Acknowledgements}
We thank Mark Hallworth, David Page-Croft, Trevor Parkin, Rob Raincock and Jamie Partridge from the GK Batchelor Laboratory for help with the experimental setup and Keith Moffatt for useful discussions. This work  was funded in part by the EPSRC (J. D.) and by an ERC Consolidator Grant (E. L.).  
%%%%%%%%%%%%%%%%%%%%%%%%%%%%%%%
%%%%%%%%%%%%%%%%%%%%%%%%%%%%%%%
%%%%%%%%%%%%%%%%%%%%%%%%%%%%%%%
\appendix
\section{Series expansion and Reynolds number}
\label{reynoldsn}
\subsection{A note on series expansions}
Suppose that the solution to the Stokes equations is written as a series expansion of the flow and pressure fields such that at every order the flow $\vec{u}^{(n)}$ is incompressible and satisfies no-slip boundary conditions on the walls, i.e.
\begin{align}
\vec{u}&=\sum_{n=1}^{\infty} \vec{u}^{(n)},~p=\sum_{n=1}^{\infty} p^{(n)},\\
\nabla\cdot\vec{u}^{(n)}&=0,~\vec{u}^{(n)}(\theta=\pm\alpha)=0,~n=1,2,3,4...
\end{align}
In the far field, one can write
\begin{align}
\vec{u}^{(n)}&=\vec{c_u}^{(n)}(\theta,\phi;\alpha,\beta)/L^n,\\ p^{(n)}&=c_p^{(n)}(\theta,\phi;\alpha,\beta)/L^{n+1},
\end{align}
where the spherical coordinates are denoted by $(L,\phi,\theta)$. In this case,  the incompressible  Stokes equations are satisfied for every $n$
\begin{align}
\nabla p^{(n)}=\mu\nabla^2\vec{u}^{(n)},
~\nabla\cdot\vec{u}^{(n)}&=0,~n=1,2,3,4...
\end{align}

This is however not the case for an expansion close to the corner, i.e.~$\hat{r}\ll1$, because the flow field in terms of harmonic functions scales as 
\begin{align}
u_r, u_{\theta}&\sim\phi_1,\phi_2,\frac{\phi_3}{r},\frac{\phi_4}{r},\\
u_z&\sim r\phi_1,r\phi_2,\phi_3,\phi_4,
\end{align}
but in order to have an incompressible flow one needs $u_r\sim u_{\theta}\sim r u_z$. We can still write the total flow as a superposition of flows satisfying incompressibility and boundary conditions, but only a full sum will satisfy the Stokes equations.
\subsection{Reynolds numbers}
We verify here  the validity of the Stokes approximation in the limits where the flow is measured close to the corner $r/\rho=\hat{r} \ll 1$ and far away from it $L=(\hat{r}^2+\hat{z}^2)^{1/2}\gg1$, where $(\hat{r},\theta,\hat{z})$ are cylindrical coordinates.

Suppose that the flow close to the 3D corner is of the order of $u_r=u_{\theta}=\mathcal{O}(\hat{r}^{N+1})$ and $u_z=\mathcal{O}(\hat{r}^{N})$, then Reynolds number is $\Re=\mathcal{O}(\hat{r}^{2N}/\hat{r}^{N+1})=\mathcal{O}(\hat{r}^{N-1})$. To have the finite flow close to the corner we require that $N>0$ and to have small Reynolds number $N>1$, i.e.~$u_r,u_\theta>\mathcal{O}(\hat{r}^2),~u_z>\mathcal{O}(\hat{r})$. However, if $u_r=u_{\theta}=\mathcal{O}(\hat{r}^{K})$ and $u_z=0$ which can happen close to the corner then $\Re=\mathcal{O}(\hat{r}^{2K-1}/\hat{r}^{K})=\mathcal{O}(\hat{r}^{K-1})$. And we require $K>1$, i.e.~$u_r,u_\theta>\mathcal{O}(\hat{r})$.

In the far field, suppose that the flow is of the order of $\mathcal{O}(1/L^M)$ then the Reynolds number is $\Re=\mathcal{O}(1/L^{M-1})$. For the flow to decay at infinity we require $M>0$ and to have small Reynolds number $M>1$. Therefore, the flow in the far field will have small Reynolds number if it decays faster than the Stokeslet in the free-space which behaves as $\mathcal{O}(1/L)$.

\section{Point force above an infinite wall at $z=0$}
\label{B}
We recall here the classical result of Blake for the flow due to point force  above an infinite wall. 
Consider a Stokeslet located at $\vec{y_0}=(0,0,h)$. Define $\vec{r}=\vec{x}-\vec{y_0}=(x,y,z-h)$ and $\vec{R}=(x,y,z+h)$. 
The solution for a Stokeslet in the vicinity of a stationary plane boundary is \cite{blake1974fundamental}
\begin{align}
 u_i = \frac{F_j}{8 \pi \mu}\left[\left(\frac{\delta_{ij}}{r}+\frac{r_i r_j}{r^3}\right)- \left(\frac{\delta_{ij}}{R}+\frac{R_i R_j}{R^3}\right)\right]+
 \\+\frac{F_j}{8 \pi \mu}\left [2h(\delta_{j \alpha}\delta_{\alpha k}-\delta_{j 3}\delta_{3 k}) \frac{\partial}{\partial R_k}\left\{\frac{h R_i}{R^3}-\left(\frac{\delta_{i3}}{R}+\frac{R_i R_3}{R^3}\right) \right\}\right].
\end{align}

For a Stokeslet  parallel to the wall we can  choose $\vec{F}=(F_1,0,0)$, leading to the flow
\begin{align}
 u_1 &=\frac{F_1}{8 \pi \mu}\left[\frac{12x^2 zh}{(x^2+y^2+z^2)^{5/2}}+\frac{6z^2 (-4x^2+y^2+z^2)h^2}{(x^2+y^2+z^2)^{7/2}}+\mathcal{O}(h^3)\right],\\
 u_2&=\frac{F_1}{8 \pi \mu}\left[\frac{12x y zh}{(x^2+y^2+z^2)^{5/2}}-\frac{30 x y z^2 h^2}{(x^2+y^2+z^2)^{7/2}}+\mathcal{O}(h^3)\right],\\
 u_3 &=\frac{F_1}{8 \pi \mu}\left[\frac{12x z^2 h}{(x^2+y^2+z^2)^{5/2}}-\frac{30 x  z^3 h^2}{(x^2+y^2+z^2)^{7/2}}+\mathcal{O}(h^3)\right].
\end{align}
For the flow due to the Stokeslet perpendicular to the wall  we have instead
\begin{align}
 u_1&=\frac{F_3}{8 \pi \mu}\left[\frac{-12h^2x z}{(x^2+y^2+z^2)^{5/2}}+\frac{30h^2 x  z^3}{(x^2+y^2+z^2)^{7/2}}+\mathcal{O}(h^3)\right],\\
 u_2&=\frac{F_3}{8 \pi \mu}\left[\frac{-12h^2y z}{(x^2+y^2+z^2)^{5/2}}+\frac{30h^2 y  z^3}{(x^2+y^2+z^2)^{7/2}}+\mathcal{O}(h^3)\right],\\
 u_3&=\frac{F_3}{8 \pi \mu}\left[\frac{-18h^2 z^2}{(x^2+y^2+z^2)^{5/2}}+\frac{30h^2   z^4}{(x^2+y^2+z^2)^{7/2}}+\mathcal{O}(h^3)\right].
\end{align}
%%%%%%%
\section{Salient corner with a point force parallel to both walls}
\label{salient}
\subsection{Exact solution}
\begin{figure}[t]       
 \includegraphics[width=0.75\textwidth]{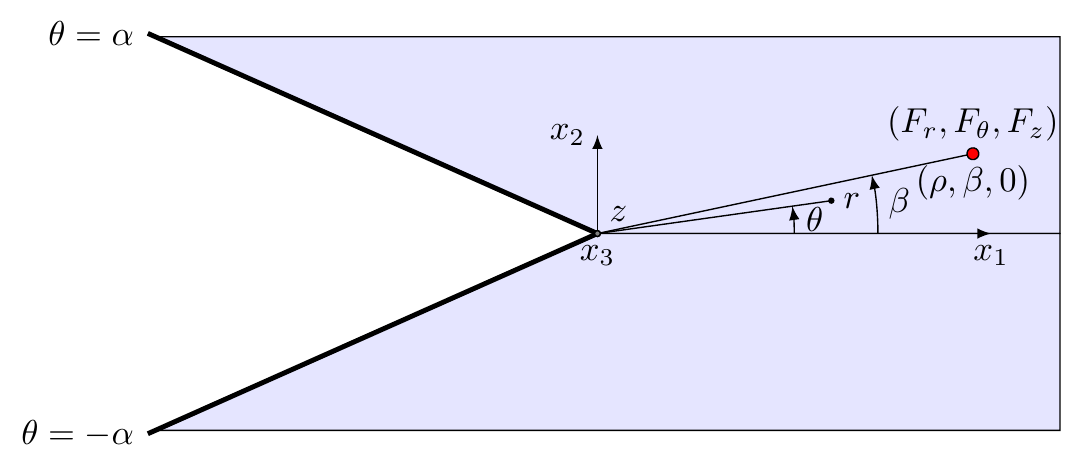}
       \caption{Geometry  for a point force $\vec{F}=(F_r,F_{\theta},F_z)$ located at $(r=\rho, \theta=\beta, z=0)$ and two plane walls at $\theta=\pm \alpha$. Cartesian co-ordinates are denoted  $(x_1,x_2,x_3)$ while  $(r,\theta,z)$ are  cylindrical co-ordinates.}
   \label{fig:salient1}
\end{figure}

%here
The solution for the flow due to a point force in a salient corner was derived by   Kim~\cite{kim1983effect}. We use the same notation as before, see Fig.~\ref{fig:salient1} for the geometry.
The solution is given by $\phi_n=\phi_n^{(1)}+\phi_n^{(2)}$ where $\phi_n^{(1)}$ is the solution in an unbounded domain and $\phi_n^{(2)}$ is the correction which is obtained using integral transforms. The correction $\phi_n^{(2)}$ is decomposed into $\phi_n^{(2)}=\phi_n^{(2)*}+\phi_n^{(2)**}$ where $\phi_n^{(2)*}$ is the solution similar to the acute and obtuse corner cases and $\phi_n^{(2)**}$ is the adjustment such that the flow is not singular the corner. In this case
\begin{align}
\phi_1^{(1)}&=\phi_2^{(1)}=\phi_4^{(1)}=0,~
\phi_3^{(1)}=c_z/R,\\
R^2&=\rho^2+r^2-2\rho r \cos(\theta-\beta)+z^2,~
c_z=-F_z/(8\pi\mu)=1,
\end{align}
and
\begin{align}
\label{eq:srf1}
\phi_1^{(2)}=&\phi_1^{(2)*}+\phi_1^{(2)**},\\
\phi_1^{(2)*}=&\frac{16}{\pi^2}\int_{0}^{\infty} S_{i\nu}g_1(\alpha,\beta,\theta,\nu)\sin{\alpha}\sinh{(\pi \nu)}\, d\nu,~\phi_1^{(2)**}=h(\hat{r},\hat{z})\cos{\left(1-\frac{\pi}{2\alpha}\right)\theta},\\
\label{eq:srf2}
\phi_2^{(2)}=&\phi_2^{(2)*}+\phi_2^{(2)**},\\
\phi_2^{(2)*}=&-\frac{16}{\pi^2}\int_{0}^{\infty} S_{i\nu}g_2(\alpha,\beta,\theta,\nu)\cos{\alpha}\sinh{(\pi \nu)}\, d\nu,~\phi_2^{(2)**}=h(\hat{r},\hat{z})\sin{\left(1-\frac{\pi}{2\alpha}\right)\theta},\\
\label{eq:srf3}
\phi_3^{(2)}=&-\frac{4}{\pi^2}\int_{0}^{\infty}I_{i\nu}
g_3(\alpha,\beta,\theta,\nu)\, d\nu,\\
\phi_4^{(2)}=&0,
\end{align}
where the functions $g_n$ are given by
\begin{align}
\label{eq:sf1}
g_1=&\frac{\sinh{(\theta \nu)\cosh{(\alpha \nu)q^A}}}{\Delta^-(\nu;\alpha) D^-(\nu;\alpha)}+\frac{\cosh{(\theta \nu)\sinh{(\alpha \nu)q^S}}}{\Delta^-(\nu;\alpha) D^+(\nu;\alpha)},\\ 
\label{eq:sf2}
g_2=&\frac{\sinh{(\theta \nu)\cosh{(\alpha \nu)q^S}}}{\Delta^+(\nu;\alpha) D^+(\nu;\alpha)}+\frac{\cosh{(\theta \nu)\sinh{(\alpha \nu)q^A}}}{\Delta^+(\nu;\alpha) D^-(\nu;\alpha)},\\ 
\label{eq:sf3}
g_3=&\frac{\sinh{(\theta \nu)\sinh{(\beta \nu)\sinh{[(\pi-\alpha)\nu]}}}}{\sinh{(\alpha \nu)}}+\frac{\cosh{(\theta \nu)\cosh{(\beta \nu)\cosh{[(\pi-\alpha)\nu]}}}}{\cosh{(\alpha \nu)}},\\
h(\hat{r},\hat{z})=&\frac{8}{\pi(4\alpha-\pi)}\sin{\frac{\pi^2}{2\alpha}}\cos{\frac{\pi\beta}{2\alpha}}S_{i\nu^{*}},~\nu^{*}=\left(\frac{\pi}{2\alpha}-1\right)i.
\end{align}
with $I_{i\nu},~S_{i\nu},~\xi,~q^A,~q^S, D^{\pm}(\nu;\alpha),~\Delta^{\pm}(\nu;\alpha)$ as defined in the equations before. The functions $f_n$ given in the equation~\ref{eq:f1} and $g_n$ only differ by $\Delta^{\pm}(\nu;\alpha)$.
\subsection{Leading-order flow}
The leading-order flow in the far field is following the same branch as for the obtuse corner.  
\begin{figure}[t!]
 \includegraphics[width=0.49\textwidth]{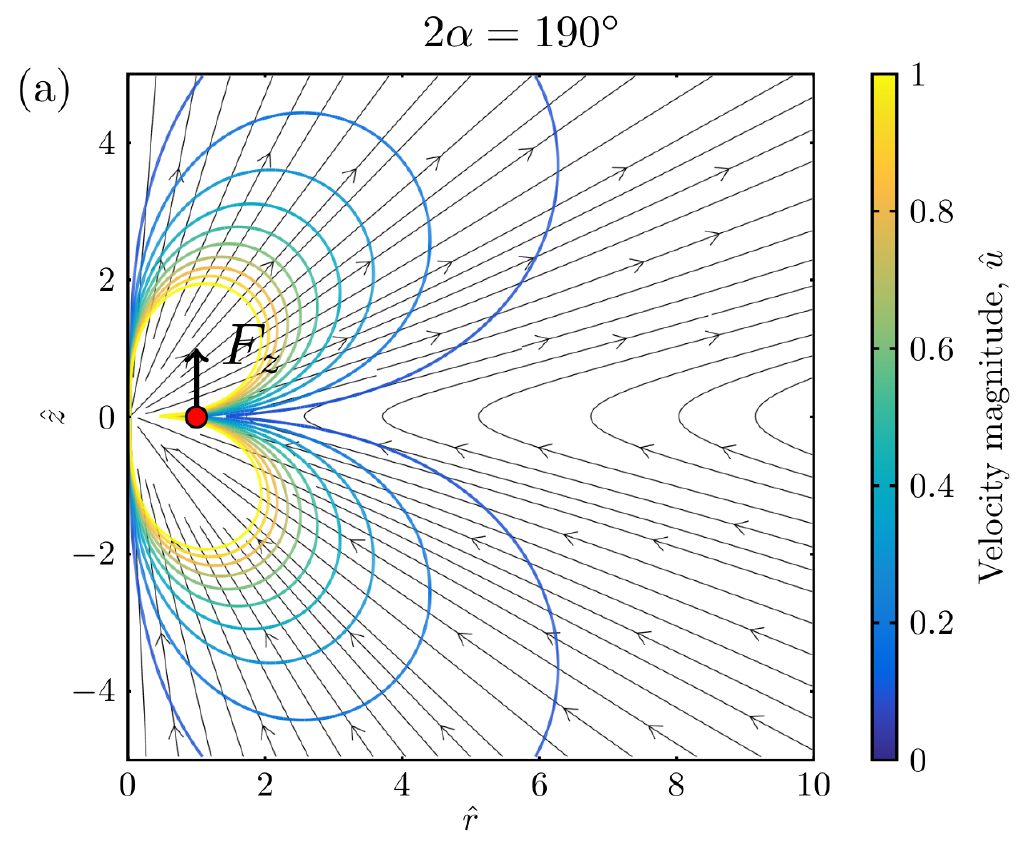}
 \includegraphics[width=0.49\textwidth]{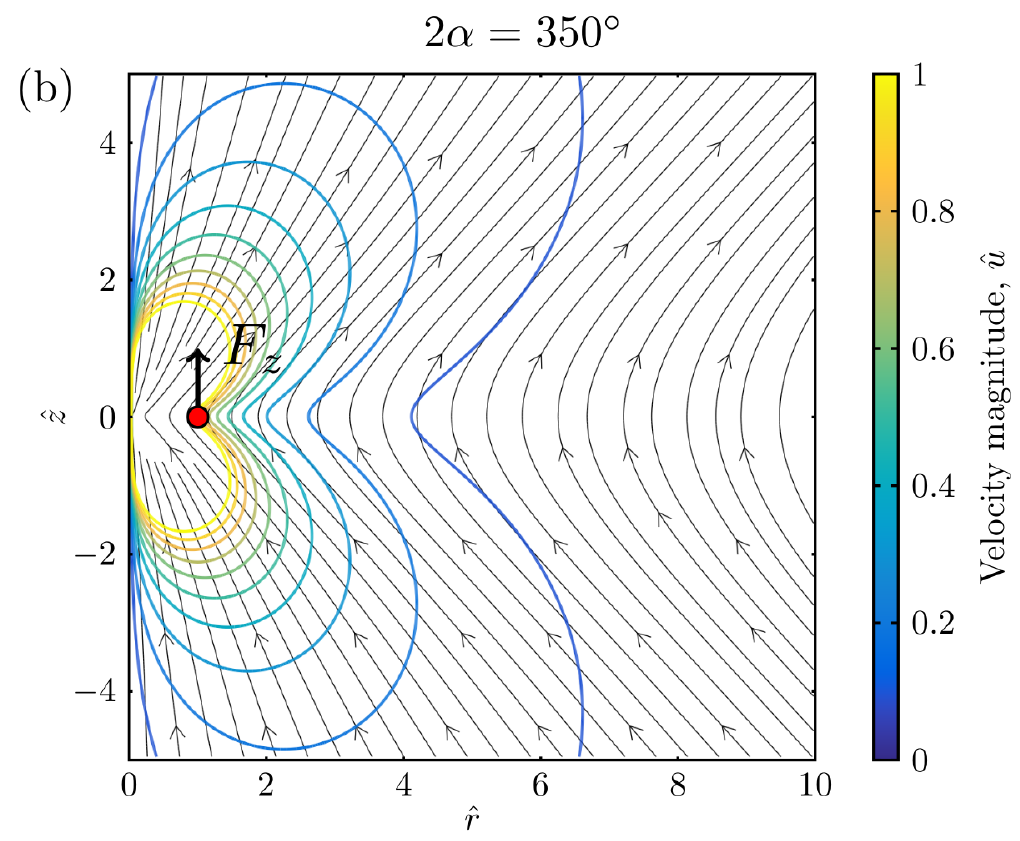}  
              \caption{Streamlines and contours of velocity magnitude, $\hat{u}=8\pi\mu\rho u/F_z=\hat{u}_r^2+\hat{u}_z^2$,  for the leading-order 2D flow and an axial point force in a salient corner. (a) $2\alpha=190^{\circ}$, (b) $2\alpha=350^{\circ}$. For all cases $\theta=\beta=0$. The flow is valid in the far field, i.e.~for $\hat{r}^2+\hat{z}^2\gg1$.}
   \label{fig:sal_2}
\end{figure}
It is thus characterise by the same streamfunction
\begin{align}
\Psi&=\frac{\rho f(\alpha,\beta,\theta)\hat{r}^{1+\pi/(2\alpha)}}{(\hat{r}^2+\hat{z}^2)^{1/2+\pi/(2\alpha)}},
~\hat{r}^2+\hat{z}^2\gg1,\\
f(\alpha,\beta,\theta)&=\frac{F_z}{8\pi\mu\rho}\frac{16\pi^{1/2}\alpha\Gamma \left(\frac{3}{2}+\frac{\pi}{2\alpha}\right)\cos \left(\frac{\pi  \beta}{2 \alpha}\right) \cos\left(\frac{\pi  \theta}{2 \alpha}\right)}{(\alpha+\pi)(4 \alpha-\pi)\Gamma \left(1+\frac{\pi}{2\alpha}\right)}\cdot
\end{align}
\begin{figure}[t!]      
	\includegraphics[width=0.59\textwidth]{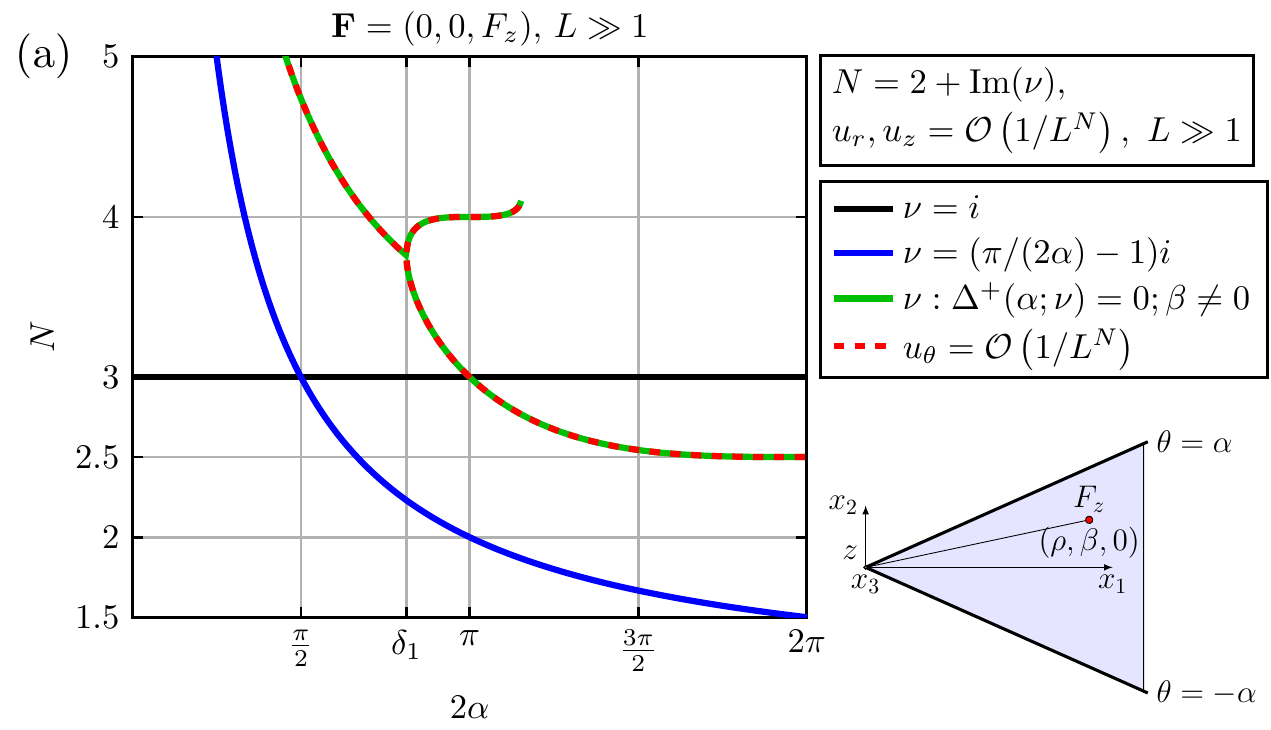}
	\includegraphics[width=0.39\textwidth]{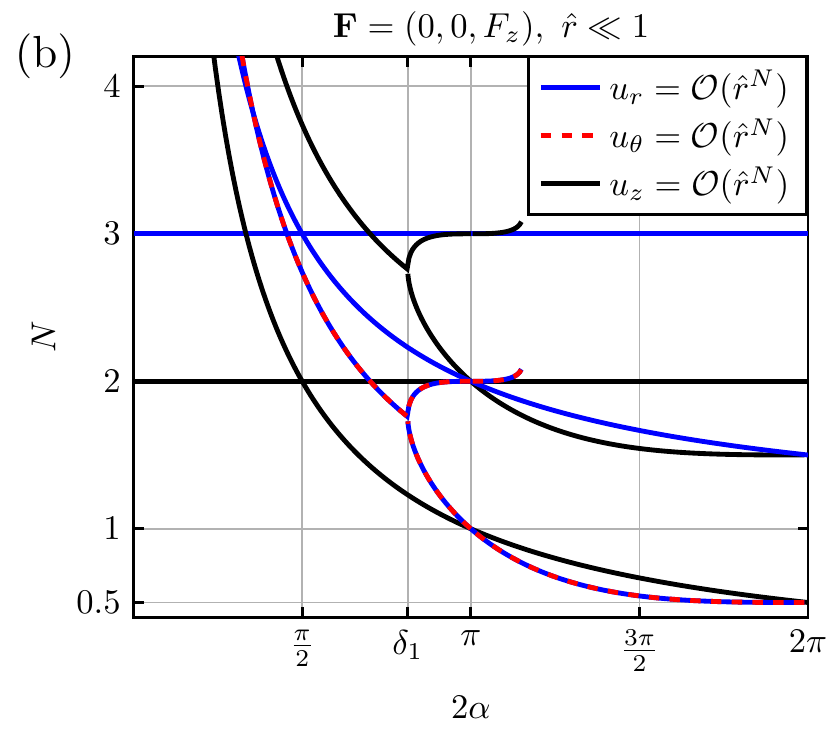}
	\caption{Decomposition of the full flow due to a point force in the axial direction, $\vec{F}=(0,0,F_z)$. (a) Velocity components  $u_r,u_{\theta},u_z$ at large distance from the point force, $L\gg1$, decay as $\mathcal{O}(1/L^{2+\Im(\nu)})+\mathcal{O}(1/L^{4+\Im(\nu)})+...$ where   $\Im(\nu)$ is the imaginary part of the pole $\nu$. The decay power $N$ is plotted against the full corner angle, $2\alpha$.  (b) Flow decay close to the corner $\hat{r}\ll1$. We show bifurcating branches only for antisymmetric flow and the top branch stops because $\Re \nu\neq0$ from that point.}
	\label{fig:sal_3}
\end{figure}
This flow is incompressible and it satisfies the no-slip boundary condition at $\theta=\pm \alpha$. The streamlines and velocity magnitude contours are shown in  Fig.~\ref{fig:sal_2}a-b in the case where $\theta=0$, $\beta=0$. 
Notice that in this case there is no recirculation of the flow and the streamlines are almost straight lines. The flow structure  depends on the corner angle, $2\alpha$. The flow is quasi-two-dimensional, because $u_r,u_z\gg u_{\theta}$, and it decays algebraically as $\mathcal{O}(1/L^{1+\pi/(2\alpha)})$ for $L=(\hat{r}^2+\hat{z}^2)^{1/2}\gg1$. The  power decay rate of the flow is $N=1+\pi/(2\alpha)$, see Fig.~\ref{fig:sal_3}a. It is equal to $N=2$ at $2\alpha=\pi$ and it approaches $N=3/2$ as $2\alpha$ tends to the full angle $2\pi$ (the Stokes approximation is still fully valid).

Close to the corner, the extended solutions are shown in Fig.~\ref{fig:sal_3}b. We see that the flow decays slower than $\mathcal{O}(\hat{r})$ for angles $2\alpha>\pi$, but this means that the Stokes approximation will break down, see Appendix \ref{reynoldsn} and the  inertial terms will be important close to the corner no matter how small the point force is. Note that this is not the case for the 2D corner flows because the diffusive terms $\partial^2 u_r/\partial z^2$ and $\partial^2 u_\theta/\partial z^2$ in the Laplacian $\mu\nabla^2\vec{u}$ are exactly zero there.

\bibliographystyle{unsrt}
\bibliography{justas}

\end{document}